\documentclass{emulateapj}
\usepackage{soul}
\usepackage{amssymb}
\usepackage{amsmath}
\usepackage{mathtools}
\usepackage{apjfonts}

\newcommand{\p}{\ensuremath{\mem{p}}}
\newcommand{\czw}{\ensuremath{^{12}\mem{C}}}

\newcommand{\ndr}{\ensuremath{^{13}\mem{N}}}
\newcommand{\ose}{\ensuremath{^{16}\mem{O}}}
\newcommand{\komma}{\mathrm{ \hspace*{0.1cm},}}
\newcommand{\punkt}{\mathrm{ \hspace*{0.1cm}.}}
\newcommand{\apleq}{\ensuremath{\stackrel{<}{_\sim}}}
\newcommand{\apgeq}{\ensuremath{\stackrel{>}{_\sim}}}
\newcommand{\ipr}{\mbox{$i$-process}}

\newcommand{\mem}[1]{\ensuremath{\mathrm{ #1}}}
\newcommand{\glt}[1]{Eq.\,(\ref{#1})}
\newcommand{\glp}[1]{(Eq.\,\ref{#1})}
\newcommand{\kap}[1]{Sect.\,\ref{#1}}
\newcommand{\abb}[1]{Fig.\,\ref{#1}}
\newcommand{\tab}[1]{Table\,\ref{#1}}
\newcommand{\stiffness}{\ensuremath{\mathcal{S}}}
\newcommand{\lsun}{\ensuremath{\, L_\odot}}
\newcommand{\msun}{\ensuremath{\, M_\odot}}
\newcommand{\jahre}{\ensuremath{\, \mathrm{yr}}}
\newcommand{\natlog}[2]{\ensuremath{#1\times 10^{#2}}} 
\newcommand{\hevi}{\ensuremath{^{4}\mem{He}}}

\slugcomment{Submitted: July 15, 2013; Accepted: August 12, 2014}

\usepackage{natbib}

\shorttitle{Shell convection entrainment}
\shortauthors{Woodward etal.}

\begin{document}

\title{Hydrodynamic simulations of 
H entrainment at the top of  He-shell flash convection}

\author{Paul R. Woodward\altaffilmark{1},
Falk Herwig\altaffilmark{2}, 
Pei-Hung Lin\altaffilmark{1}
}
\altaffiltext{1}{LCSE \& Department of Astronomy, University of
  Minnesota, Minneapolis, MN 55455, USA \email{paul@lcse.umn.edu}}
\altaffiltext{2}{Department of Physics \& Astronomy, University of
  Victoria, Victoria, BC V8P5C2, Canada \email{fherwig@uvic.ca}}

\begin{abstract}
  We present the first 3-dimensional, fully compressible gas-dynamics
  simulations in $4\pi$ geometry of He-shell flash convection with
  proton-rich fuel entrainment at the upper boundary. This work is
  motivated by the insufficiently understood observed consequences of
  the H-ingestion flash in post-AGB stars (Sakurai's object) and
  metal-poor AGB stars. Our investigation is focused on the
  entrainment process at the top convection boundary and on the
  subsequent advection of H-rich material into deeper layers, and we
  therefore ignore the burning of the proton-rich fuel in this
  study. We find that, for our deep convection zone, coherent
  convective motions of near global scale appear to dominate the flow.
  At the top boundary convective shear flows are stable against
  Kelvin-Helmholtz instabilities. However, such shear instabilities
  are induced by the boundary-layer separation in large-scale,
  opposing flows. This links the global nature of thick shell
  convection with the entrainment process. We establish the
  quantitative dependence of the entrainment rate on grid
  resolution. With our numerical technique simulations
  with $1024^3$ cells or more are required to reach a numerical
  fidelity appropriate for this problem. However, only the result from
  the $1536^3$ simulation provides a clear indication that we approach
  convergence with regard to the entrainment rate. Our results
  demonstrate that our method, which is described in detail, can
  provide quantitative results related to entrainment and convective
  boundary mixing in deep stellar interior environments with very
  stiff convective boundaries. For the representative case we study in
  detail, we find an entrainment rate of $4.38 \pm 1.48 \times
  10^{-13}\msun\mathrm{/s}$.
\end{abstract}

\keywords{ stars: AGB and post-AGB, evolution, interior 
       --- physical data and processes: turbulence, hydrodynamics, convection}

\section{Introduction}
\label{sec:intro}  Convection in stars is the most important
  mixing process. It is essential for heat transport and mixing of
  elements. Therefore, convection is an essential ingredient in the
  evolution of stars and in the formation of the elements in stars and
  stellar explosions \citep{woosley:02,herwig:04c}. The treatment of
  convective boundaries is a key uncertainty in present stellar
  evolution calculations. In particular, mixing at the rather stiff
  (see below) boundaries of convection in the deep stellar interior
  has been related to numerous observational properties of
  stars. Determining reliable quantitative mixing properties of
  convection boundaries remains an unsolved problem. Overshooting at
  the boundary of H-core convection has been investigated for the past
  40 years
  \citep[e.g.][]{Maeder:1976wg,Schaller:1992vq,ventura:98,Deupree:2000jd,VandenBerg:2006cn}. Evidence
  for convective boundary mixing (CBM) at the bottom of the solar
  convection zone has been presented based on helioseismology
  \citep{ChristensenDalsgaard:2011fr} and hydrodynamic simulations
  \citep[e.g.][]{freytag:96,Rogers:2006ks}. For thermal-pulse
  asymptotic giant branch (AGB) stars CBM is important for the
  occurrence of the third dredge-up \citep[for
  example][]{herwig:97,Herwig:2000ua,Mowlavi:99,Weiss:2009kh}, the
  reproduction of observed properties of AGB stars
  \citep{Karakas:2010dk}, and in post-AGB stars
  \citep{Herwig:1999uf,MillerBertolami:2006dr,Werner:2006bf}.  More
  recently it has been shown that CBM can explain the observed
  enrichment of nova ejecta \citep{Denissenkov:2012cu}. Furthermore,
  CBM at the bottom of C- and Ne-burning shells in super-AGB stars and
  transition mass objects is critical for the progenitors of supernova
  Ia and possible core-collapse supernova
  \citep{Denissenkov:2013dd,Jones:2013iw}.

 In this paper we present results for simulations of He-shell
  flash convection in AGB stars, including convective boundary mixing
  at the upper boundary. We would like to
  investigate the global properties of convective flows in this type
  of convection, and the convective boundary mixing in general for a
  case in which the boundary is very stiff. Specifically we focus here
upon the entrainment of hydrogen-rich material into the convection
zone above a helium burning shell.  Such situations have been
addressed by \citet{Mocak:2010kj}. \citet{mocak:11a} and
\citet{stancliffe:11}, and similar entrainment events by the
convection zone above an oxygen burning shell and the H-burning core
convection have been treated by \citet{meakin:07b}.
\citet{viallet:13} also discuss such simulation results with a goal of
producing improved mixing length models for use in 1-D stellar
evolution simulations.  Our own work has to this point focused on the
entrainment problem
\citep{herwig:06a,herwig:07a,woodward:08a,woodward:08b}, separated
from the complicating factors of nuclear burning of the ingested fuel.
We wish to establish that our computational methods are capable of
accurately simulating the entrainment process before we add in the
nuclear burning and back reaction on the flow of the energy liberated
in that process.  Nevertheless, we have produced a detailed analysis
using 1-D stellar evolution techniques with a parameterized treatment
of convective boundary mixing that indicates the dynamical behavior
that a proper 3-D analysis of hydrogen ingestion and burning is likely
to yield in the specific, well observed case of Sakurai's object
\citep{herwig:10a}.  A follow-on to this article will present results
obtained using our PPMstar code for that particular case
\citep[see][for preliminary results]{Herwig:2013vf}.

The present paper is intended to describe the challenges that
convective boundary mixing poses to numerical simulation, set out in
the context of a representative hydrogen ingestion problem.  The upper
convective boundary of a He-shell flash convection zone about to make
contact with the H-rich layer above is much stiffer compared, for
example, to the cases investigated by \citet{Meakin:2007dj}, and
consequently we find a much smaller entrainment rate in this
situation.  We describe how we meet the challenge of converged
entrainment simulations for a very stiff convective boundary by a
combination of advanced numerical techniques and high grid resolution.
We present evidence that for the aspect of these problems that we
judge to be the most critical -- the hydrogen entrainment rate at the
upper convective boundary -- our simulations are able to produce
results that converge upon grid refinement.  Because this ingestion is
expected to have dramatic effects upon the star by releasing more
energy than the shell-burning source producing the convection that
causes it, it is very important that it be computed accurately.  We
cannot comment upon the accuracy of simulations by others, but we note
that the results that are obtained vary greatly.  In work by others,
the conditions of the convective boundary mixing differ substantially
from our own and from each other, so that direct comparisons are not
possible.  Nevertheless, we have observed in our own work that the
all-important entrainment rate depends strongly upon the grid
resolution until a sufficiently fine grid is used.  All investigators
have of course reported results obtained on different grids.
\citet{stancliffe:11} find luminosities that differ by 3 orders of
magnitude for simulations of the same problem performed on grids
differing in linear resolution by a factor of 2. \citet{mocak:11a}
present results for only a single 3-D grid. It is arranged in a
$45\deg$ wedge geometry that precludes the development of large-scale
or global modes. They establish their grid requirement from
simulations performed at different resolutions in 2-D.
\citet{viallet:13} present results for a series of simulations on
progressively finer grids, which cover a sector of the convection zone
above an oxygen burning shell.  They state that they do not observe
convergence under grid refinement in the mean fields in the narrow
region of steep gradients at the base of their oxygen burning shell.
Because we will base our subsequent studies of hydrogen ingestion
flash events upon our PPMstar code, discussed here, and because
previously reported results establish the difficulty of accurate
simulation of such events, our goals in this paper are twofold. In
addition to presenting our quantitative results on entrainment for our
particular problem and the general properties of He-shell flash
convection in $4\pi$ geometry, we set out in this paper our
computational techniques and the evidence that they are equal to the
task we have set for ourselves in our subsequent work.

Regions in stars are convectively unstable if the radiative energy
transport is less efficient than convective transport
\citep{kippenhahn:90}. The Schwarzschild condition determines
instability against convection if the radiative temperature gradient
is larger than the adiabatic temperature gradient. 
The radiative temperature
gradient is given by
\begin{equation}
  \label{eq:Trad}
 - \frac{dT}{dr} \Biggm \vert_\mem{rad} = \frac{3}{16 \pi a c}\frac{\kappa L \rho}{r^2 T^3}
\end{equation}
where $\kappa$ is the Rosseland mean opacity or the electron
conduction coefficient, $L$ is the luminosity, $T$ the temperature,
$r$ the radius, $\rho$ the density, $a$ is the radiation-density
constant and $c$ the speed of light. 

\glt{eq:Trad} shows that convective instability can be caused by
either large opacity or by large luminosity \citep[see, for example,
Fig.\,3 in][]{paxton:11}. The former is the case for surface
convection, for example in the shallow surface convection of A-type
stars and white dwarfs \citep{freytag:96}, in the sun
\citep[e.g.][]{stein:98,misch:00,robinson:03,miesch:08}, or in the
deep convective envelopes of giant stars
\citep[][]{porter:00a,porter:00b,freytag:08}. Convection in the deep
interior is usually driven by high luminosity, as for example in
He-shell flashes in thermal pulse Asymptotic Giant Branch stars
\citep{herwig:06a,herwig:07a}, in oxygen and carbon shell burning in
the advanced evolution phases of massive stars
\citep{asida:00,meakin:07b} or in He-core flashes
\citep{mocak:08,mocak:09}.

Convection in the deep interior at high densities is usually very
efficient and the temperature gradient is therefore nearly
adiabatic. However, in surface convection, where in some cases a
significant fraction of the energy is transported by radiation, the
actual temperature stratification is often super-adiabatic, reflecting
the inefficiency of convection. Along with this difference between
near-surface and envelope convection on the one side, and deep interior
convection on the other side goes generally speaking a marked
difference in stiffness of the convective boundary. This stiffness
reflects the ratio of the degree of acceleration in the convectively
unstable region to the degree of deceleration in the stable
region. 

The stratification can be approximated piecewise by polytropes with
\begin{equation}
  \label{eq:poly}
P=K_\mem{s} \rho^{\gamma_\mem{s}}  
\end{equation}
 where $K_\mem{s}$ is a constant,
$P$ and $\rho$ are the pressure and the density and $\gamma_\mem{s}$
is given by the polytropic index
$n_\mem{s}=\frac{1}{\gamma_\mem{s}-1}$.  The stiffness \stiffness, or
relative stability, can then be expressed in terms of the polytropic
index for the adiabatic stratification ($n_\mem{ad}=3/2$ for
$\gamma_\mem{ad}=\frac{5}{3}$), the convectively stable ($n_\mem{1} >
n_\mem{ad}$) and the convectively unstable ($n_\mem{2}< n_\mem{ad}$)
stratification by \citep{hurlburt:94}:
\begin{equation}
  \label{eq:stiffness}
  \stiffness = - \frac{n_\mem{ad}-n_\mem{1}}{n_\mem{ad}-n_\mem{2}} \punkt
\end{equation}
The stability of the stratification of a monatomic ideal gas in terms
of the polytropic index $n_\mem{s}$ follows from the definition of the
entropy
\begin{equation}
  \label{eq:entropy}
  S = c_\mem{v} \log (p/\rho^\gamma) + \mem{constant} 
\end{equation} considering that the entropy gradient
$\frac{\mem{d}S}{\mem{d}r}$ is zero for an adiabatic stratification,
$>0$ for a stable (subadiabatic) and $<0$ for an unstable
(superadiabatic) stratification. 

As may be expected, the degree of penetration and overshooting of flows across the
convective boundary and the associated degree of mixing of
thermodynamic quantities and species concentrations is larger for smaller stiffness
\citep{brummel:02,rogers:05}. At small stiffness, such as in shallow
surface convection \citep{freytag:96}, coherent convective systems will
cross the convective boundary and only start to decelerate on the
stable side due to buoyancy effects.

In the deep stellar interior convection zones, high density implies
effective convective transport. The denominator in \glt{eq:stiffness}
is small ($n_\mem{ad}-n_\mem{2}\ll 1$) and therefore $\stiffness \gg
1$. In this situation the convective flows nearing the boundary are
``feeling'' via the building pressure the almost impenetrable boundary
already from a distance \citep{rogers:05}, and in order to obey mass
continuity will start to turn around already inside the convectively
unstable layer.  A requirement for this behavior is also of course
convective velocities well below the local speed of sound.  In the
He-shell flash convection situations of interest to us, this
requirement is certainly fulfilled, at least in the absence of
positive feedback from the burning of ingested hydrogen.  Even while
no coherent convective systems cross the convective boundary, mixing
at the boundary will still occur.  In this case the shear flows
induced by turning-around convection flows will induce entrainment
\citep{meakin:07b}, mostly via the Kelvin-Helmholtz instability in the
boundary layer.

In summary, a range of hydrodynamic processes and instabilities
contribute to mixing and entrainment at convective boundaries
\citep[see][for a similar but more detailed
  discussion]{viallet:13}. For deep stellar interior mixing the
CBM layer is very small.  Another example
of this type of CBM is found in the context of the
thermonuclear-runaway-driven convection in novae
\citep{casanova:11a,denissenkov:12}. While averaged properties
pertaining to the inside of the convection zone seem to converge
already at modest numerical resolutions
\citep[e.g.][]{herwig:06a,viallet:13} the narrow convective boundary
layers constitute a challenge in terms of reaching numerical
convergence \citep{viallet:13}.

A narrow and very stiff convective boundary layer can be found at the
top of the He-shell flash convection zone in AGB and post-AGB stars at
the moment when they make contact with the H-rich material above. Such
a situation has been encountered in stellar evolution simulations of
AGB stars with very low metal content with [Fe/H]$\leq-2$
\citep[e.g.][]{fujimoto:00}, post-AGB stars \citep{iben:95b}. It
leads to violent convective-reactive events involving rapid nuclear
energy release from $\czw + \p$ reactions in convective regions with
large mass fractions of primary \czw. Numerous investigations of this
scenario have been subsequently carried out in the spherically
symmetric stellar evolution framework
\citep[e.g.][]{Herwig:1999uf,lawlor:03,Herwig:2003wk,iwamoto:04,campbell:08,Lau:2009bd,Suda:2010em,Cristallo:2009cu,campbell:10,Suda:2010em}. However,
protons ingested into the He-shell flash convection zone on the
convective time scale will be advected downward to increasing
temperatures until the nuclear reaction time scale of the
$\czw(\p,\gamma)\ndr$ rate equals the hydrodynamic flow time
scale. The resulting significant energy feedback on the convective
flow time scale violates the assumptions of the mixing length theory,
concerning time and spatial averages, and a treatment in spherical
symmetry adopted in one-dimensional stellar evolution codes may become
problematic. Evidence for the failure of mixing-length theory based
stellar evolution models arises from the analysis of the light-curve
\citep{herwig:01a} and the highly non-solar, anomalous abundance
distribution \citep{herwig:10a} in the H-ingestion, very-late thermal
pulse post-AGB object Sakurai's object.

This post-AGB star belongs to the $\sim 20\%$ of all single post-AGB
stars that suffer a final He-shell flash after they have already left
the AGB. Among those, the very-late thermal pulse stars
\citep{werner:06}, but especially Sakurai's object, provide a unique
opportunity to investigate the H-\czw-combustion regime in stellar
evolution, because it is a nearby object that has been observed in
real-time \citep[][and references therein]{duerbeck:96,hajduk:05} so that
detailed information about the light curve is available. Even more
importantly, abundance information of the post-flash evolution has
been obtained at a time after the outburst \citep{asplund:99} and this
has been shown to provide strong constraints on the hydrodynamic
processes of combined convection and rapid nuclear burning, i.e.\ the
physics of H-\czw-combustion in stellar evolution
\citep{herwig:10a}. We are therefore fortunate to have a very powerful
validation case for simulations of H-\czw\ combustion, and therefore
our attention is directed to carefully check if our simulations pass
this validation test.

As mentioned earlier, hydrodynamic simulation results of H-ingestion
presented so far are either unlikely to be converged and therefore
must be considered questionable \citep{stancliffe:11}, or it is not
clear if they indeed constitute a convective-reactive H-ingestion
case. We suspect the latter to be the case for the simulations
presented by \citet{mocak:11a} who adopt an initial setup with an
artificially shifted H-profile. The emerging H-burning convection
features a long nuclear time scale for the $\czw(\p,\gamma)\ndr$
reaction due to the relatively low temperature in that zone. The
resulting Damk\"ohler number is $D<0.01$, much lower than the value
around unity that characterizes a convective-reactive regime.  We may
therefore conclude that previous work on convective boundary mixing as
it pertains to the H-ingestion events, which are our interest in this
paper, provides only limited guidance.

Our strategy is to attack the problem of H-ingestion into He-shell
flash convection zones by constructing accurate 3-D hydrodynamic
simulations.  To verify these simulations we show that they converge
upon grid refinement in important respects, such as the entrainment
rate studied here.  In the future we will take advantage of the validation
opportunity provided by Sakurai's object to compare simulation results
with observations.  Our goal at this point is not to develop models of
mixing and burning in H-ingestion that can be applied in stellar
evolution calculations to all possible realizations of the H-ingestion
regime. As a first step to improve upon the rather qualitative
simulations presented in \citet{herwig:10a}, we would rather like to
investigate specifically the entrainment process at the upper boundary
of He-shell flash convection that is just about to connect to the
H-rich envelope.

In H-ingestion cases, like in Sakurai's object, the onset of the
He-shell flash expands the shell layers before the upper boundary
reaches the H-rich layers. Typically the aspect ratio $\Delta r /
r_\mathrm{bot} > 1$, where $\Delta r$ is the geometric width of the
convection zone and $r_\mathrm{bot}$ is the radius of the bottom of
the convection zone. We found in our previous studies
\citep{porter:00a,porter:00b} that the
convective flows are dominated by large scale and global modes for
such large aspect ratios. We therefore consider it necessary to choose
a full $4\pi$ geometry for our simulations. One may also expect that
the interaction of H-enriched downflows with high-temperature
conditions and the ensuing nuclear burning will depend on the
entrainment process and vice versa.

For this study we set up our initial conditions to produce shell
convection that is typical of the conditions encountered during
initial mixing of H-rich material from above the He-shell flash
convection zone into the material of the convection zone. We expect
the entrainment rate that develops in such a simulation to depend upon
this initial base state, on the heating rate corresponding to the
He-burning driving the convection, as well as, of course, on the grid
resolution of the simulation. We might expect the size of the velocity
of the fluid of the convection zone near the location of H-rich
material entrainment to play an important role in determining the
entrainment rate. Therefore, we also investigate the dependence of
these velocities in our simulations upon the grid resolution. 

With this study we wish to introduce our updated star simulation
methodology and establish that simulations of this type can provide
good estimates of the entrainment rate at affordable cost before we
proceed to augment these simulations with a treatment of the burning
of the ingested H-rich material and the back reactions upon the flow
dynamics that this burning produces \citep[see][for preliminary
  results]{Herwig:2013vf}.

This paper is organized as follows. In \kap{sec:method} we will
briefly describe the numerical techniques of the PPMstar hydrodynamics
code and the initial setup used for the simulations presented
here. The properties of the hydrodynamic flow and results for
entrainment as a function of grid size will be described in
\kap{sec:results}. We close with a discussion (\kap{sec:concl}). The
results presented in this paper depend critically on the deployment of
the PPB multifluid advection scheme, which is described in detail in
the appendix (\kap{sec:app_method}).
\begin{figure}[t]         
   \includegraphics[width=0.5\textwidth]{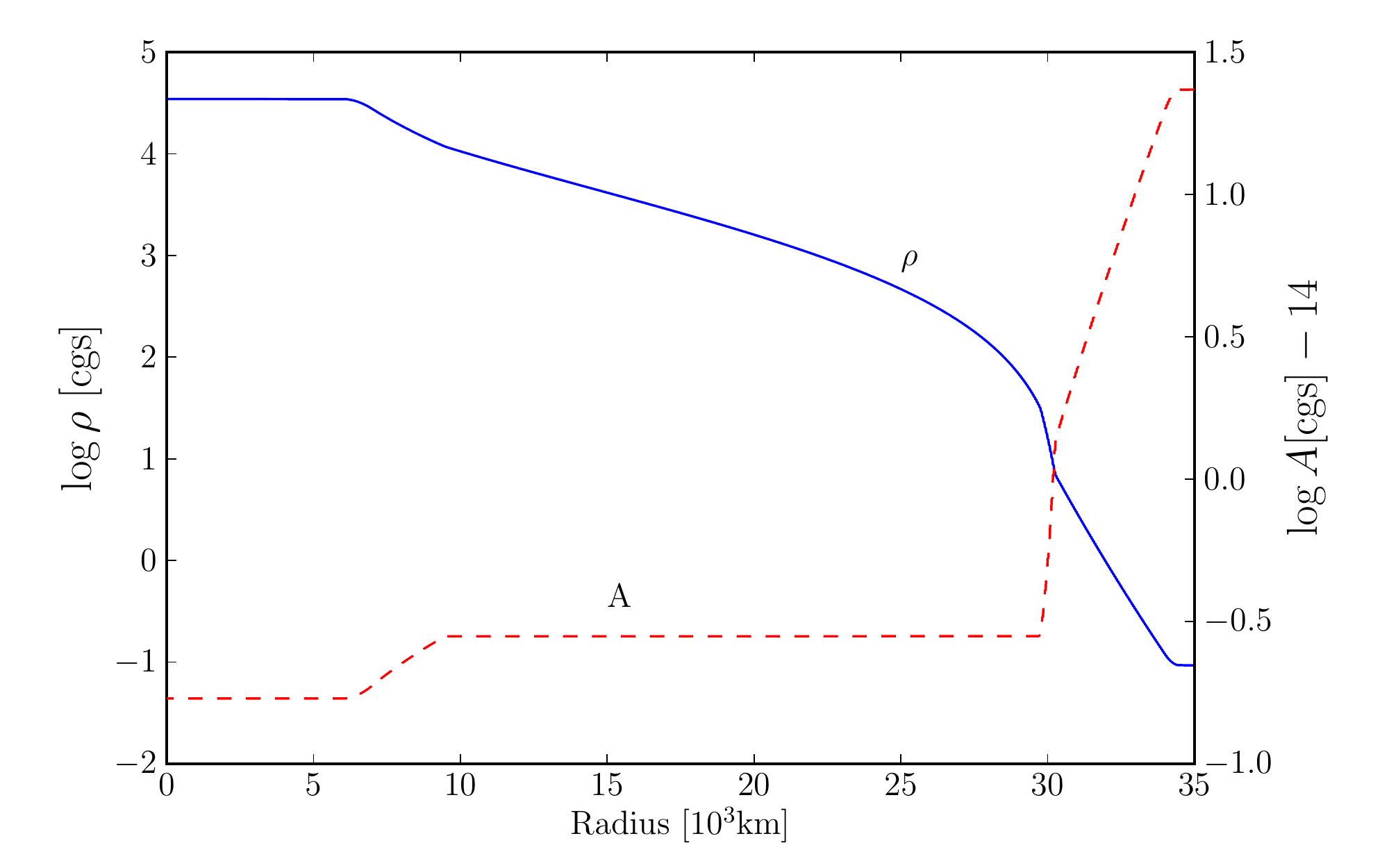}
   \caption{Initial stratification for all simulations presented in
     this paper. The entropy-like quantity $A$ is defined in \glt{eq:entropy-like}.}
   \label{fig:initial_star}
\end{figure}

\section{Method}
\label{sec:method}
\subsection{The challenge}
\label{sec:challenge}
 To properly simulate the hydrogen ingestion flash, our numerical
techniques must be capable of very accurately treating the dynamics
within the small range in radius where the Kelvin-Helmholtz shear
instabilities act  to entrain the stably stratified, more buoyant gas
above the helium shell flash convection zone into the convection flow.
These instabilities result in breaking waves that cause small
puffs of more buoyant gas to become incorporated into the convection
flow and subsequently dragged downward into the convection zone.
These breaking waves are observed as trains of small eddies
 near the
top of the convection zone.  These trains of eddies turn downward
where opposing horizontal flows associated with adjacent large-scale
convection cells meet  (see also the detailed discussion of the
entrainment process in \kap{sec:entrainment}).
Here buoyant fluid from above the convection
zone that has become incorporated into the eddies is pulled downward
into the convection zone with the descending sheets of cooler gas
(cf.\ \kap{sec:general_properties}).  The final panel of \abb{fig:FV_half}
\begin{figure*}[tb]   
   \includegraphics[width=0.5\textwidth]{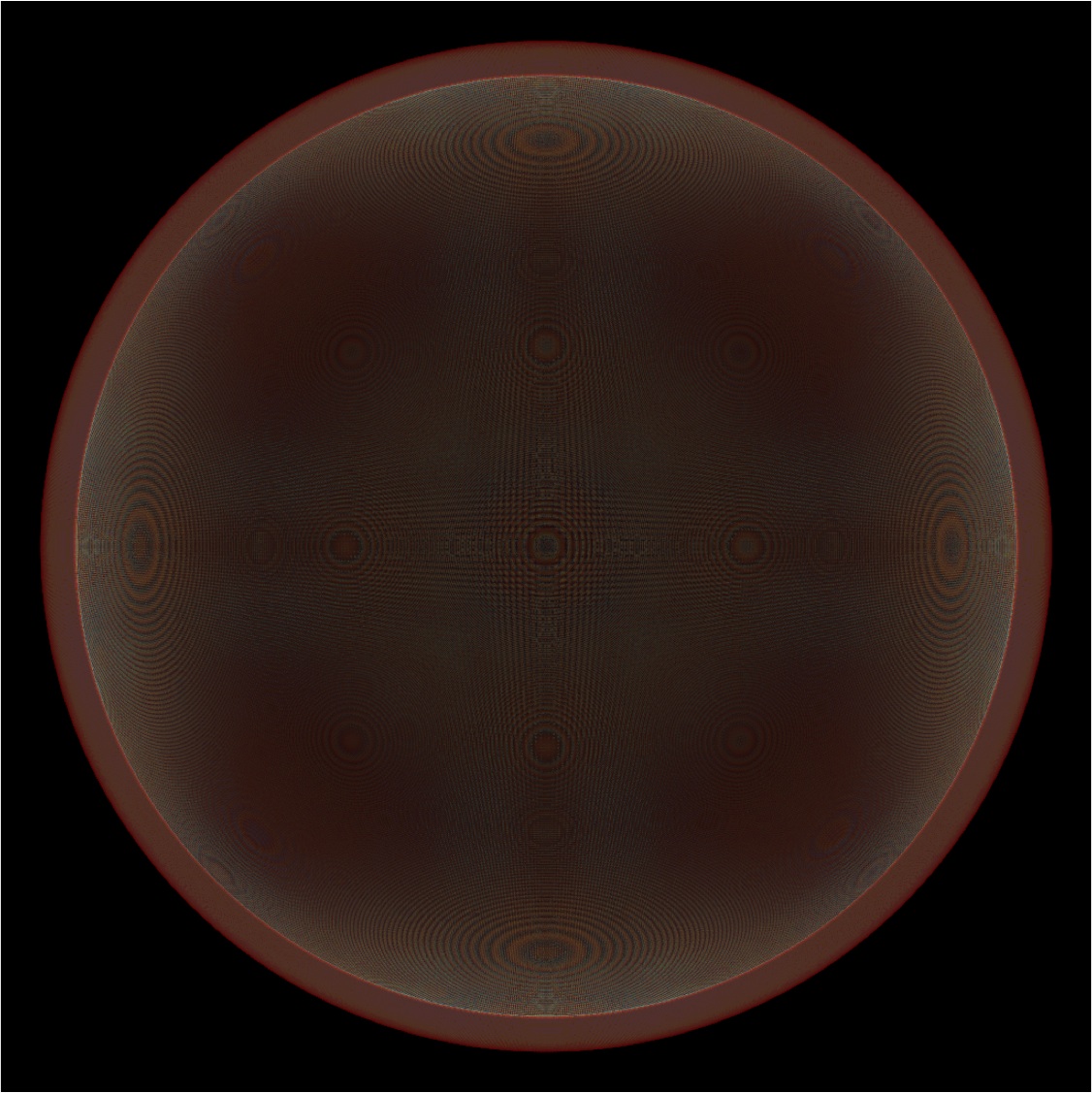}
   \includegraphics[width=0.5\textwidth]{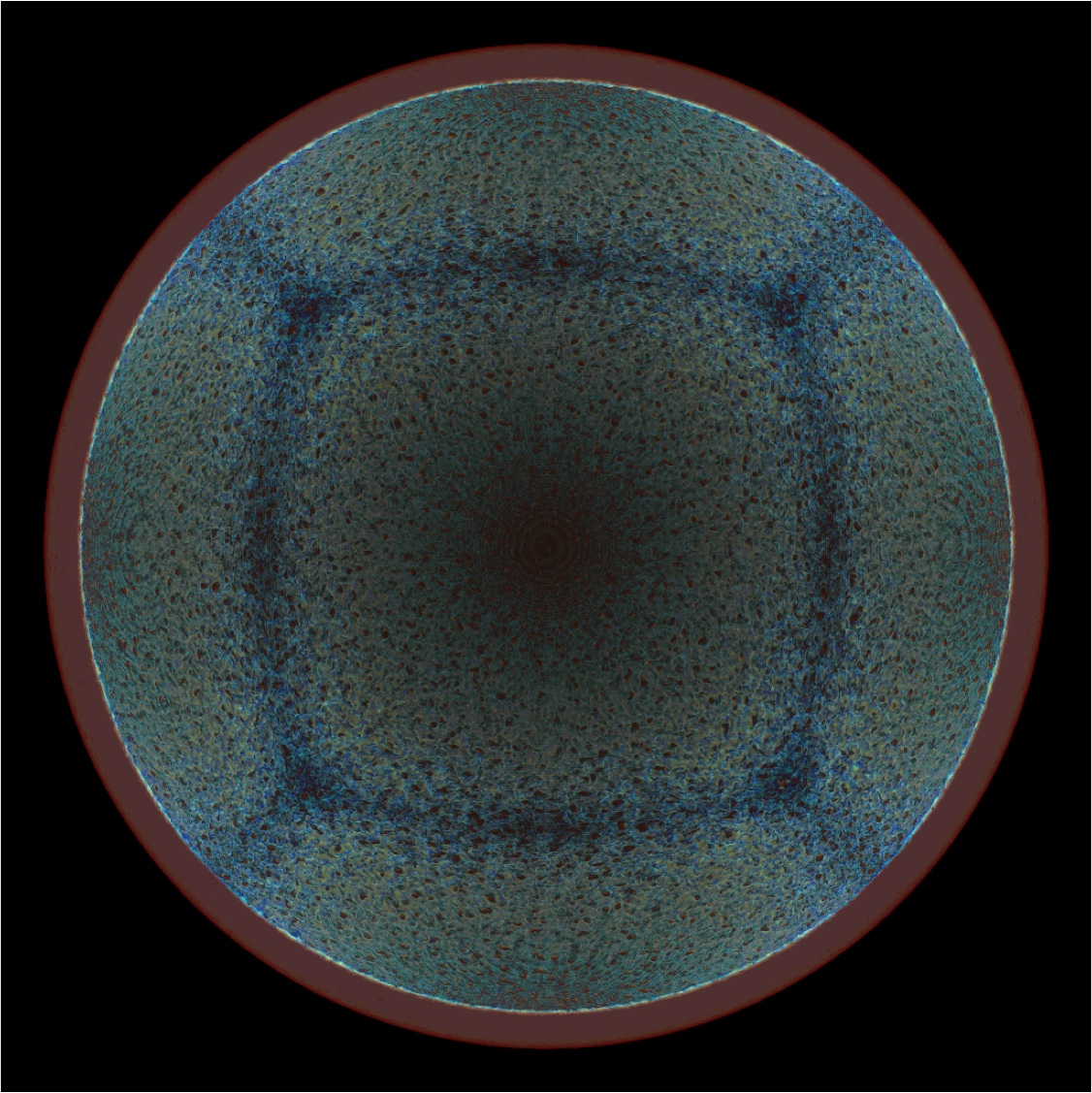}
   \includegraphics[width=0.5\textwidth]{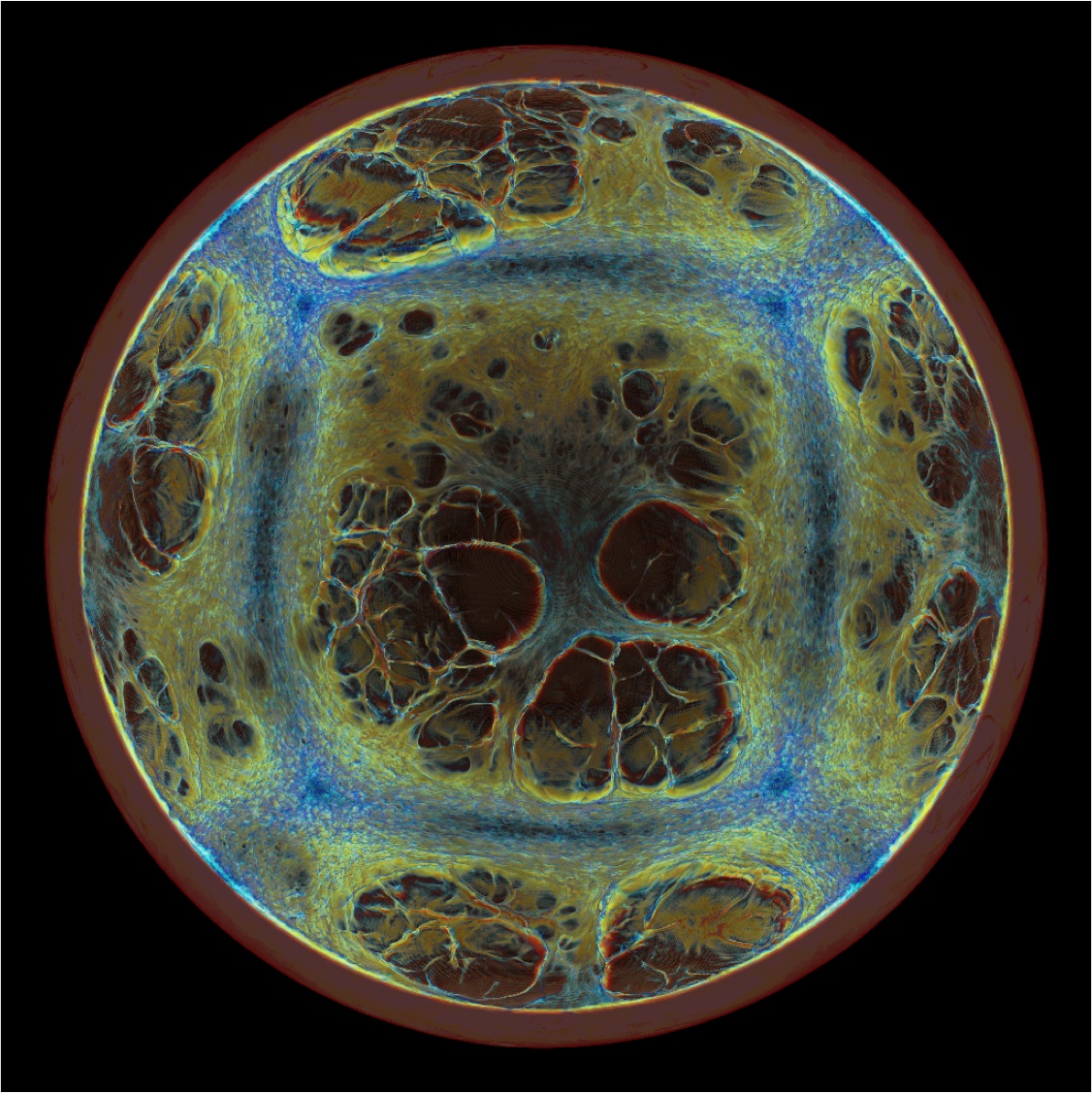}
   \includegraphics[width=0.5\textwidth]{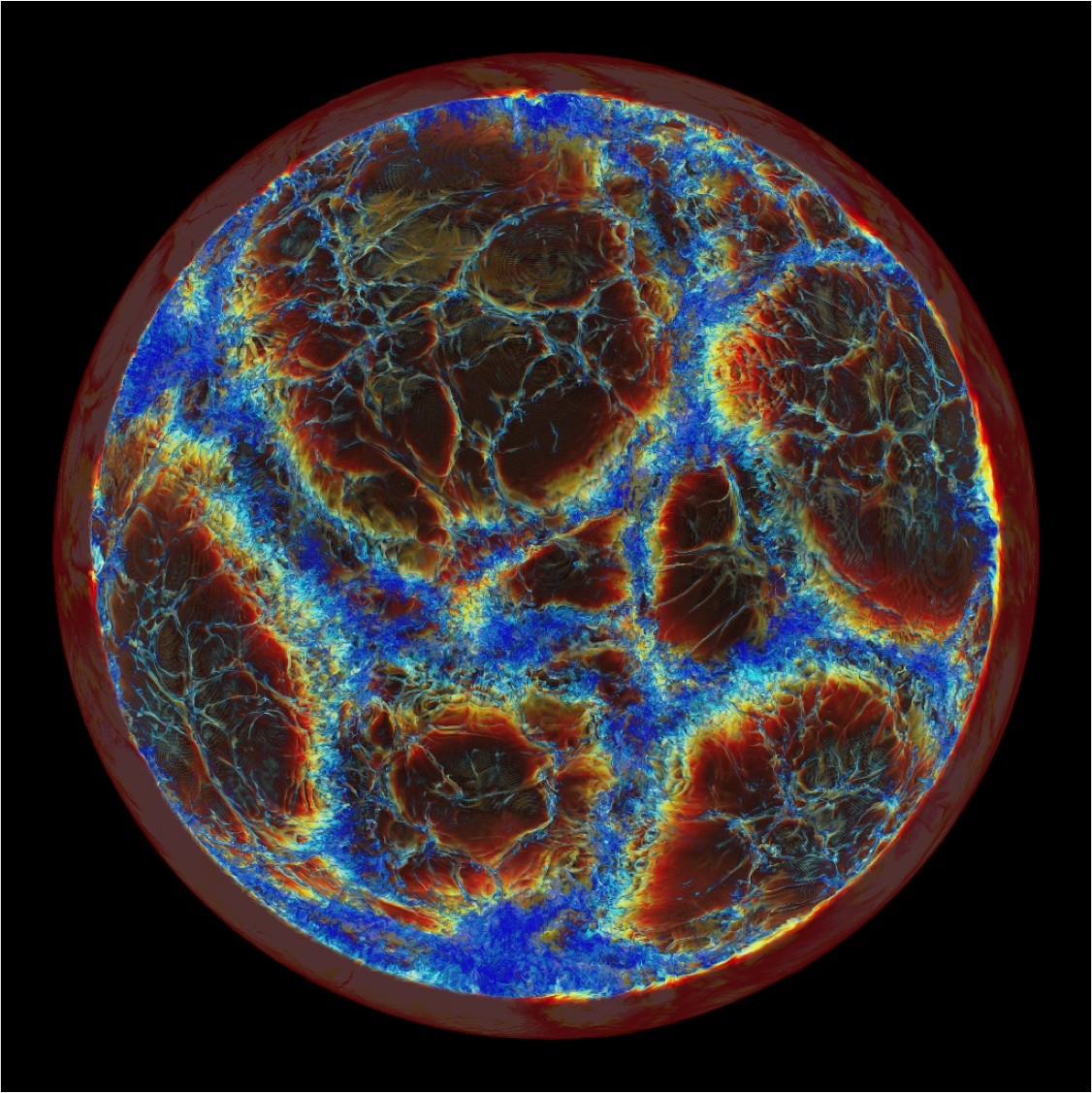}
   \caption{Evidence of initial grid imprints and how they are
     overwhelmed when the physical flow gets established in the
     $1536^3$ grid simulation. Shown is the logarithm of the
     fractional volume of the 'H+He' fluid originally located only in
     the stable layer above (\kap{sec:setup}), in the rear $75\%$ of
     the simulation domain (the $25\%$ of the front are cut away). The
     opacity for the volume rendering is chosen such that levels above
     $\sim 10^{-4}$ are transparent. The observer is therefore looking
     into the open shell and the top convection boundary and the
     mixing interface at that location is seen from the inside. Left
     top: $t=1\mem{min}$, right top: $t=20\mem{min}$, left bottom:
     $t=30\mem{min}$, right bottom: $t=75\mem{min}$. Each of the
     visualizations appears to have a red ring around it's outer
     circumference. This is due to the fact that the portion of the
     $4\pi$ convection zone is a bit more than one half. The red ring
     represents the outside of the convection zone in the small extra
     portion of the sphere that is closest to the observer.   }
   \label{fig:FV_half}
\end{figure*} 
 shows these large convection cells separated by regions shown
in blue of descending gas carrying entrained, buoyant fluid along
with it.  Trains of eddies peeling off from the top of the convection
zone can also be seen in \abb{fig:vort_480} , which shows the magnitude of
the vorticity in a thin slice through the star.
\begin{figure*}[tb]                       
  \includegraphics[width=1.0\textwidth]{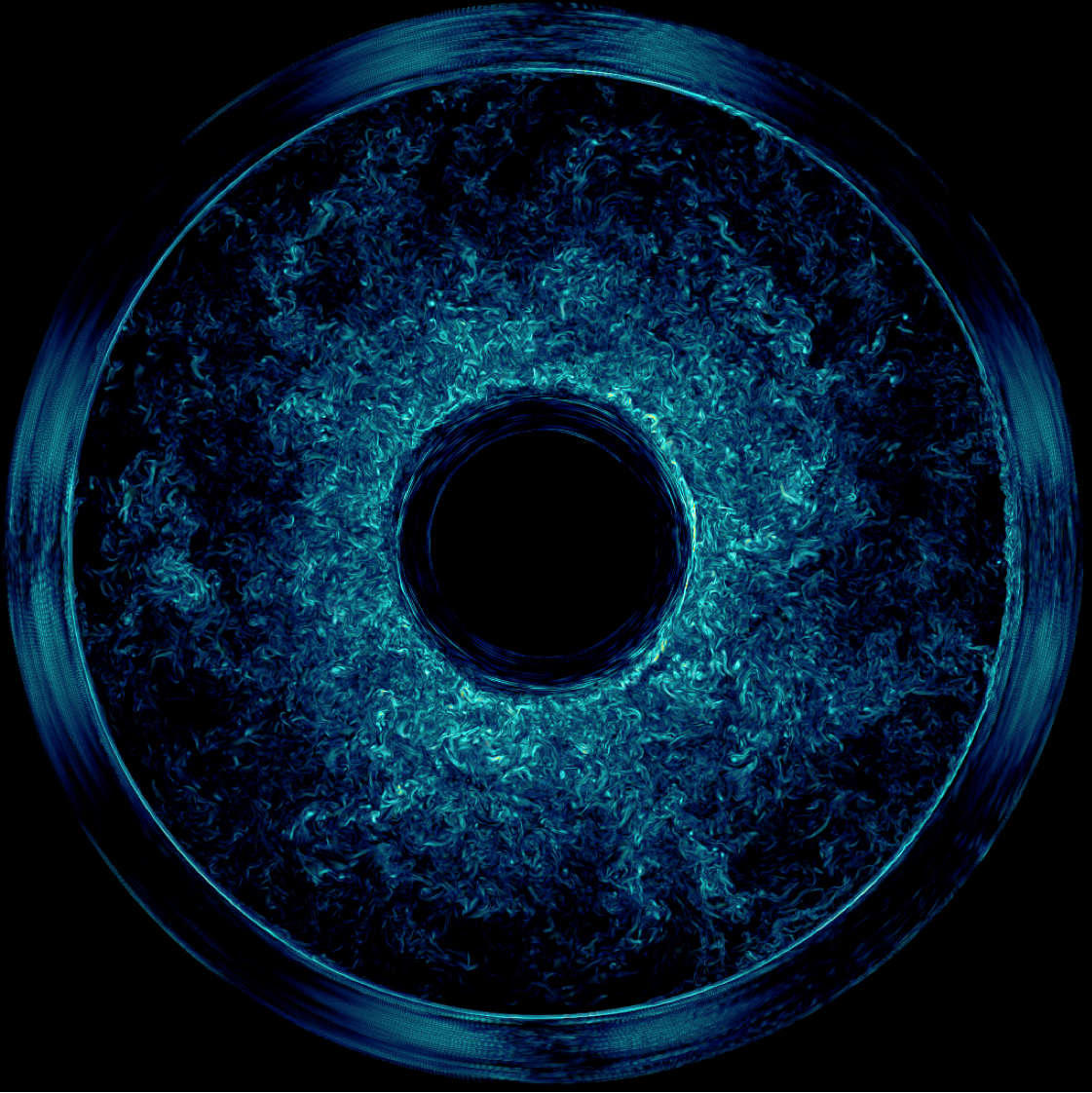}
  \caption{Vorticity of simulation with $1536^3$ grid at
    $t=480\mem{min}$. The volume rendering shows a thin slice of the
    central $\sim 5 \%$ of the full $4\pi$ simulation domain. The
    inert core appears as a black disk in the center.}
   \label{fig:vort_480}
\end{figure*}

Not only must our method be able to describe the growth and breaking
of these Kelvin-Helmholtz waves at the top of the convection zone, but
it must also be capable of carefully tracking the entrained buoyant
fluid as it is dragged downward and progressively mixed into the
surrounding gas of the convection zone due to turbulence. This is no
small feat, because the boundary between the convectively unstable and
stable gas at the top of the convection zone is quite thin.  This is
clear from \abb{fig:vort_480}, where this transition region appears as
an almost perfectly circular line, at which the behavior of the flow
changes suddenly and radically, for reasons that were discussed
earlier.  Although the wavelengths of Kelvin- Helmholtz shear modes
will generally be much larger than the thickness of this transition
layer, the incorporation of buoyant material at this transition is
hard to describe accurately in a numerical treatment unless the
thickness of the layer can be resolved on the computational
grid. But even deciding what the thickness of this layer would be is
not straightforward. The physical thickness of the transition region where the concentration
of hydrogen falls to essentially zero is initially determined by the character
of hydrogen burning in the star before the helium shell flash occurs. As an example that we have adopted here, based
on 1-D simulations \citep{herwig:01a} this thickness is found to be
about $500 \mathrm{km}$ (\abb{fig:entrained_profiles}) for the
post-AGB model before the H-shell flash convection zone.  In the lower
panel of that figure, the transition region is shown in a zoomed-in
view, and individual grid cell intervals are also indicated.  On our
finest grid of $1536^3$ cells, this transition occurs over $11$ grid
cell widths (along the grid direction -- we use a uniform Cartesian grid
to describe this region, which has the topology of a thin spherical
shell). However, the hydrodynamic properties are not only described by the
mean-molecular weight gradient. The entropy profile matters as
well. This cannot be known for the convective boundary region from
one-dimensional stellar evolution calculations. Our initialization strategy is
described in \kap{sec:setup}. This is an important point because of
the concern that our initial setup may somehow be special and not
representative of the conditions in a real star. 
\begin{figure}[tb]                  
  \includegraphics[width=0.5\textwidth]{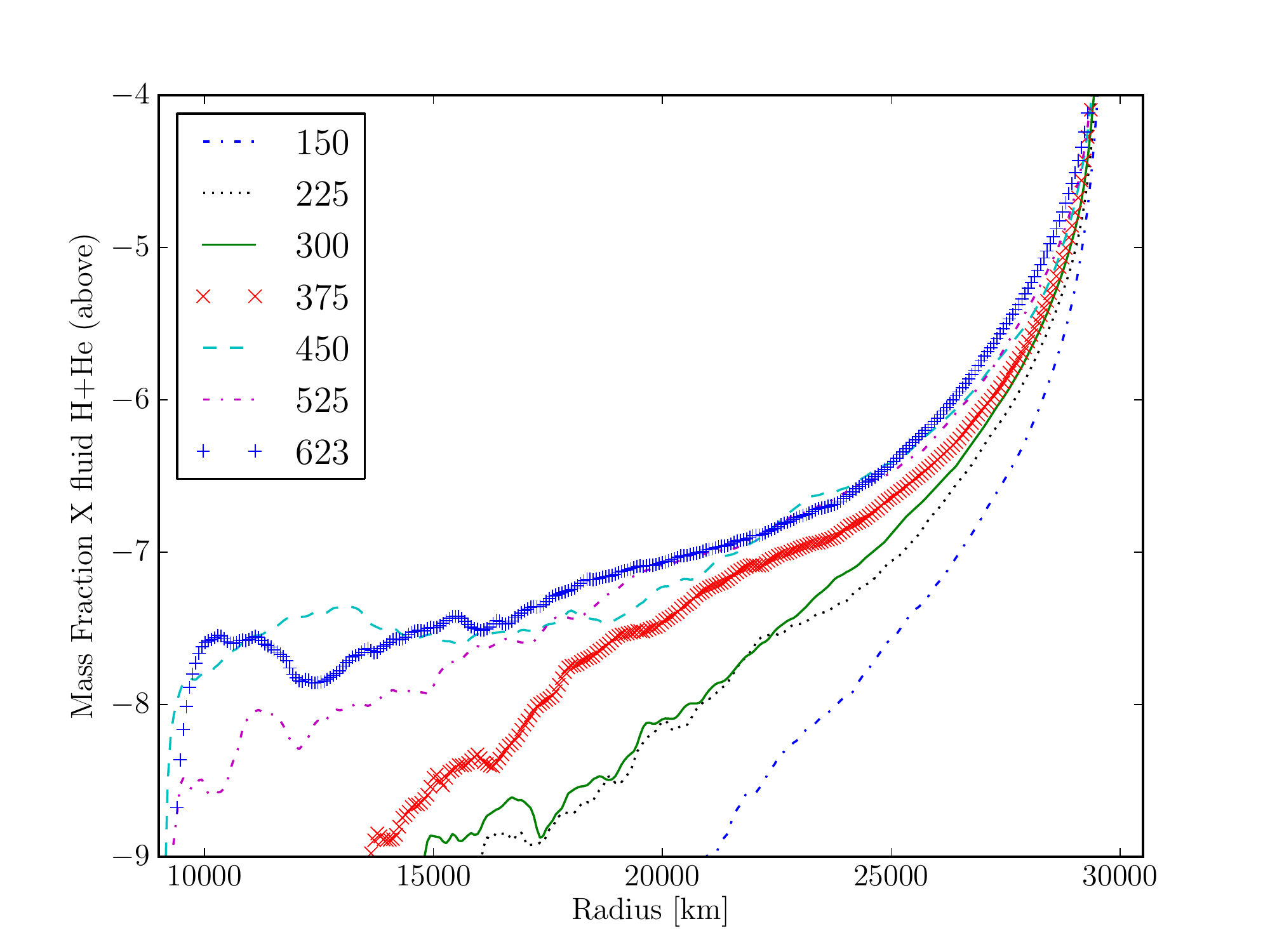}
  \includegraphics[width=0.5\textwidth]{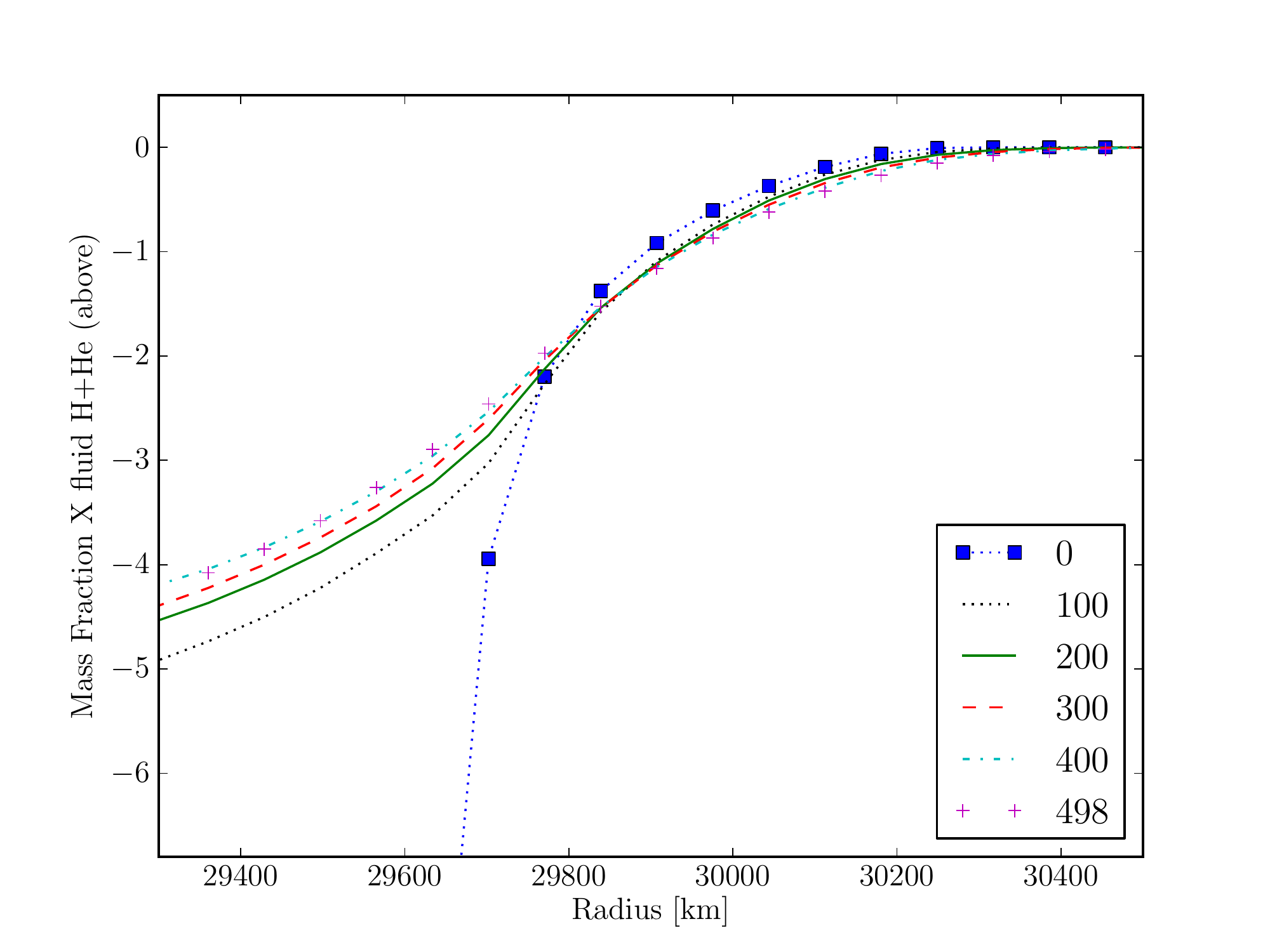}
   \caption{Spherically averaged abundance profiles in the convection
     zone for a series of dumps.  Each dump corresponds to $ \approx 1
     \mathrm{min}$ star time. The markers correspond to grid
     zones. \emph{Top:} $1024^3$-grid simulation for the entire
     convection zone.  \emph{Bottom:} Spherically averaged abundance
     profiles at top convection boundary for the $1536^3$-grid
     simulation, at different times, line labels correspond to minutes
     of simulated time. Minute $0$ represents the initial profile
     reflecting the smooth transition according to the former H-shell
     burning at that location (see text).  }
   \label{fig:entrained_profiles}
\end{figure}

We face an additional computational challenge, because our simulations lie in a regime
that is right about at the break-even cost Mach number for implicit
relative to explicit numerical methods.  The peak Mach numbers in our
flows range around $0.03$.  We have addressed this flow regime with an
explicit gas dynamics scheme, relying on the high computational
efficiency of explicit schemes and the $12\%$ to $23\%$ of the machine's
peak floating point computation rate (depending upon the machine) that
our code achieves \citep{woodward:10a,woodward:12a} to make this
approach competitive with implicit methods.

\subsection{The code}
Because the entropy plays such a key role in the convective stability
or instability of the gas in our problem, we find that to obtain
accurate results, it is necessary to solve a conservation equation for
the entropy rather than for the total energy.  In the absence of
shocks and nuclear reactions, the entropy of our gas is conserved along streamlines.  At
our low flow Mach numbers we do not expect shocks, and we can
represent nuclear reactions by source terms in an entropy conservation
law.  With our gamma-law equation of state, with $\gamma =5/3$, we use
a conservation law for the adiabatic constant $A = p/{\rho }^{\gamma }$
\glp{eq:entropy-like}.  This is a technique employed in
meteorological codes, where a quantity called the potential
  temperature takes the place of the entropy.  In weather prediction,
Mach numbers in our same general range are typically encountered, so
it is natural that similar numerical techniques are useful.

 The explicit gas-dynamics code is the same as the one we used in
 \citet[][appendix A.2]{herwig:10a} and is described
 in full in \citet{woodward:06b}.  Of particular importance for the
 entrainment properties at the stable-unstable interface at the top of
 the convection zone with a mean-molecular weight gradient is the PPB
 moment-conserving advection scheme
 \citep[see][]{woodward:08a}\nocite{woodward:08b}.

 Details of the features of our version of the Piecewise-Parabolic
 Method (PPM)
 \citep{woodward:81,woodward:84,colella:84,woodward:86,woodward:06b}
 that make this explicit scheme highly accurate in this flow regime
 are given in \kap{sec:app_method}. We also describe the
 Piecewise-Parabolic Boltzmann (PPB) scheme
 \citep{woodward:86,woodward:05,woodward:10b} as well as the features
 that enable our PPM scheme to be coupled in a natural way to the PPB
 scheme to describe multifluid fractional volume advection there.

 As in previous work \citep[e.g.][]{porter:00a,herwig:06a} we adopt a
 monatomic ideal gas equation of state, which represents the
 conditions in advanced He-shell flash convection at high densities
 well. In particular, this equation of state provides a good
 representation of the conditions in the post-AGB He-shell flash that
 has occurred in Sakurai's object. In some other cases, such as
 He-shell flashes in the first thermal pulse of low-mass and low
 metallcity stars, radiation pressure may be important.  The code uses
 appropriately scaled code units for all physical quantities.

\subsection{Setup of star simulations}
\label{sec:setup}
We consider a two-fluid setup of a convectively unstable shell and a
stable layer below with initially one fluid, and a stable layer above
the convection zone with a second fluid with lower mean molecular
weight.  The simulations are performed on a uniform Cartesian grid
with a range of grid sizes up to $1536^3$. The full $4\pi$ shell of
the convectively unstable layer is included in the simulation, which
ensures that any global or large-scale motions can be captured. The
stratification described below covers $8.3$ pressure scale heights in
the entire simulation domain, and $4.9$ pressure scale heights in the
convection zone.

As in many other multi-dimensional investigations of the
hydrodynamics of convection \citep[e.g.][and many of the works
  mentioned in \kap{sec:intro}]{porter:94,hurlburt:94,herwig:06a}, we
construct the radial stratification with a set of connected, piecewise
polytropic layers in which the polytropic constant and the polytropic
index are chosen in such a way that the overall representation
resembles a typical situation of a luminosity-driven shell convection
layer bounded by a stable layer above and below, just like He-shell
flash convection. This approach has the advantage that the initial
state is in very good numerical hydrostatic equilibrium, and initial
transients are minimized. Mapping a 1-D profile from a stellar
evolution code \cite[as done, e.g.\ by][]{Meakin:2007dj} would provide
little in additional accuracy in our case where the microphysics is
exceptionally simple.  In \abb{fig:initial_star} we show the resulting
initial stratification in terms of $\rho$ and in terms of the
adiabatic constant $A$, which is related to the entropy
\glp{eq:entropy}
\begin{equation}
  \label{eq:entropy-like} 
\log A =
\frac{1}{c_\mem{v}} (S + constant) = \log (p/\rho^\gamma) \punkt
\end{equation}
with $\gamma = 5/3$ for the monatomic ideal gas.  The sound speed
decreases from $\sim 1500\mem{km/s}$ at the bottom of the convection
zone to $\sim 200\mem{km/s}$ at the top of the convection zone.

The dimension and stratification details of He-shell flash convection
in AGB stars are variable, as a function of mass, metallicity, early
vs.\ late thermal pulse and AGB vs.\ post-AGB thermal pulse. In
addition, and quite obviously, the conditions also change dramatically
for a given thermal pulse as a function of time due to the large
energy deposit from the He-shell flash \citep[see, for example,
Fig.\,1 and 2 in][]{herwig:06a}. The variety of conditions is also
reflected in the degree in which radiation pressure contributes to the
total pressure. Initially, when the flash starts, the density is so
high that the entire pressure is provided by the gas. As the He-shell
layer absorbs the peak-flash luminosity of several $10^7\lsun$ it
rapidly expands. The temperature decreases more slowly and the
radiation pressure contribution increases. The lowest value for the
gas pressure fraction $\beta = P_\mem{gas}/P$ is found at the bottom
of the He-shell flash convection zone where it decreases to $\beta
\sim 0.94$ at the time when the He-burning luminosity reaches its
maximum for thermal pulses in a stellar evolution sequence of
$M_\mem{ini} = 2\msun$ and $Z = 0.02$ with core masses ranging from
$0.5$ to $0.6\msun$.

\citet{herwig:06a} adopted a plane-parallel modeling approach, which
is appropriate only for a small ratio of the geometric thickness of
the convection shell and the radius of the underlying core. They have
therefore chosen a time of the flash just before the peak luminosity
is reached. We simulate the full $4\pi$ geometry of the shell. This is
necessary, because the entrainment starts only when the peak
luminosity has been reached, or just passed. At that later time in the
thermal pulse the convection zone has already greatly expanded and is
now thicker than the radius of the underlying core.  We have therefore
chosen to reproduce in our setup the geometric dimensions encountered
at $t=0.25\jahre$ of the model sequence in Fig.\,1 and 2 shown in
\citet{herwig:06a} where $t=0\jahre$ corresponds to the time when
He-burning has reached the flash-peak luminosity. At this time the
density would have further decreased, but still $\beta_\mem{min}
\apgeq 0.8$.

Thus, we are not simulating here a specific metal-poor AGB star, or a
post-AGB star. Rather than trying to reproduce the specific conditions
of a particular case we investigate more generally the typical
behavior of shell flash convection with very stiff convective
boundaries, which do not depend on the exact numerical value of any of
the setup parameters.

\paragraph{The unstable layer}
The density, the pressure, the gravity and the radius at the bottom of
the convection zone are $\rho_\mem{bot} =
\natlog{1.174}{4}\mem{g/cm^3}$, $P_\mem{bot} =
\natlog{1.696}{20}\mem{g/(cm\, s^2)}$, $g_\mem{bot} =
\natlog{4.9545}{7}\mem{cm/s^2}$ and $r_\mem{bot} =
\natlog{9.5}{8}\mem{cm}$ which imply a core mass below the convection
zone of $M_\mem{bot} = 0.337\msun$.  This mass is lower than the
typical core mass of thermal pulse AGB stars, but it allows us to
accommodate our choice of an ideal gas equation of state and the
geometric dimension of the convection zone while ignoring for the
moment a moderate fraction of radiation pressure.

The top of the convection zone is located at $r_\mem{top} =
\natlog{3.00}{9}\mem{cm}$.  The equations for hydrostatic equilibrium,
mass conservation and \glt{eq:poly} with $\gamma_\mem{s} =
\gamma_\mem{ad}$ and $K_\mem{s} $ given by $\rho_\mem{bot} $ and
$P_\mem{bot}$, are numerically integrated to provide the density and
pressure stratification in the convection zone.

\paragraph{The upper boundary and the stable layers}
The layer above the convection zone is stable with
$\gamma_\mem{s-top}= 1.01$. This region is populated by a fluid
representing the H- and He-dominated envelope of AGB and post-AGB
stars with a mean molecular weight $\mu_\mem{H+He}= 0.7$ similar to a
mostly unprocessed envelope abundance distribution that consists
predominantly of H and He. This fluid is therefore labeled 'H+He'.
This compares to $\mu_\mem{conv}= 1.58$ of the fluid in the convection
zone (labeled 'conv') which contains a mix of \hevi, \czw\ and \ose\
that is typical for the He-shell flash convection zone in AGB or
post-AGB stars.

The abundance interface between these two layers is not a
discontinuity, but instead it reflects the smooth transition resulting
from H-shell burning with varying efficiency across the temperature
profile at that location. The resulting H-profile of a post-AGB
stellar evolution model shown in Fig.\,~5 in \citet{Herwig:1999uf} is
representative for this situation. 
Therefore, the H-profile at the bottom of
the H-rich envelope in a radiative layer, before the He-shell flash
convection approaches from below, will vary smoothly from a mass
fraction of $X(\mem{H}) \sim 0.7 $ to zero over a width of the former
H-burning shell of $\Delta r_\mem{H-shell}\sim\natlog{5}{7}\mem{cm}$
(bottom panel, \abb{fig:entrained_profiles}). The width of this
gradual transition from H-rich to H-free depends on the temperature
gradient in the H-burning shell, and may vary depending on the
particular type of thermal pulse as well as the mass of the underlying
degenerate core and the metallicity.

When continuing the integration of the stratification into the stable
layer we let both $\gamma$ and the concentration of the two fluids
vary smoothly over a width of $\Delta r_\mem{H-shell}$ centered at
$r_\mem{top}$. That means that the constant entropy region of the
convective unstable zone ends at $r_\mem{top}-\Delta r_\mem{H-shell}/2
= 29,750\mathrm{km}$ where the entropy gradient becomes positive.  In
the $1536^3$ grid simulation the interface width $\Delta
r_\mem{H-shell}$ corresponds to $11$ cells along a grid axis. The shape
of the resulting profile is shown in the bottom panel of
\abb{fig:entrained_profiles}.

When integrating the initial stratification downward from the bottom
of the convection zone into the lower stable layer we adopt
$\gamma_\mem{s-bot}= 1.2$ and a transition layer of
$\natlog{2.5}{7}\mem{cm}$ centered at $r_\mem{bot}$. The fluid type is
the same as in the convection zone. 

\paragraph{Luminosity driving the convection}
The flash constitutes a thermonuclear
runaway that is the result of an unstable balance between thermodynamic,
hydrodynamic and nuclear physics components. When mapping a 1-D profile into
a 3-D code we cannot expect the thermonuclear runaway to continue on the
same trajectory as in 1-D, with a multitude of small differences
between the two simulation approaches. Instead, we use the
piecewise-polytropic stratification as a background and add a constant
volume heating  at the bottom of the He-shell flash convection zone,
that has a total luminosity equivalent to the He-burning luminosity in
a 1-D stellar evolution model. 
\begin{figure}[tb]                                   
  \includegraphics[width=0.5\textwidth]{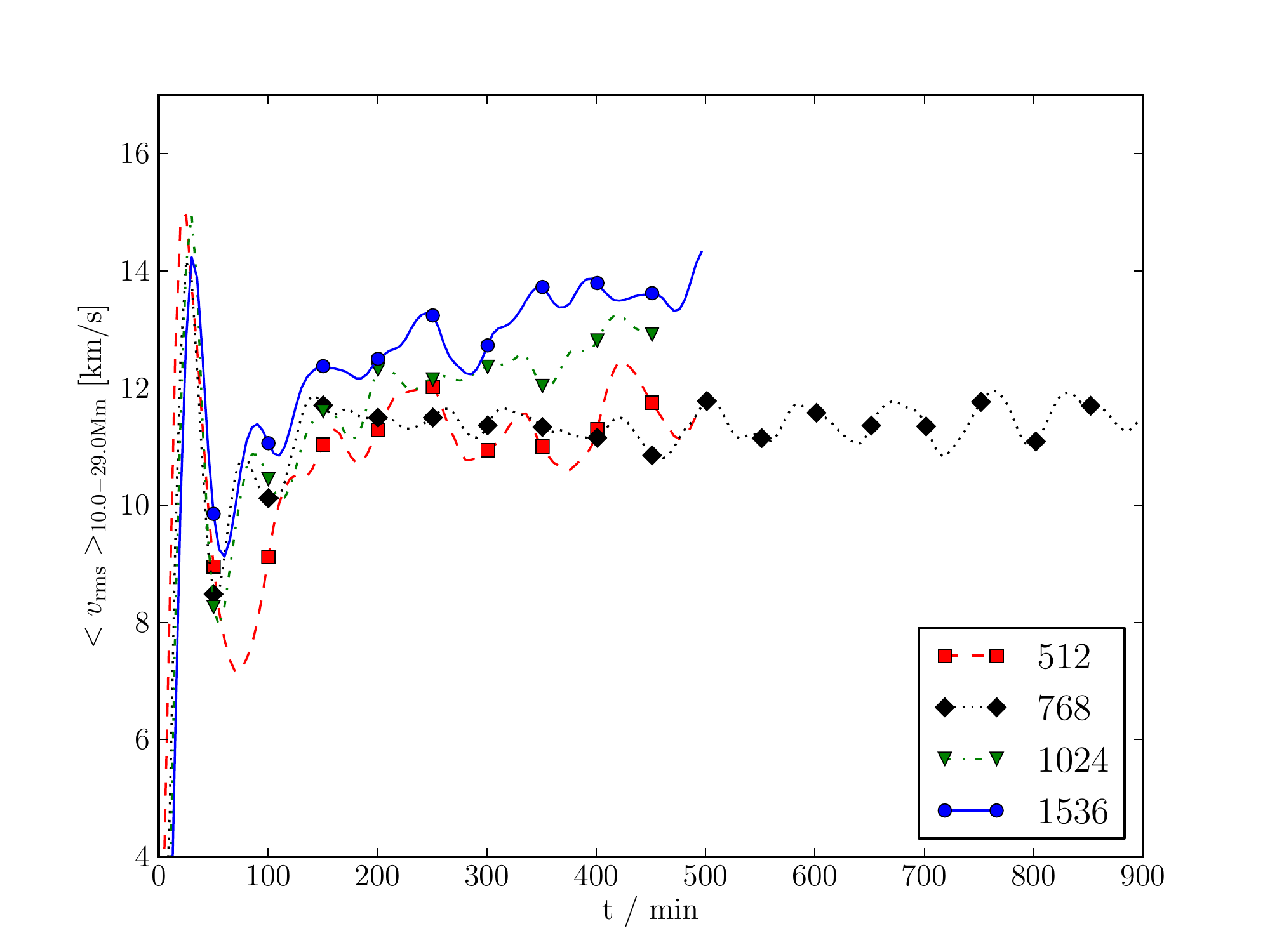}
   \caption{Spherically and radially averaged rms-velocity (averaged over
     the sphere and over convection region between $r=\natlog{10}{6}$
     and $\natlog{29}{6}\mathrm{m}$ ) as a function of time for runs
     with different grid sizes $n^3$. The number of grid points $n$ in
     one direction is shown in the legend.}
   \label{fig:Ekmax}
\end{figure}
The constant heating is applied over a heating region of width
$\Delta r_\mem{heat} = 10^8\mem{cm}$, the lower boundary of which is
offset from the bottom of the convection zone by
$\natlog{5}{7}\mem{cm}$ to avoid any entrainment of low-entropy
material from below the convection zone. The heating is applied in the
$4\pi$ shell with a smooth bell-shaped radial intensity
distribution. This arrangement is providing a robust heat source that
is independent of the resolution and represents the He-burning
luminosity from the triple-$\alpha$ reaction. The heating rate in the
standard case corresponds to $L_\mem{He}=\natlog{4.20}{7}\lsun$.

\paragraph{On the absence of radiation in our simulations}
Unlike in envelope convection and shallow surface convection
simulations, the convection is not driven by a subtle degree of
superadiabaticity that is only reached in a quantitatively correct
sense over the rather long thermal time scale due to the effect of
radiative transport at the bottom and top layer of the convection
zone. Instead, the convection is very efficient and driven by the
large luminosity from the underlying thermonuclear runaway.

In the deep interior case of He-shell flash convection, radiation
transport is not important on the dynamic or convection turn-over time
scale (but radiation pressure may be). The mean free path of a photon
is $l_\mem{ph} = 1/\rho \kappa$ where the $\rho \sim 10^{4}$ to $10
\mem{g/cm^3}$ in the convection zone. The opacity is of order
unity, which means that $l_\mem{ph} < 0.1\mem{cm}$. The characteristic
length scale for radiation transport can be expressed as $l_\mem{rad}
= 2\sqrt{(D t)}$ where $D = \frac{1}{3} c l_\mem{ph} \sim
10^9\mem{cm^2/s}$. With a radial domain size of
$R=\natlog{3.5}{9}\mem{cm}$ the highest resolution runs presented in
this paper (with $1536/2$ radial zones) have a grid size of $\mem{d}x
= \natlog{4.6}{6}\mem{cm}$. If we take as a convective timescale the
radial extent of the convection zone divided by the typical velocity
in the convection zone (\abb{fig:Ekmax}) we obtain $t_\mem{conv} \sim
1500\mathrm{s}$ and $l_\mem{rad} \sim \natlog{2}{6}\mem{cm}$.

Therefore $l_\mem{rad} \approx \mem{d}x$, or in other words, within
the time that it takes convection flow radially across the entire
convection zone radiation transport crosses one simulation cell in the
highest resolution case. However, in outermost layers of the
simulation domain the density decreases to $0.1 \mem{g/cm^{-3}}$ and
here indeed radiation diffusion will eventually become relevant over
the entire course of the simulated time (typically $\apleq
1000\mem{min}$). If these lowest density regions will become important
in future simulations, or if the simulated time will further increase
or if we further increase the resolution, then indeed diffusive
radiation transport should be included. For the present set of
simulations this physics aspect can, however, safely be neglected
\citep[see][for a similar conclusion for their O-shell burning
  simulations]{Meakin:2007dj}.

\subsection{Tests and simulations}
We started this investigation with numerous test simulations at
various resolutions, experimenting, for example, with different
assumptions for the treatment of the convection boundaries, the
properties of the heating zone, peculiarities of different Fortran
compilers, and the effect of using single vs.\ double precision
arithmetic. For the simulations shown here we are adopting the Intel
12 Fortran compiler and double precision arithmetic. We have also run
test simulations on different hardware platforms. 

After we were satisfied with the results of these tests, we performed
a grid of simulations for the following grid resolutions\footnote{We
  did tests with a $384^3$ grid, which however showed obvious numerical
  artefacts and were not further considered. A particularly
  troublesome feature were entropy-shelves that formed above and below
  the proper convection zone. These were small constant-entropy
  spheres, detached from the big convection zone that formed and
  survived in the initially stable layers. These features went
  entirely away once we adopted higher grid resolutions and
  double-precision arithmetic.}: $512^3$, $768^3$, $1024^3$,
$1536^3$. Movies of these simulations can be found at 
\url{http://www.lcse.umn.edu/movies}.
The simulations on the coarsest grids were performed on the
25-node workstation cluster at the Laboratory for Computational
Science \& Engineering (LCSE) at the University of Minnesota.  Those
on $768^3$ grids were performed at either the Minnesota Supercomputing
Institute (MSI) or on the Canadian WestGrid Orcinus and Lattice
high-performance computers, runs on $1024^3$ grids were performed at
MSI, while runs at the highest resolution were performed on the NSF's
Kraken supercomputer at the National Institute for Computational
Sciences (NICS).  Our run on a grid of $1536^3$ cells used 98,312 CPU
cores on Kraken and consumed 5.2 million CPU-core hours.  The
delivered performance was about 120 Tflop/s in 64-bit precision.

We will demonstrate in the next section that the $512^3$ grid gives an
entrainment rate that is a factor of almost ten larger than the
asymptotic entrainment rate, while the $768^3$ grid provides enough
resolution to reproduce the asymptotic entrainment almost within a
factor of two. We therefore consider the latter grid size the absolute
minimum to perform, for example, differential investigations of the
effects of various input assumptions.

\section{Results}
\label{sec:results}

\subsection{General properties of the
  convection simulations}
\label{sec:general_properties}

\paragraph{Initial transients and grid effects}
We do not seed the initial setup with perturbations. Instead, the grid
provides asymmetries that allow initial perturbations, fueled by the
continuous heating at the bottom of the convection zone, to form and
grow along the axis directions of the grid
(\abb{fig:vort_inital}). The initial start-up phase of any simulation
like this is prone to artificial, initial transients that are also
observed in the time evolution of global properties, such as the
spherically and radially averaged rms velocity (\abb{fig:Ekmax}). The
velocity maximum at early times is due to preferential rise of the
first plumes along the grid axis. These initial transients do not
matter since we run the simulations for long enough, until we reach a
convective steady-state, which is dominated by the physically relevant
and realistic fluid motions. The amount of H-rich material entrained during the
initial transients is also very small.  Even if its burning were to be
included in the simulation, this small amount of entrainment would not
generate a dynamically relevant amount of energy.
\begin{figure}[tb]             
   \includegraphics[width=0.48\textwidth]{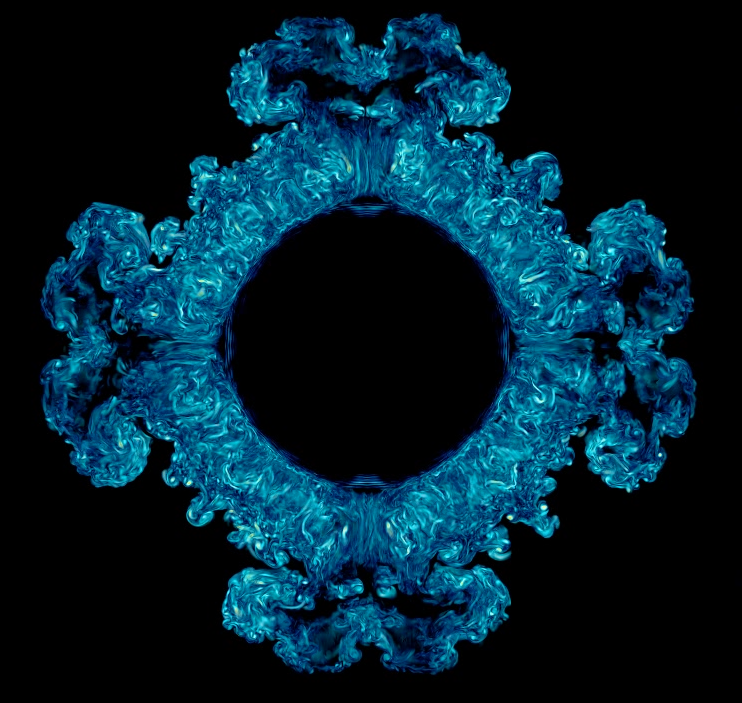}
   \caption{Zoomed in view of the vorticity during the
     initial transient period ($t = 20\mathrm{min}$) of the run with
     $1536^3$ grid. Same view as in \abb{fig:vort_480} except that a
     smaller portion of the simulation domain is shown. The heads of
     the upwelling plumes along the grid axis have advanced
     approximately half the distance between the bottom and top of the
     convection zone.}
   \label{fig:vort_inital}
\end{figure}

During the initial transient the first convective plumes are launched
along the grid directions. This is shown in 
\abb{fig:vort_inital}, which shows that at time $20\mathrm{min}$ these
plumes have traveled through about half of the radial distance to the
upper convective boundary. In all cases an initial peak of the
transient velocity subsides within $\approx 50\mem{min}$, shortly
after the first plumes have first reached the upper convection
boundary (bottom, left panel \abb{fig:FV_half}). Velocities
adopt a local minimum before the heating luminosity takes over the
fluid flow dynamics and the rms velocities increase again. Abundance
profiles at $100\mathrm{min}$ (\abb{fig:entrained_profiles}) show the action of the initial
convective turn-overs

In some previous investigations such initial transients have been
damped before or while the convection driving luminosity was turned on
\citep[e.g.][]{herwig:06a}. We do not do anything in particular to
dampen the initial transients, but instead study how they become
smaller in magnitude with increasing resolution. Indeed, close
inspection of \abb{fig:Ekmax} shows that for the highest resolution
run ($1536^3$), the peak rms velocity of the initial transient is about
as high as the steady-state value approached after $200\mathrm{min}$. 
Another concern may be that the initial transient is shaking the
interface so much that we get artificial smearing. The profile
evolution at the top convection boundary (bottom panel,
\abb{fig:entrained_profiles}) shows that this is not the case.

The initial transients can also be observed in the entrainment
visualization shown in \abb{fig:FV_half}.  These images show the
'H+He' fluid that initially occupies only the region above the
convection zone (see bottom panel,  \abb{fig:entrained_profiles}). Abundances
$\apgeq 10^{-4}$ are assigned to be transparent in the volume
rendering. Levels down to $\approx 10^{-7}$ are assigned with
opacities of different color, from red to yellow to blue in decreasing
order of concentration. The upper left panel in
\abb{fig:FV_half} shows the initial state in which by design the
visible interface is represented by only about one grid zone (cf.\
\abb{fig:entrained_profiles}) because we only show concentrations below
$\approx 10^{-4}$. Because the thickness of the visible interface is
comparable with the size of a grid zone interference, rings and
patterns appear at the location of the Cartesian grid directions.

During the initial transient phase (top-right and bottom left panel,
\abb{fig:FV_half}) rectangular grid imprint patterns are clearly
visible. They correspond to very small perturbations at this interface
when the convectively driven motions at that boundary are not
fully established yet. However, already at $t=75\mathrm{min}$ any evidence
of grid-imposed patterns has vanished. This time corresponds to the
end of the initial transient phase when the radially and spherically
averaged rms velocities have a minimum
(\abb{fig:entrained_profiles}). From this point onward the
luminosity-driven convective motions are strong enough to dominate any
grid-imposed numerical noise in our simulations. 

Once the convective motions dominate the flow in the entire
constant-entropy shell the patterns at the upper boundary imply
large-scale features. Typically two to four coherent patches (or
granules) can be identified per hemisphere. These represent areas in
which fluids are collectively upwelling.  The situation shown in the
lower-right panel in \abb{fig:entrained_profiles} is representative
for later stages (see also \abb{fig:FV_ring}), which rather tend to
show fewer coherent systems and therefore a more global nature of the
convection flows. These global modes have as their limit the global dipole
mode reported for non-rotating, fully convective configurations
\citep{porter:00a,kuhlen:06} and are correlated with the aspect ratio
$\Delta r_\mathrm{shell}/r$ 
of shell convection. Large aspect ratios lead to larger cells, while
small aspect ratios have smaller cells. In any case, convection cells
fill the entire radial extent of the convection zone, and
approximately the radial extent determines the horizontal extent. 
\begin{figure*}[tb]                    
   \includegraphics[width=0.5\textwidth]{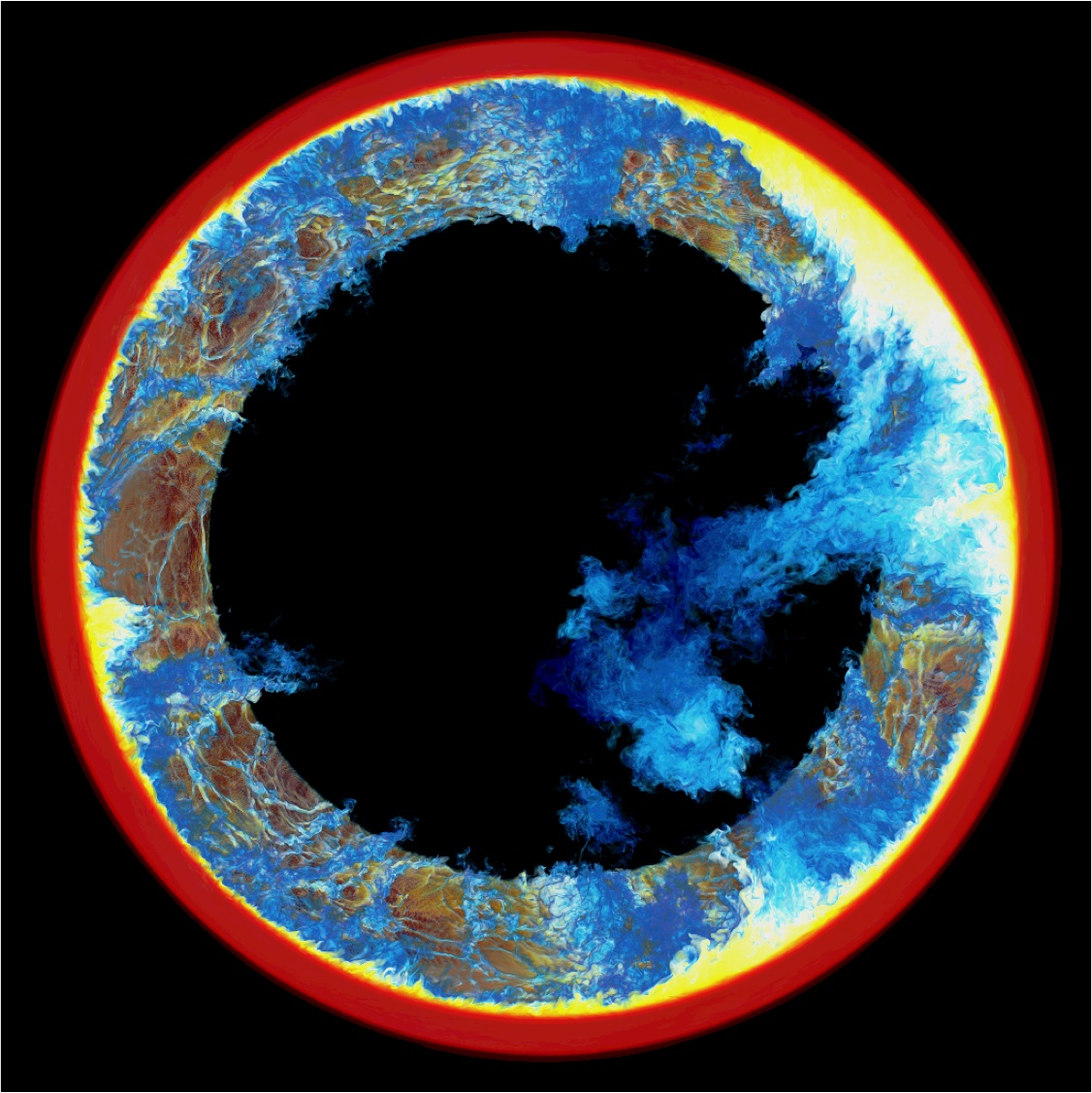}
   \includegraphics[width=0.5\textwidth]{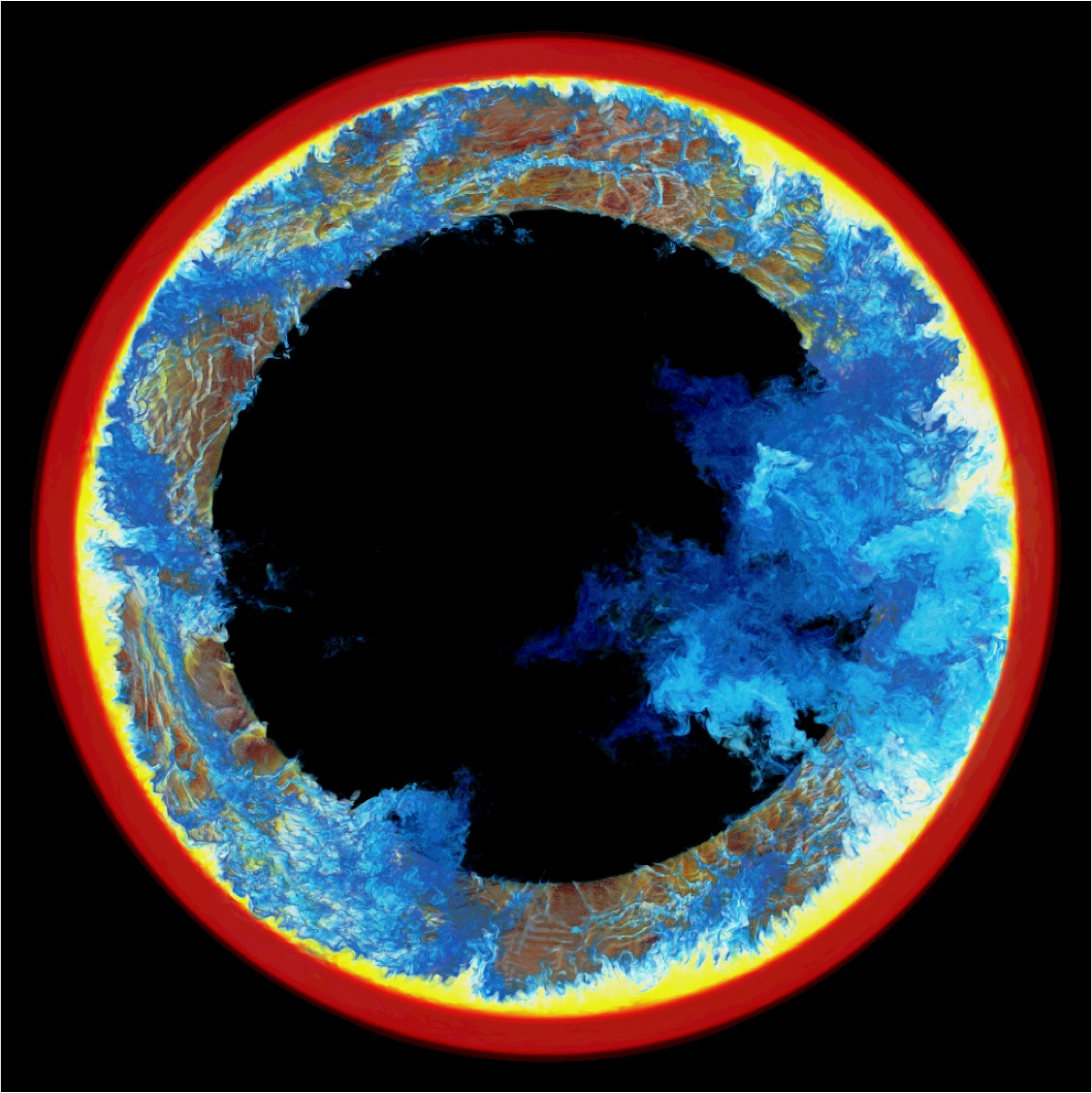}
   \includegraphics[width=0.5\textwidth]{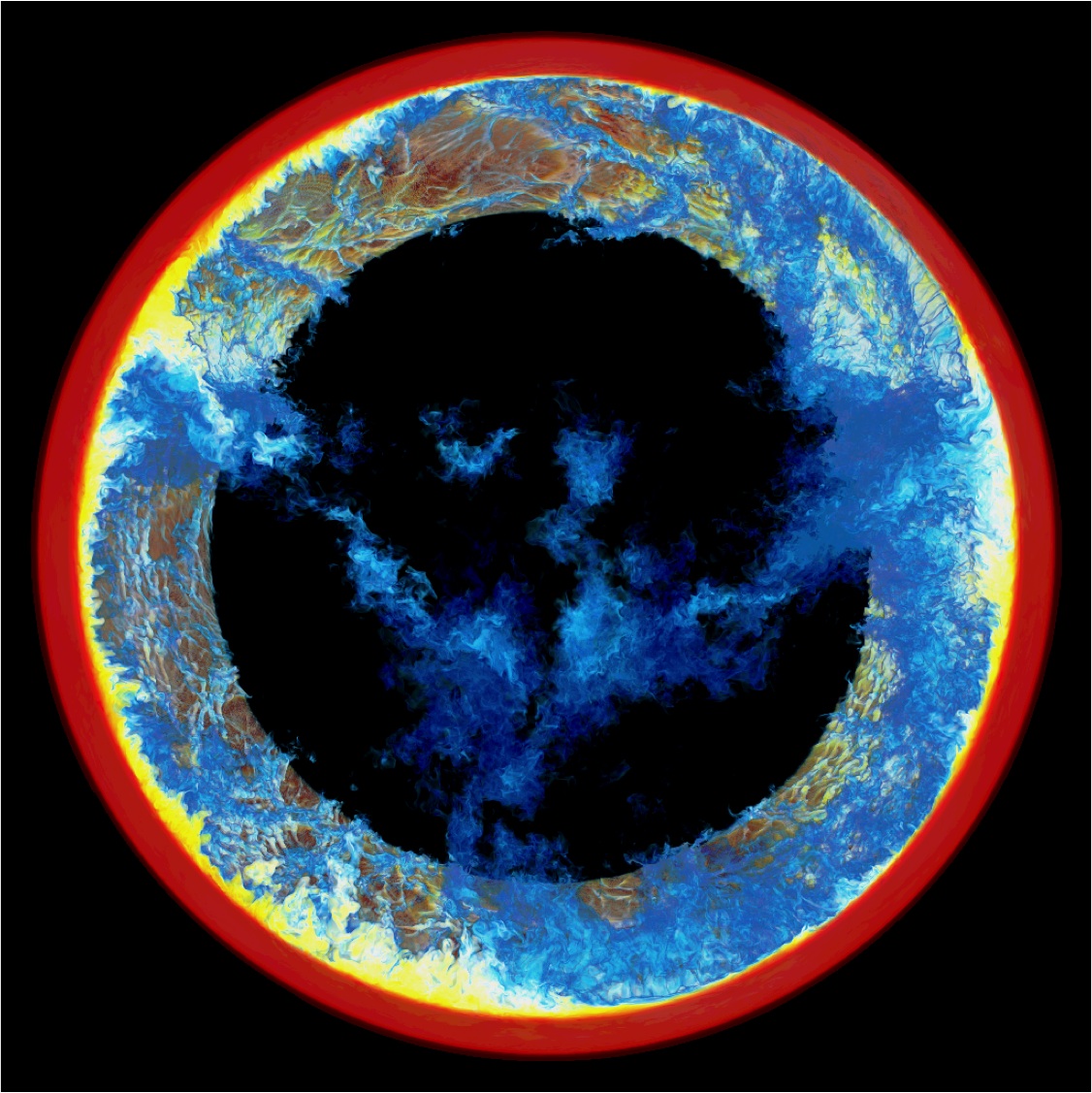}
   \includegraphics[width=0.5\textwidth]{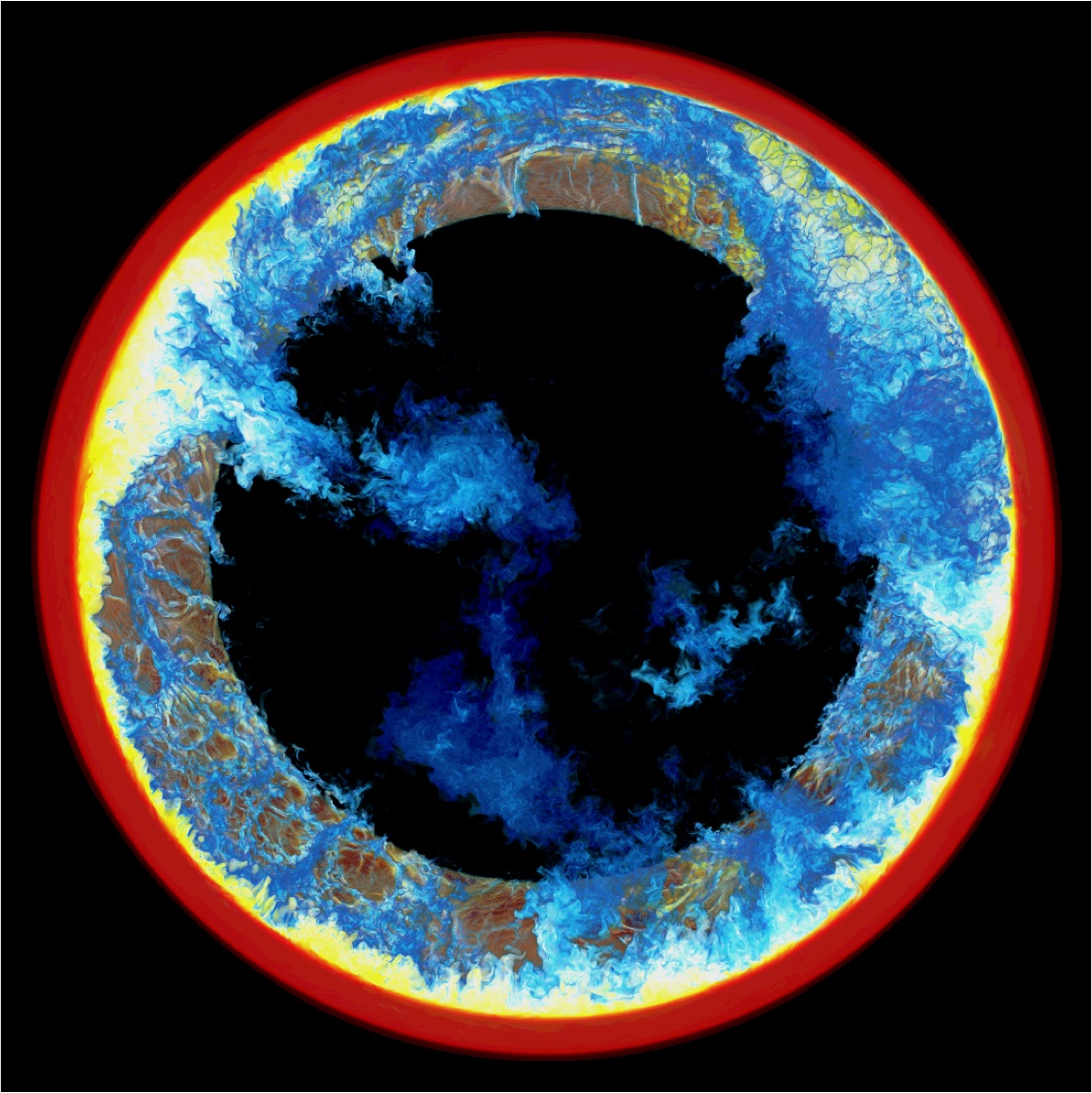}
   \caption{As in \abb{fig:FV_half} four snapshots of the the logarithm of the
     fractional volume of the 'H+He' fluid is shown as it is entrained
     into the convection zone. The lighter and darker blue color
     corresponds to concentration of $10^{-5}$ to $10^{-6}$. The front
     and rear $25 \%$ are cut away in order to show the interior
     distirubtion of entrained material. Left
     top: $t=510\mem{min}$, right top:  $t=555\mem{min}$, left bottom:
     $t=625\mem{min}$ , right bottom:  $t=640\mem{min}$    }
   \label{fig:FV_ring}
\end{figure*}

\paragraph{Aspect ratio and geometry of simulation domain}
Shell-flash convection typically starts out with a small aspect
ratio. This regime is suitable even for plane-parallel simulations
\citep{herwig:06a}. As the thermo-nuclear runaway proceeds, much of the
energy released goes into expansion work and lifts the outer layers of
the convection zone and above leading to an increasing aspect
ratio. Stellar evolution models suggest that the energy release from
ingested hydrogen may induce a further dramatic expansion of the
convection zone and increase of the  aspect ratio. For large aspect
ratios geometric properties of the simulation domain become
important. We would expect that entrainment processes be driven by
the velocity field at the upper convective boundary, which depends on
the realistic simulation of the large-scale cells. In this situation a
3-D 4$\pi$ simulation domain is therefore likely to give the
most realistic representation of convection flows and therefore
entrainment. 

\paragraph{Scales and turbulence}
Images of the abundance distribution clearly show the large-scale
nature of convectively-driven entrainment shear flows at the upper
convective boundary (bottom-right panel, \abb{fig:FV_half}) and the
subsequent downward advection in coherent cloud systems
(\abb{fig:FV_ring}). The picture of convective cells, however, has
limits. The downward directed H-rich plumes are highly turbulent and
have significant substructure. This is supported by the vorticity
visualization (\abb{fig:vort_480}), which shows that the convection
zone is highly turbulent throughout, and that the large-scale features that
can be identified in the mixing images (\abb{fig:FV_ring}) carry along
with them a full range of smaller-scale motions.  Vorticity, arising
from derivatives of the velocity field, accentuates the smallest
scales, and therefore, looking at the vorticity alone would make the
notion of dominant large-scale structures less
assertive. Nevertheless, the largest features are those corresponding
to the bulk convective motion from the bottom to the top of the
convection zones, and can be identified to span the radial extent of
the convection zone ($\approx 19,000 \mathrm{km}$). The largest
turbulent scale can be determined by finding the largest scale at
which the direction of the convective large-scale motions cannot be
identified. This scale is $\approx 2,800\mathrm{km}$.

\begin{figure}[tb]                    
   \includegraphics[width=0.5\textwidth]{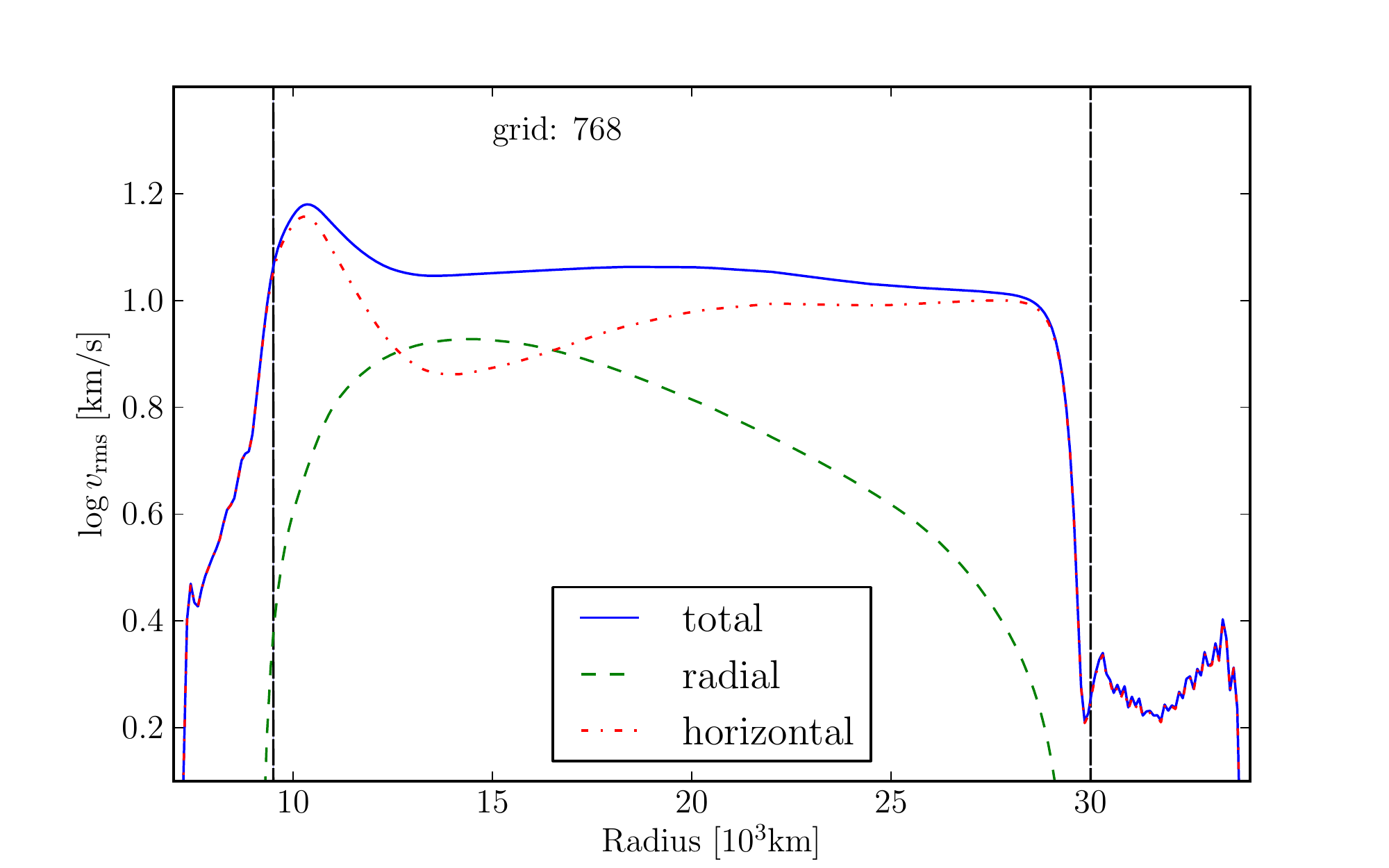}
   \includegraphics[width=0.5\textwidth]{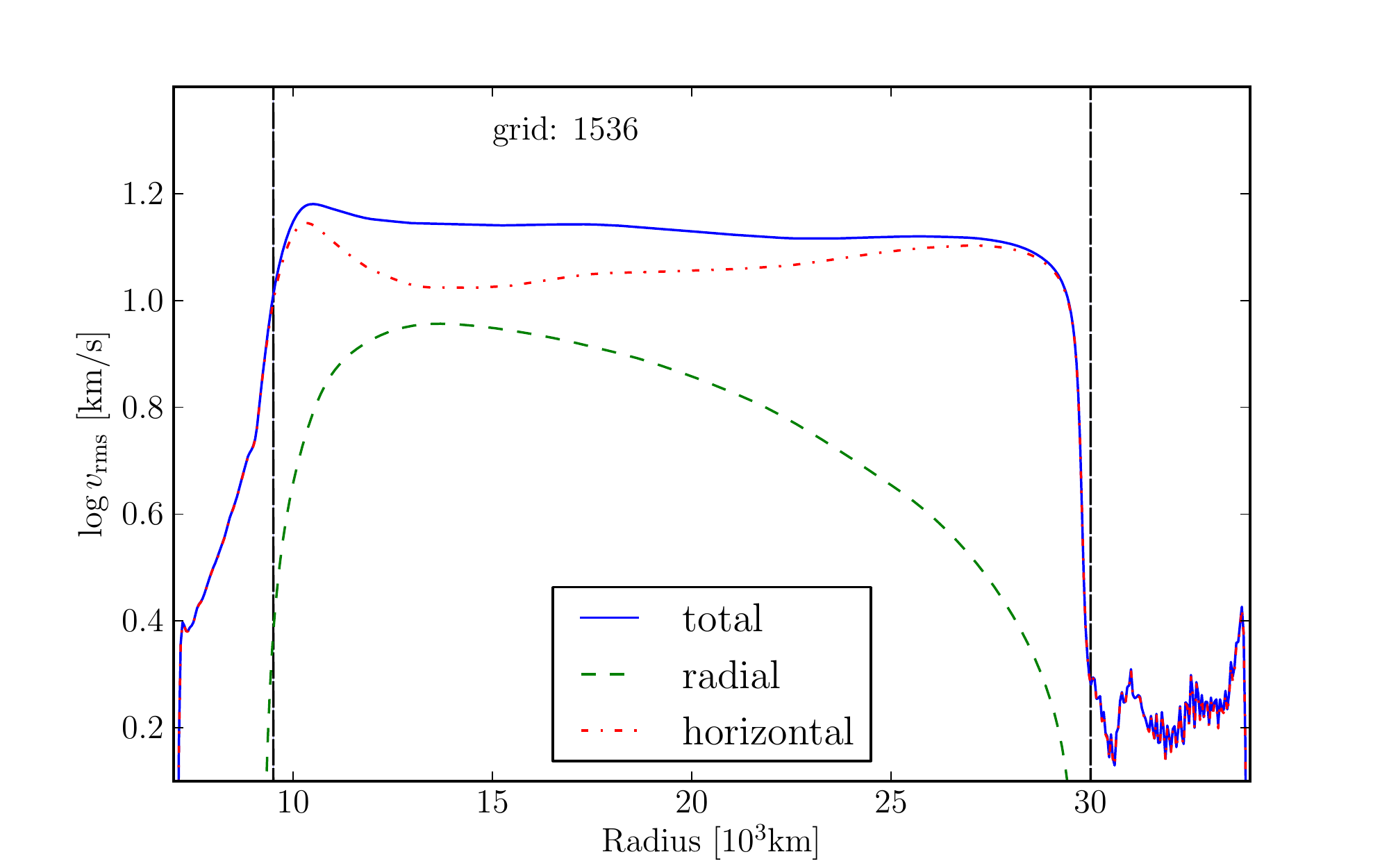}
   \caption{Logarithm of spherically averaged rms velocity, averaged
     from minute $400$ to $430$, for two resolutions. The dashed
     vertical lines are placed at the convective boundaries according
     to the initial setup.}
   \label{fig:profiles_velocity_averaged}
\end{figure}

\paragraph{The velocity field and the lower convective boundary}
We return to the discussion of velocities in our simulations. We
follow the simulations for several hundred minutes past the initial
transient phase (see above), where the convection time scale is
$\approx 25\mathrm{min}$ (\abb{fig:Ekmax}). From the spherically and
radially averaged rms velocity, we obtain an overall velocity scale for
our particular convection setup of $v_\mathrm{rms} \approx 11$ to
$13.5\mathrm{km/s}$ depending on resolution.

The spherically averaged profiles of the rms velocity, as well as the
vertical and horizontal components
(\abb{fig:profiles_velocity_averaged}), reveal the dominance of the
horizontal over the vertical convective velocities near the convective
boundary. This was described for the upper convection zone in the
previous paragraph, but it applies equally to the bottom convection
zone for the same reasons. Also here the radial velocities decrease
well inside the convective boundaries while the horizontal velocities
that are associated with the turn-around of fluid elements peak right
above the convective boundaries.
\begin{figure}[tb]               
   \includegraphics[width=0.5\textwidth]{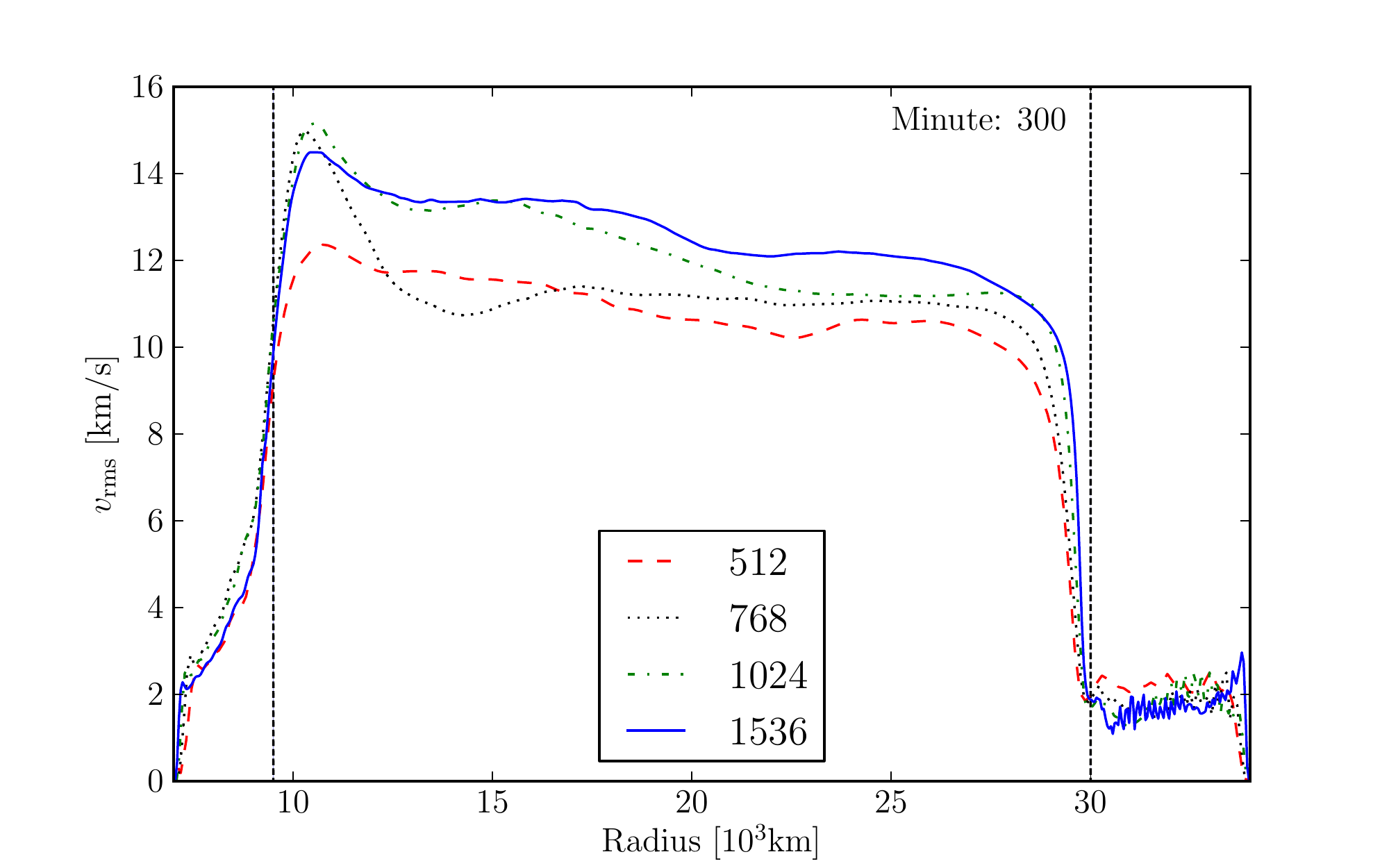}
   \includegraphics[width=0.5\textwidth]{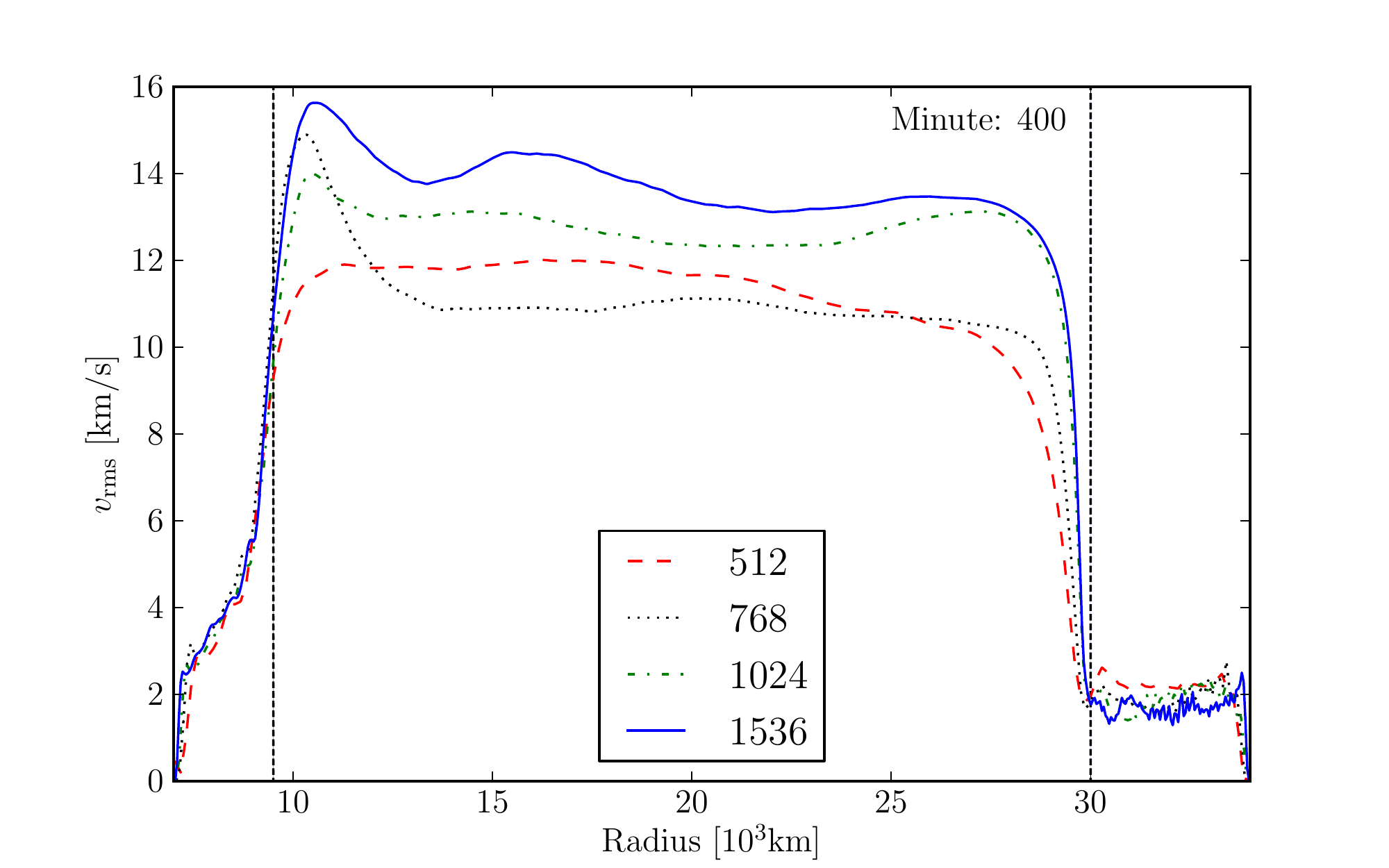}
   \includegraphics[width=0.5\textwidth]{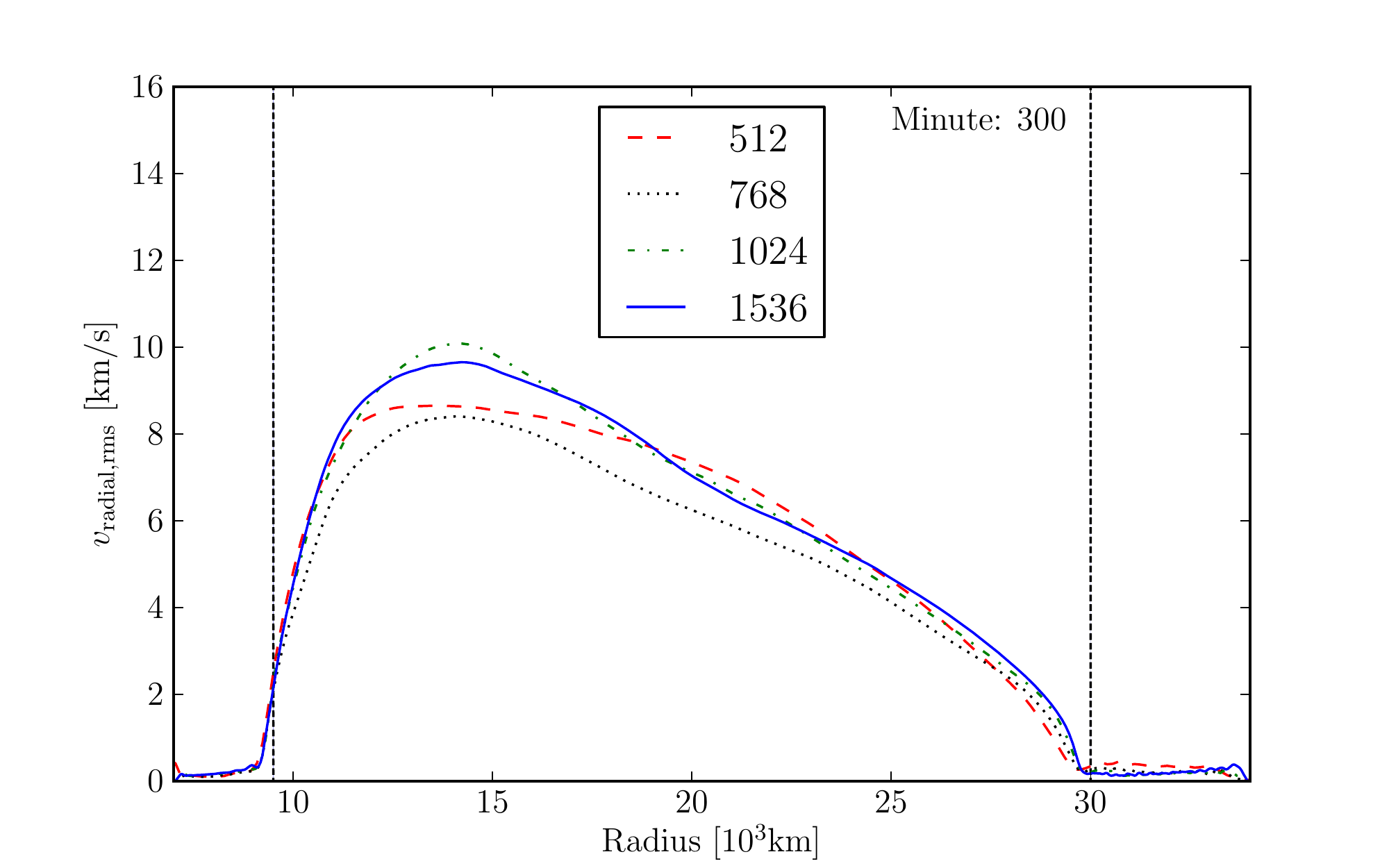}
   \caption{Spherically averaged rms velocity for $300$ and
     $400\mathrm{min}$ (not averaged in time), and just the radial
     component for $300\mathrm{min}$, for all four grid size
     simulations. The dashed vertical lines are placed at the
     convective boundaries according to the initial setup.}
   \label{fig:profiles_velocity_resolution}
\end{figure} 

\subsection{The entrainment process}
\label{sec:entrainment}
We noted above the importance of the large-scale motions for the
entrainment process. Our simulations show for this setup consistently
that large upwelling features occupy a full quadrant or a larger
fraction of an entire hemisphere. When these upwelling fluid elements
approach the upper boundary they are turning around. The
vertical velocities are decreasing while the horizontal
motions increase (top panels
\abb{fig:profiles_velocity_averaged}). The contniuity equation demands
that fluid elements start turning around some distance away from the convection boundary (which
they cannot cross in highly efficient convection in the deep stellar
interior, cf.\ \kap{sec:intro}). Thus, the large-scale flows cause
convection-induced shear flows at the convective boundary. 
The Richardson number is
\begin{equation}
  J = -\frac{g}{\rho}\frac{d\rho/dz}{(dU/dz)^2}
\label{eqn:Ri}
\end{equation}
where for our case $U$ is the horizontal velocity, $z$ is the radial
coordinate and $g$ and $\rho$ have their usual meaning.  For shear
flows to be unstable against Kelvin-Helmholtz instabilities it is
necessary that \[ J < \frac{1}{4} \] \citep{chandrasekhar:61}. From a
1D profile it is not obvious how large the transition region should be
assumed to be. In the region of the upper convection boundary a
pressure scale height is $H_\mem{p}\approx 1000\mathrm{km}$. The
actual velocities of the shear flows will have a range around the
value given by the spherically averaged 1D profile. Therefore, we show
in \abb{fig:Ri-number} the Richardson number as a function of an
assumed shear velocity difference $\Delta U$ for positions $r$ within
the region of a pressure scale height below the upper convective
boundary at $r_\mem{top} = \natlog{30}{6}\mathrm{km}$. We approximate
$dz$ in \glt{eqn:Ri} with $\Delta r = r - r_\mem{top}$ (where $r$ is
the radius), $d\rho$ with $\Delta \rho = \rho(r) - \rho( r_\mem{top})$,
$dU$ with $\Delta U = U(r)- U( r_\mem{top})$. We also provide the
profile of the spherically averaged velocities at time
$300\mathrm{min}$. Even taking into account that the local horizontal
velocities are fluctuating around the spherical average the Richardson
number indicates that the convectively-driven horizontal shear flows
are stable against Kelvin-Helmholtz instabilities. 

This is consistent with the flow images of the convective boundary
where the horizontal flows take the form of coherent gusts that sweep
along over the inside of the large upwelling patches visible at the
top convection boundary.
Larger patches of such apparently stable and very thin boundary layers
can be seen in \abb{fig:FV_compare} in locations (here, e.g. the lower
left) where radially upwelling flows first encounter the convective
boundary, and then are deflected.

\abb{fig:composite} shows at the bottom left a sliced view of another
Kelvin-Helmholtz stable part of the horizontal flow.
In the region near the center of this image, two such horizontal flows
directed toward each other meet and are deflected downward.  This is
the region where the entrainment of H-rich fluid occurs.  The
deflection of the flow downward, away from the top of the convection
zone, is akin to the well-known phenomenon of 
boundary-layer separation.  However, in this instance, we do not have
a solid wall, as in engineering flows, but a strongly stable
transition layer from one fluid to another. Boundary-layer separation
flows are highly non-linear and cannot be described by simple
relations \citep[e.g][]{Simpson:1989ef}. Quantitative characterizations
of such flows do therefore rely entirely on sufficiently resolved
simulations or suitable experiments. 

In \abb{fig:boundaryKHzoomed} we see zoomed-in views of the vorticity
magnitude in the flow near the top of the convection zone, taken from
the larger images in \abb{fig:composite}.
In these zoomed-in views, we indicate with prominent white double
arrows the 500 km thickness (11 grid cell widths) of the initial
transition region from the fluid of the convection zone to the more
buoyant, stably stratified 'H+He' fluid above it.  In visualizing the
concentration of the 'H+He' fluid in the images at the bottom of
\abb{fig:boundaryKHzoomed}, we have used the 10 moments of this
variable (see Appendix) that are updated by the PPB advection scheme
to generate 8 octant averages in each grid cell.  This permits the
full resolution of the PPB scheme, with its subcell information,
to be seen in this double resolution visualization process.  For
the image of the 'H+He' fluid concentration, the white double arrow
is thus 22 voxels wide.
In both zoomed-out and zoomed-in views we see that the transition
layer at the top of the convection zone is strongly stable
to the Kelvin-Helmholtz instability.  This is due to the great
strength of gravity and the Rayleigh-Taylor stable stratification
of this layer.  Nevertheless, at the tip of the wedge of turbulent
gas that separates the two large convection cells and that serves
to deflect their flows downward, we clearly see that Kelvin-Helmholtz
roll-up indeed occurs.  

At these locations, we no
longer have the parallel flow in a single direction assumed in the
text-book Kelvin-Helmholtz stability analysis. Instead we have two
colliding flows that must, at these locations, generate vertical
motions perpendicular to the transition layer that marks the top of
the convection zone, in order to satisfy the continuity equation for
this essentially incompressible flow.  A classic series of breaking
Kelvin-Helmholtz waves is formed at each tip of this wedge of
relatively stagnant gas between the large convection cells.  Not only
is low-concentration H-rich gas entrained into the gas of this wedge,
but also along the shear layer between the wedge and convection cell
gases entrainment into the rapidly moving convection cell gas occurs.
This process is perhaps more clearly seen in the zoomed-in views of
the concentration of H-rich gas in \abb{fig:boundaryKHzoomed}.  In
these zoomed-in images of the H-rich gas concentration, the slice
through the domain is 3 times thicker, namely $0.6\%$ of the domain,
and therefore the thickness of the red region in these images of more
concentrated H-rich gas is exaggerated slightly by being seen in
projection through the transparent region of pure H-rich gas.  These
images clearly show the strong decrease in H-rich gas concentration as
we go away from the wedge tip and away from the top of the convection
zone.  Nevertheless, the turbulent eddies in this region bring about
mixing, and there is also some small amount of entrainment from the
transition layer even well away from the wedge tip. Further down, as
the horizontal velocity further decreases and the inward radial
velocity component further increases, this leads to elongated downward
entrainment curtains that carry clouds of material somewhat enriched
with the 'H+He' fluid from above down into the convection zone
(\abb{fig:FV_ring}).

The results of
our measured amounts of entrainment in our series of runs with
increasing grid resolution, given in \kap{sec:entrain_grids},
indicates that the extent to which we have
resolved this process in the highest resolution simulation that is
shown in \abb{fig:boundaryKHzoomed} is likely to be sufficient to compute the time and
space averaged entrainment rate with reasonable precision.
\begin{figure}[tb]               
   \includegraphics[width=0.5\textwidth]{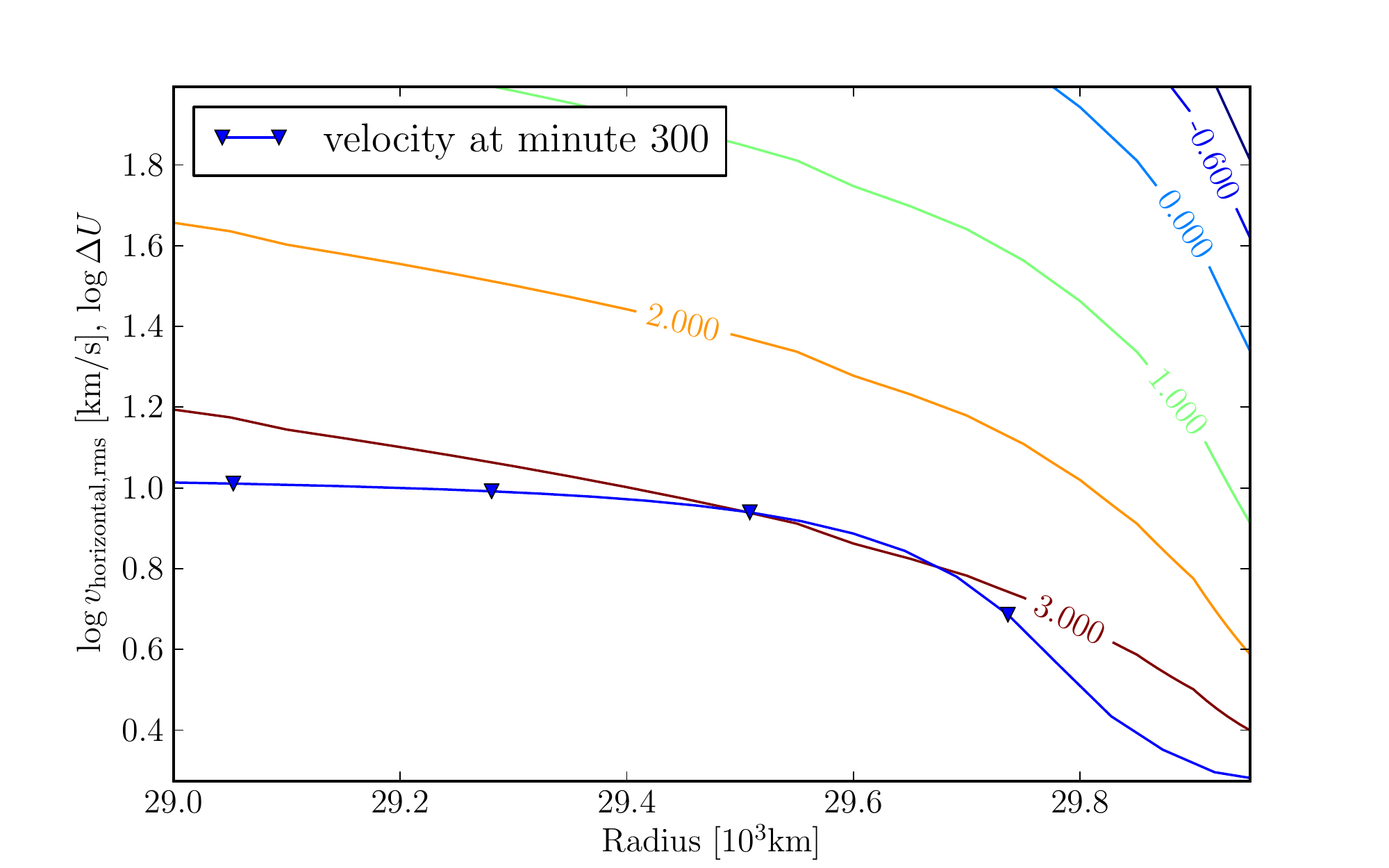}
   \caption{Lines of constant Richardson number $\log J$ as a
     function of distance from the boundary and assumed velocity
     difference $U$ between that distance and the midpoint of the
     boundary at $\natlog{30.0}{6}\mathrm{m}$. The line with triangle
     markers shows the spherically averaged profile of the horizontal
     velocity component. For stratified shear flow that is unstable against
     Kelvin-Helmholtz instabilities $\log J < -0.6$ (see text). }
   \label{fig:Ri-number}
\end{figure} 
\begin{figure*}[tb]               
\centering
   \includegraphics[width=0.4\textwidth]{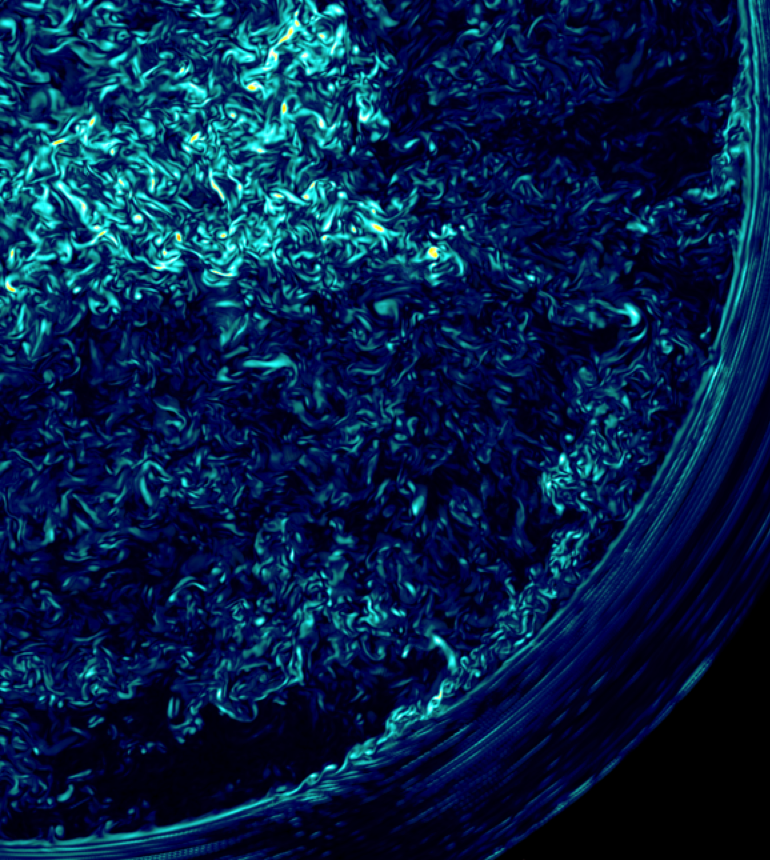}
   \includegraphics[width=0.4\textwidth]{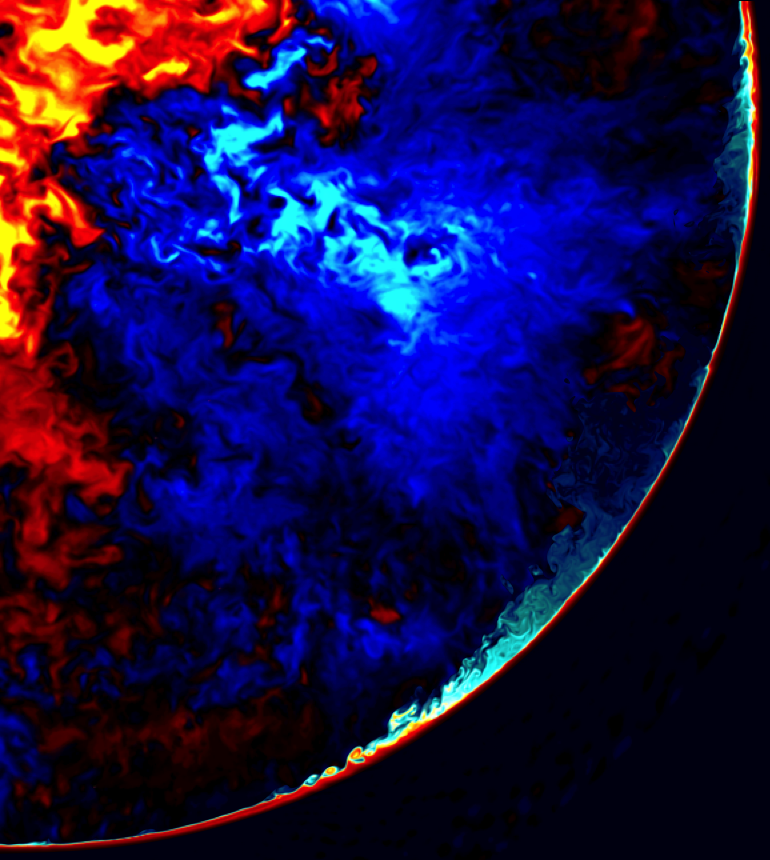}
   \caption{The vorticity magnitude (left), and (right) the radial
     (blue inward, yellow/red
     outward) velocity and the fractional volume of the H+He fluid
     (visibile only near the outer boundary of the convection zone) in
a slice through the domain with a thickness of $0.2\%$ of the domain
     for time $408\mathrm{min}$ from the $1536^3$-grid simulation.}
   \label{fig:composite}
\end{figure*} 

\begin{figure*}[tb]               
\centering
   \includegraphics[width=0.8\textwidth]{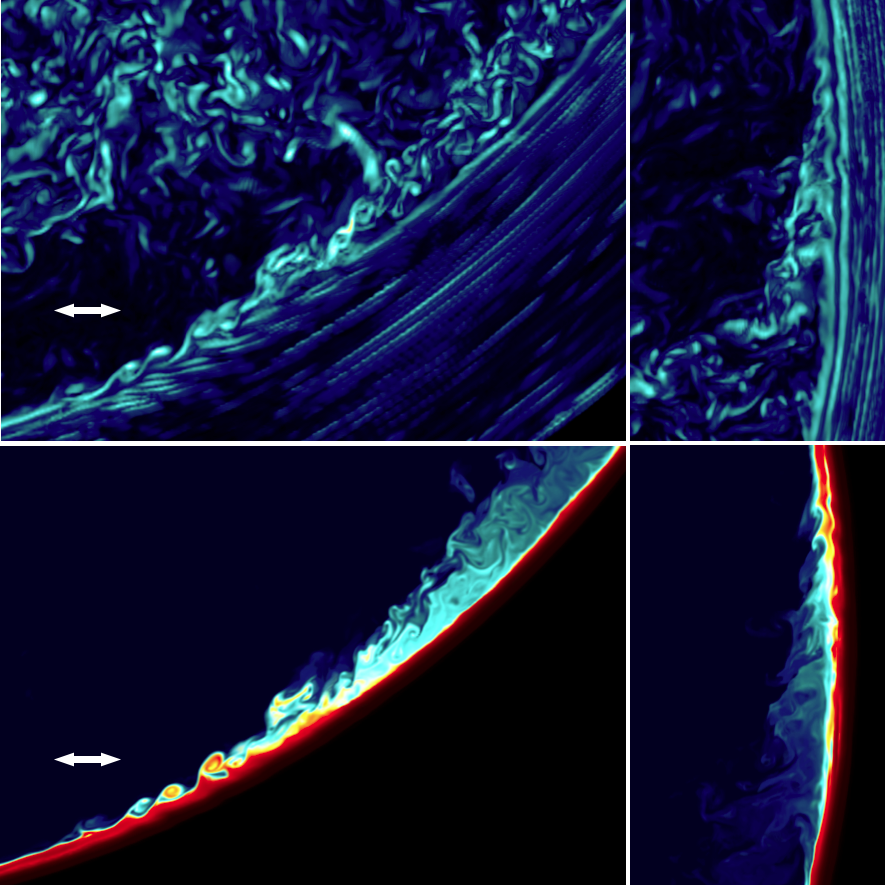}
   \caption{Zoomed-in view of Kelvin-Helmholtz unstable breaking waves
     in the boundary-separation layer shown in
     \abb{fig:composite}. The white double arrows indicate the 500 km thickness
    of the initial transition layer between the gas of the convection zone and
    the more buoyant, stably stratified 'H+He' gas above it. }
   \label{fig:boundaryKHzoomed}
\end{figure*}

As the fluid elements enriched with 'H+He' fluid are dragged down
inside the entrainment curtain, horizontal turbulence soon disperses
and dilutes the 'H+He' fluid, and its fractional volume falls below
the rendering cut-off chosen for the visualization. At that point the
'H+He' fluid becomes invisible in \abb{fig:FV_ring}. The cumulative
entrainment evolution is shown by means of spherically averaged
abundance profiles in \abb{fig:entrained_profiles} and by zoomed in
views of spherically averaged entropy profiles in
\abb{fig:A-top-boundary}.  In \abb{fig:entrained_profiles} we have
transformed the fractional volume, which is the primary quantity to
express abundances in the hydrodynamics code, into mass fractions,
which are customary in stellar evolution codes. Even at the end of the
simulation, after 12 convective turn-over times (24 convection time
scales), the abundance profiles show significant variations in the
lower part of the convection zone. These features are also highly
variable in time and represent individual clouds of 'H+He'-enriched
material to arrive at the still mostly pristine lower part of the
convection zone.

We can summarize that the entrainment process in our simulations
consists of three closely interacting components. Large-scale
convective upwelling motions create fast horizontal flows over big
patches of the convective boundary surface. The horizontal convective
flows are initially stable against
Kelvin-Helmholtz instabilities. However, where such opposing flows
meet, the boundary layer separates from the stiff convective boundary
and Kelvin-Helmholtz instabilities are encountered in this highly
unstable environment. Finally, the interacting and colliding horizontal
gusts lead to coherent downward flows that contain 'H+He'
fluid. Therefore, the entrainment process is a combination of the
local properties at the convective boundaries and the global
properties of the convection that depend on all aspects of the
particular situation throughout the convection zone. 

\begin{figure}[tb]           
   \includegraphics[width=0.5\textwidth]{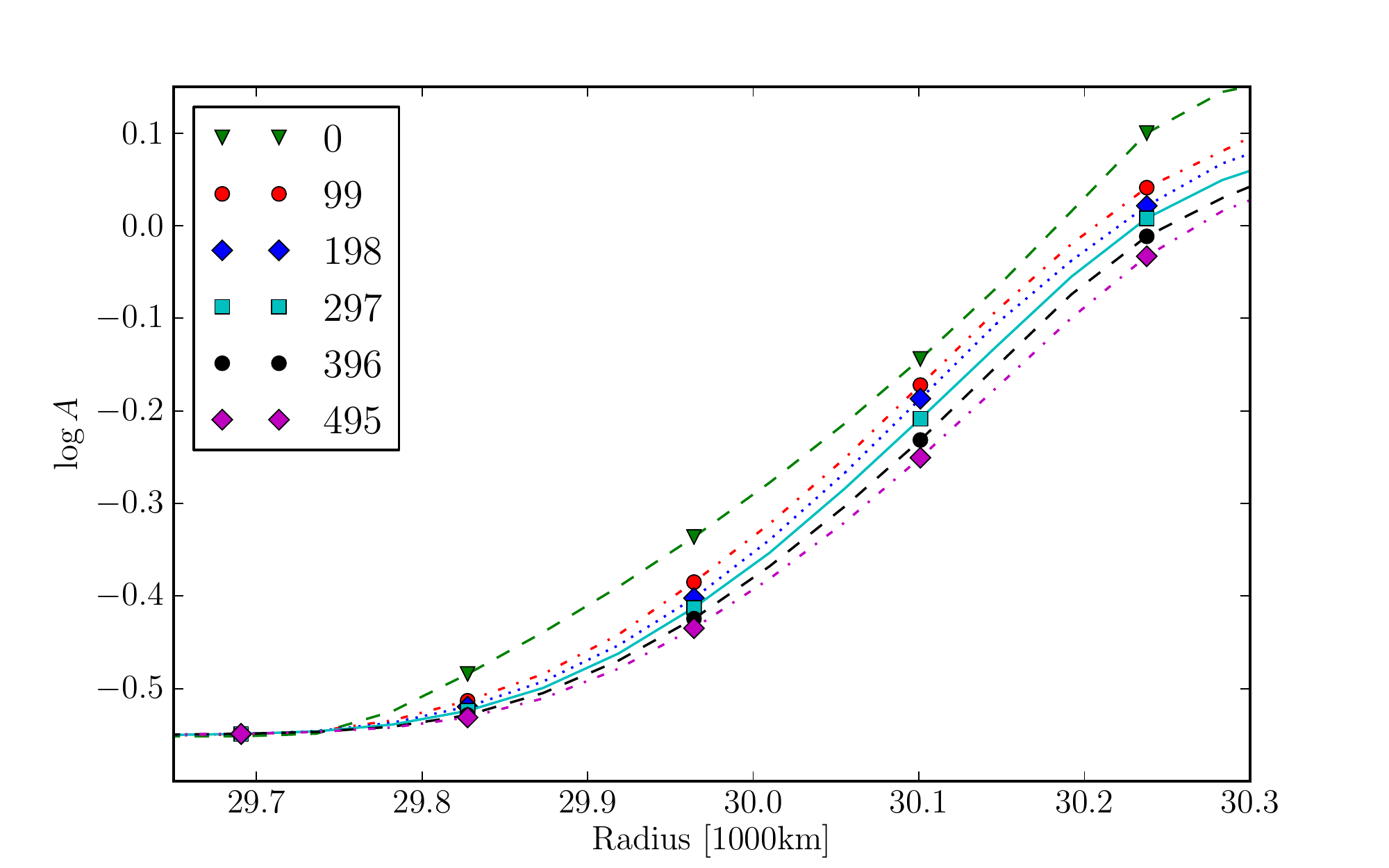}
   \caption{Entropy \glp{eq:entropy-like} evolution at the top of the convection zone for
     the $1536^3$ simulation. Logarithm of entropy in the
     transition layer for the initial setup (dump 0) and subsequent
     evolution (labels in minutes simulated time). A line mark is
     placed every third grid point.}
   \label{fig:A-top-boundary}
\end{figure}

\subsection{Entrainment simulations for a
  range of grid sizes}
\label{sec:entrain_grids}
 An important concern in any multi-dimensional simulation is the
 dependence of results on the adopted grid resolution. Here we focus
 on a qualitative discussion of the velocity field and a more
 quantitative analysis of the entrainment process. But first we would
 like to comment on the effect of numcerical viscosity.

\begin{figure*}[tbp]                                          
   \includegraphics[width=0.5\textwidth]{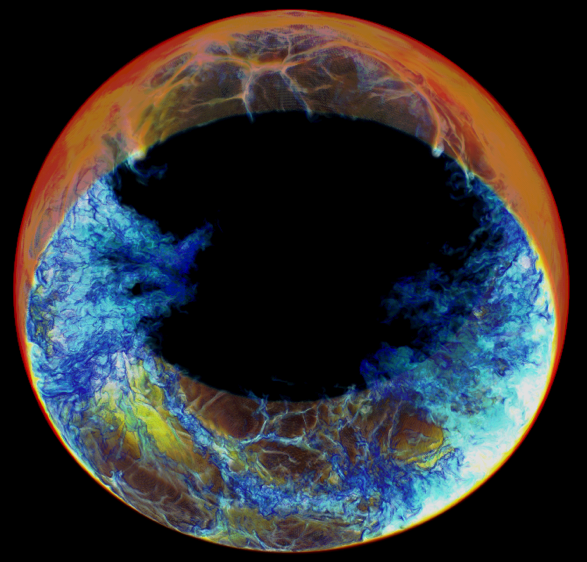}
   \includegraphics[width=0.5\textwidth]{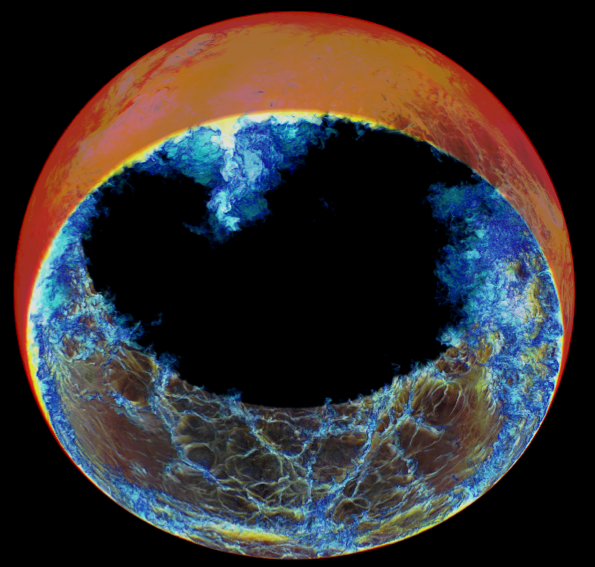}
   \caption{Logarithm of fractional volume of 'H+He' fluid shown for a
     slice of 3-D domain at $t=491\mathrm{min}$. The smallest visible
     values are $10^{-6}$ to $10^{-7}$. The blue descending plumes trace out the
     convection cells. \emph{Left:} $768^3$ grid, \emph{Right:}
     $1536^3$ grid.}
   \label{fig:FV_compare}
\end{figure*}  

\paragraph{Numerical viscosity}
One might expect the numerical viscosity in the
simulation to affect the details of the entrainment process.  The
image shown in \abb{fig:vort_480} has $1121^2$ pixels, and the
underlying simulation has a $1536^3$ grid. Smallest features that can
be identified have a scale of $~\approx 100\mathrm{km}$, which
corresponds to only a few grid zones.  The train of
  Kelvin-Helmoltz waves in the boundary-separation layer
  (\kap{sec:entrainment}) has a typical length of $1000\mathrm{km}$
  but the width of this feature is only a fraction of that. In
\abb{fig:boundaryKHzoomed} we show zoomed-in views of such
boundary-separation layers.  The double-arrows indicate $\Delta
r_\mem{H-shell} = 500\mathrm{km}$, the width of the transition layer
over which the mixing fraction of hydrogen-rich gas changes from
nearly zero to nearly unity.  From these close-up views we can see that
the simulation's $1536^3$ grid is resolving vorticity features that
are considerably smaller than $11$ grid cells.  The double resolution
views of the 'H+He' concentration, made possible by the subcell
resolution of the PPB advection scheme (see Appendix) show that
still smaller features in the advected concentration can be made out.
This subcell resolution feature of PPB advection can be seen in the simple
advection test problem shown in Fig.\,2 in \citet{woodward:08a}.

Detailed simulations of shear layer instability in isolation, as
  well as our earlier simulations of this sort of convection-driven
  shear in small sections of such a flow \citep{woodward:08a} indicate
  that the development of the unstable layer tends to converge when we
  are able to resolve the largest relevant scales as well as scales of
  wavelengths smaller by about a factor of three.  This is just a rule
  of thumb, of course.  Only quantitative convergence under grid
  refinement of a quantity of interest, which in our present
  investigation is the entrainment rate for the hydrogen-rich gas, can
  be our guide to the adequacy or inadequacy of our computational
  grid.  The appearance of resolved swirls on multiple scales in the
  figure is an indication that our grid may be fine enough, but it is
  the actual observed convergence of the entrainment rate that is our
  more reliable guide.  The principal causes of error in the
  entrainment in our numerical simulation are the limitations imposed
  by the grid cell size on the smallest resolvable overturning waves
  that can produce mixing.

  The behavior of PPM gas dynamics simulations of such complex,
  nonlinear, turbulent flows, with and without PPB advection of
  multifluid volume fractions, has been studied in comparison to
  matching simulations with Navier-Stokes viscosity terms added in
  \citet{Sytine:2000fh} and \citet{Woodward:2010us}.

\citet{porter:94}  carefully investigated the numerical viscosity of
  our PPM gas dynamics scheme and quantified it as consisting of two
  terms, the effects of each of which tend toward zero as a
  disturbance wavelength increases.  The effects of physical,
  Navier-Stokes viscosity tend toward zero as $\lambda^{-2}$, where
  $\lambda$ is the disturbance wavelength. The actual viscosity in the star
  is smaller than could be represented in our simulation.  Instead, it
  is the numerical viscosity of PPM that is dominant in the
  simulation.  This numerical viscosity tends to zero as
  $\lambda^{-3}$ or $\lambda^{-4}$, depending upon which of our two
  numerical terms is larger under the particular circumstances.  For
  the numerical viscosity, it is the disturbance wavelength measured
  in grid cell widths that matters.  When doubling the grid resolution
  the impact of the numerical viscosity on the flow, as measured by
  effective Reynolds numbers for wavelengths of physical importance,
  is diminished by a factor of four or a factor of eight, depending
  upon which of our two terms dominates.  Therefore, when our results
  converge under grid refinement, it must be true that our numerical
  viscosity no longer matters.  The physical reason why this can
  happen is that the entrainment due to breaking Kelvin-Helmholtz
  waves (\kap{sec:entrainment}) is not a viscous phenomenon.
  Viscosity determines the wavelengths at which shear instabilities
  are damped away, but it is the largest wavelengths, largely
  unaffected by viscous damping, that are chiefly responsible for the
  entrainment.  These we may hope to capture in a well-resolved
  simulation.  Our results in \kap{sec:entrain_grids} indicate that
  our finest grid of $1536^3$ cells has just marginally captured these
  important waves accurately.  One could attempt to double the grid
  resolution once more to lay any doubt on this point to rest, but we
  have judged that the cost of doing this on present computing
  equipment makes this more definite demonstration of convergence
  impractical.

 \paragraph{Radial profiles of velocity}
The overall velocity scale, shown for the total rms velocity in
\abb{fig:Ekmax} increases somewhat with resolution. Inspection of the
separate components of spherically averaged velocities reveals that
this difference is mostly in the horizontal velocities that span a
range between $12\mathrm{km/s}$ for the $512^3$ grid and
$17\mathrm{km/s}$ for the highest resolution, while the radial
velocities for all resolutions fall in a narrow band between
$v_\mathrm{radial, rms} \approx 8$ and $10\mathrm{km/s}$
(\abb{fig:profiles_velocity_resolution}). The overall difference
between horizontal and vertical velocity components is larger for
higher resolution. However, as \abb{fig:profiles_velocity_averaged}
shows, the difference between the horizontal and radial velocity
profile shape is larger for lower resolution. In particular, the
$768^3$ grid horizontal velocity profile shows a much more pronounced
peak at the lower convective boundary and a significant depression
just above the heating zone where the radial velocities have a
maximum.

 \abb{fig:profiles_velocity_averaged} also shows that the gradient of
 the horizontal velocities becomes steeper for the higher resolution
 runs. In the upper part of the convection zone the total rms velocity
 is dominated by the horizontal component. This steeper horizontal
 velocity gradient can therefore also be seen in
 \abb{fig:profiles_velocity_resolution} at different times. In the
 inner part of the convection zone the velocities at a given radius
 are not always strictly correlated with resolution. This is partly
 due to long-term convective breathing modes that are out of phase for
 the different runs, as well as chaotic fluctuations. However, near
 the top boundary, where the horizontal velocity drops sharply, we
 always found velocity profiles to be strictly ordered by resolution,
 with the highest velocity and the steepest velocity gradients found
 for the highest resolution runs.

 \paragraph{Entrainment process and rate} 
Eventually we are interested to study the burning of the entrained
H-rich fuel via the $\czw(\p,\gamma)\ndr$ reaction and the feedback of
energy from that process into the hydrodynamic flow. The amount of
entrainment will determine the time-scale and type of feedback from
the energy release of the nuclear burning. This, in turn, will
determine the nucleosynthesis in reactive-convective environments,
including the n-capture process in \ipr\ conditions
\citep{herwig:10a}. The entrainment rates will be affected by the
hydrodynamic feedback from the ingested H. The entrainment rate and
the way in which the entrained H will be advectively distributed
throughout the He-shell flash convection zone will determine the
properties of the subsequent convective-reactive H-combustion, its
feedback into the convective flow and the conditions for
nucleosynthesis. 

A first test is the visual inspection and comparison of 3-D images of
the entrainment process for simulations with different resolutions. We
find no obvious artefacts or differences in scales or patterns when
comparing a $768^3$ and $1536^3$ simulation (\abb{fig:FV_compare}). In
both cases the large-scale upwelling flows that turn around at the
upper boundary of the convection zone cover large fractions of a
hemisphere, just as described in the previous section. Also the
scraping horizontal motions  associated with boundary-layer
  separation of converging horizontal flow at the convective boundary are evident in
both cases in very similar patterns. If anything, the interfacial
structures and the entrainment curtains have a somewhat smaller scale
in the higher resolution case. This is consistent with the steeper
gradient of the horizontal velocity found in the higher resolution
runs (see above). 

\begin{figure}[tbp]      
   \includegraphics[width=0.5\textwidth]{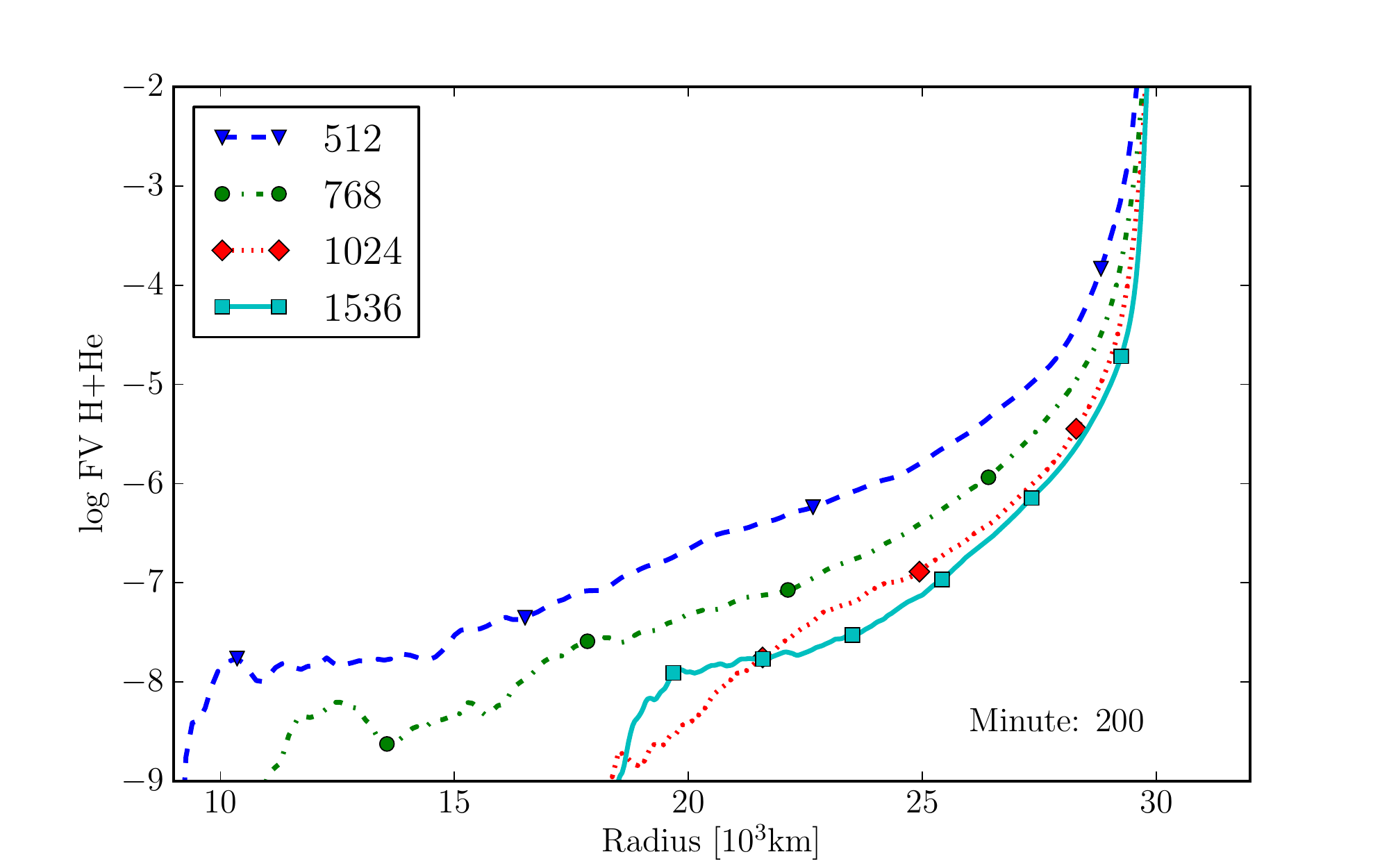}
   \includegraphics[width=0.5\textwidth]{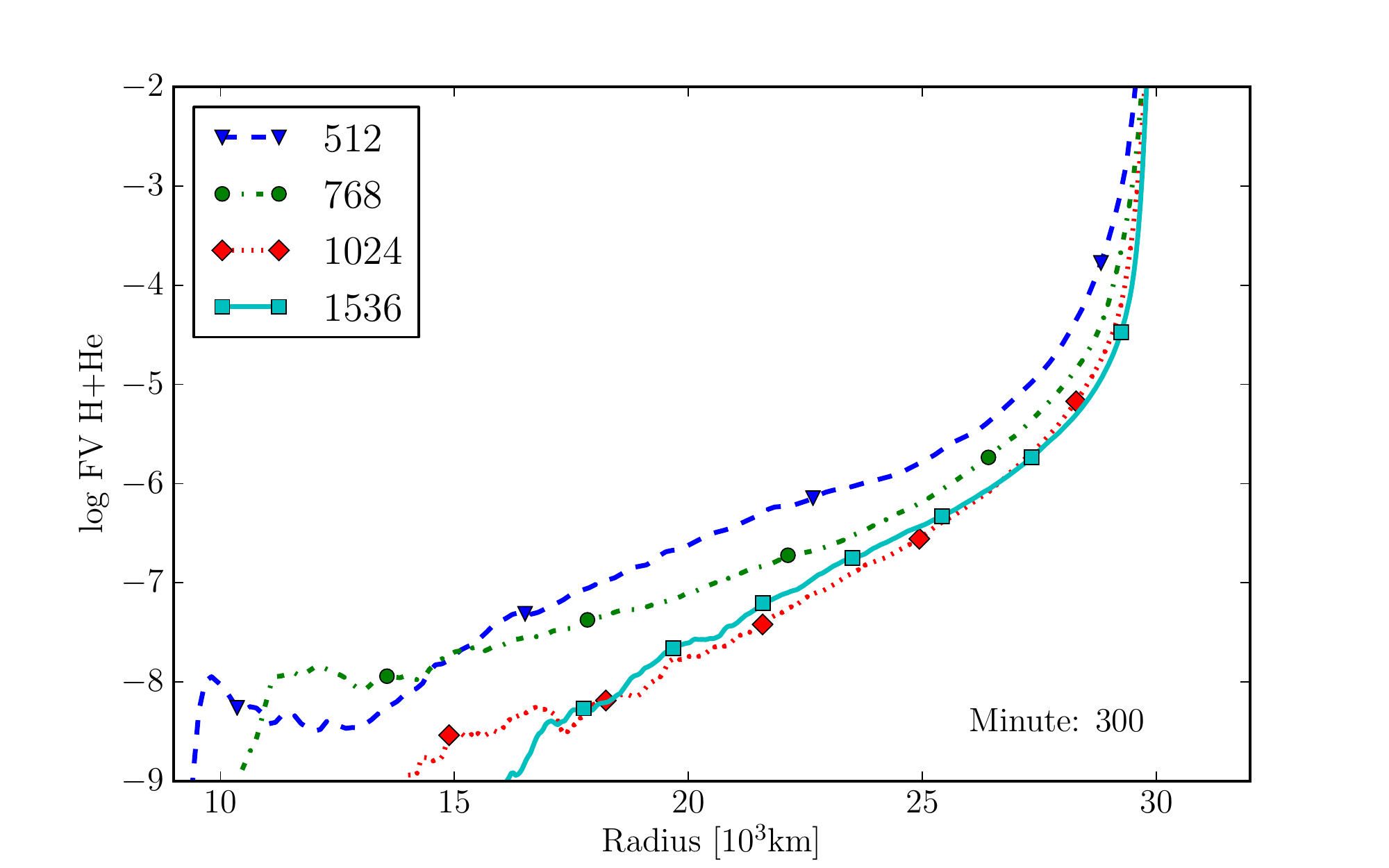}
   \includegraphics[width=0.5\textwidth]{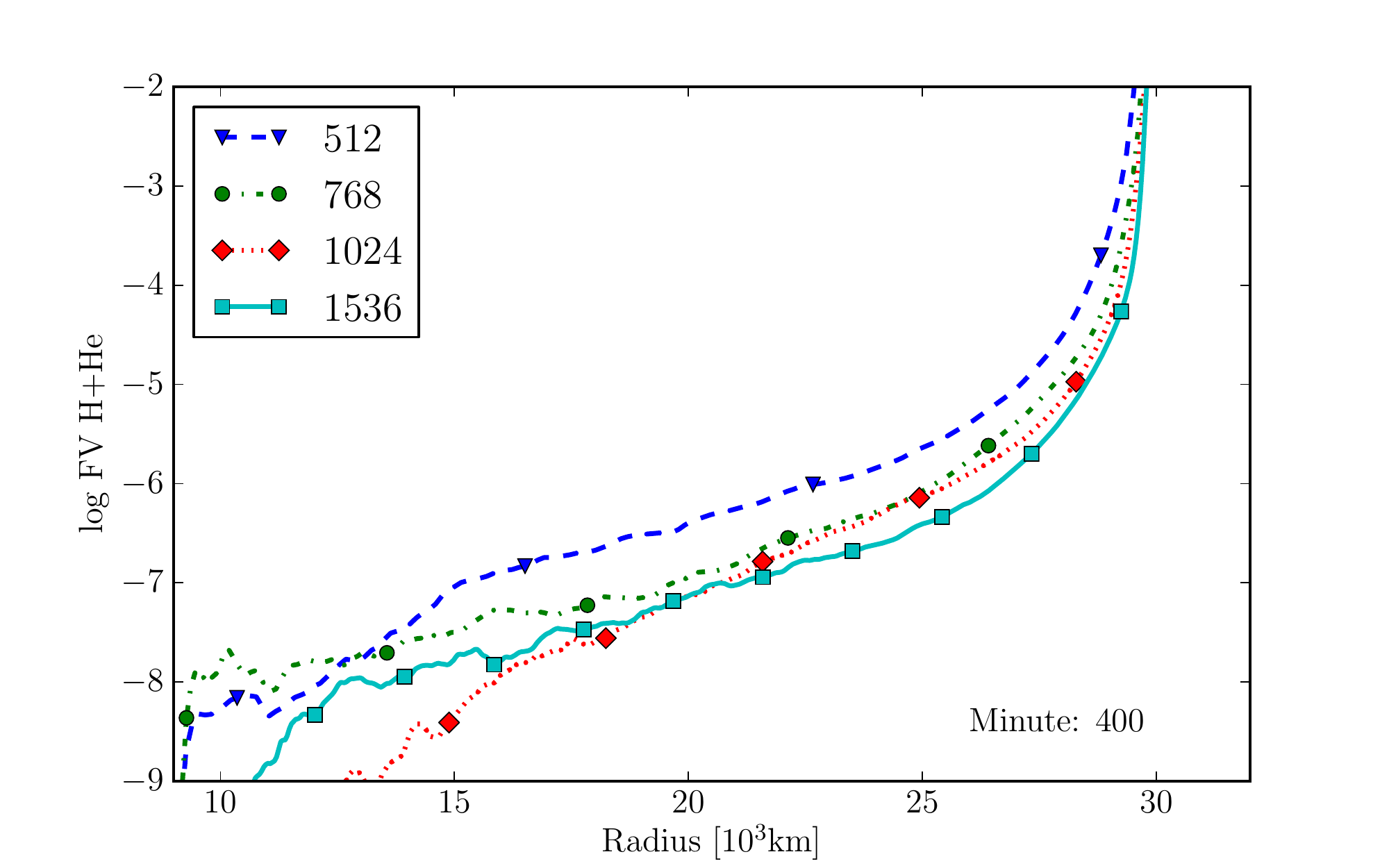}
   \caption{Profile of the entrained fractional volume of the H+He fluid for the different
     grid resolutions at (from top to bottom) $200$, $300$ and $400$ minutes. }
   \label{fig:profiles_resolution}
\end{figure} 
We have described above how the entrained 'H+He' fluid is cumulatively
filling the convection zone (\abb{fig:entrained_profiles}). In
\abb{fig:profiles_resolution} we show the spherically averaged
fractional volumes\footnote{The fractional volume $FV_\mathrm{1}$ is
  the primary variable to represent concentration of fluid 1 in the
  PPMstar hydrodynamics code. It is related to the mass fraction $X$
  by $X_\mathrm{1} = \rho_\mathrm{1} * FV_\mathrm{1} / \rho$, where
  $\rho_\mathrm{1} $ is the density of fluid 1 and $\rho$ is the
  density.} for all four grids at three times. The entrained mass can
be determined from these spherically averaged abundance profiles by
integrating over the convection zone. The profiles already reveal that
the entrained mass is similar in the two highest resolution runs, and
that the lowest grid case ($512^3$) has at all times a significantly
larger entrained mass. This pattern prevails through all times, which
is evident from the animation view of these profiles available at
\url{http://www.lcse.umn.edu/3Dstar-convection-entrainment}. The
profiles, especially in the animated presentation, are perhaps the
strongest indication of convergence of the entrainment rate in our
simulations. We can attempt to summarize this finding in the following
way.

\begin{figure}[tb]    
   \includegraphics[width=0.5\textwidth]{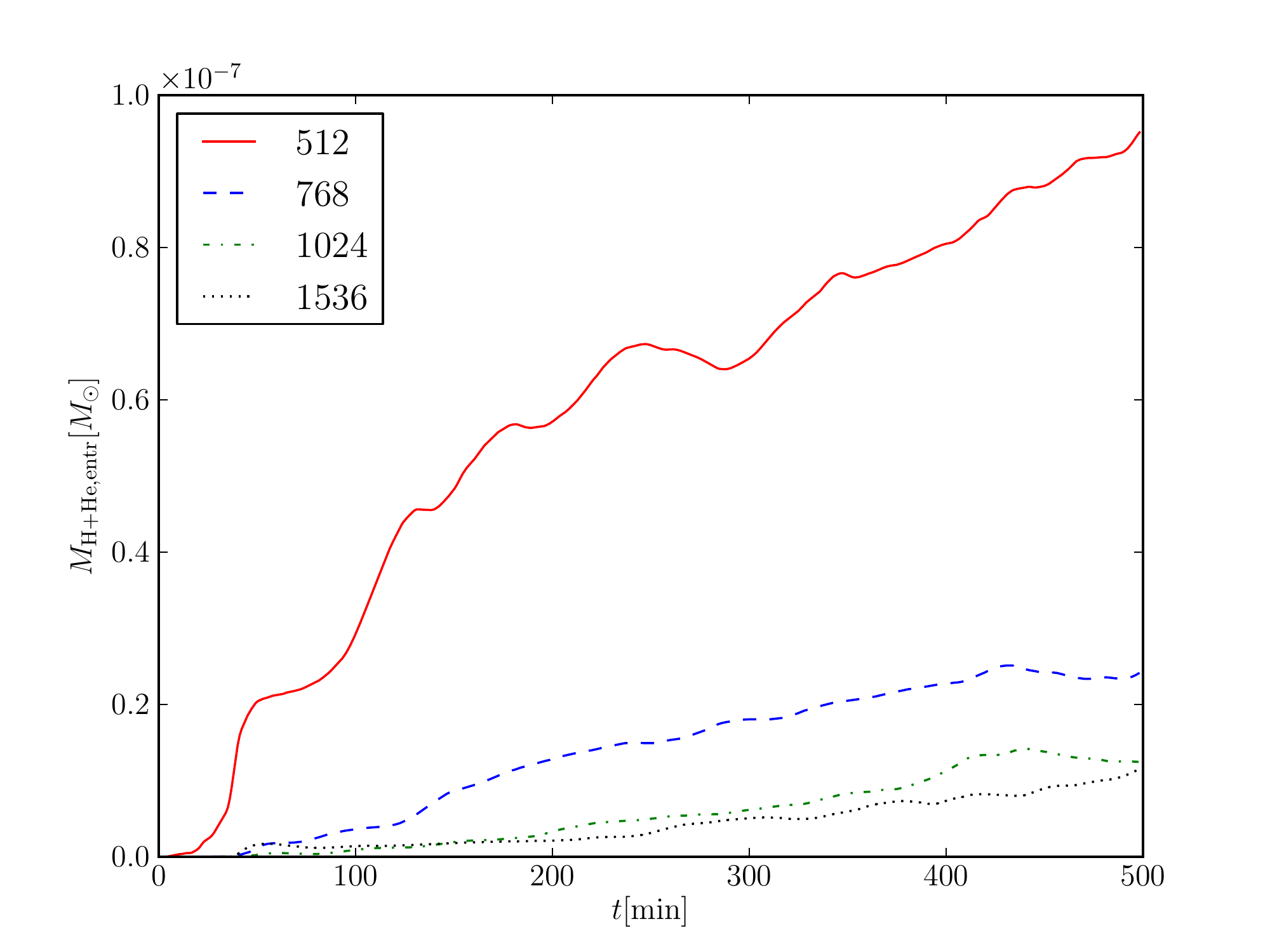}
   \caption{Entrained mass, integrated between the radial coordinates
     $10100 \mem{km}$ and $29200\mem{km}$, as a
     function of time. Simulations for different grid sized $n^3$ are
     shown. The number of grid points $n$ in one direction is shown in
     the legend.}
   \label{fig:entrained_mass}
\end{figure}
The entrainment rate is just the amount of mass of fluid 'H+He' mixed
into the convection zone per unit time. The entrained mass is shown
for the four grid choices as a function of time in
\abb{fig:entrained_mass}. The entrainment rate is therefore the
slope of these lines. 

\abb{fig:entrained_mass}  is showing in another form the same result as the
profile evolution. With the two smaller grids the
entrainment rate is much higher compared to the high-resolution
runs. A larger amount of entrained H would lead to a large H-burning
luminosity, and this seems to be the case in the simulations presented
by \citet[][Fig.\, 3]{stancliffe:11} in which the low-resolution run
has a 1000 times higher H-burning luminosity compared to the
high-resolution case.

The entrainment proceeds in a gusty or intermittent manner, with
episodes of larger than average entrainment for periods of $\approx
30$ to $60\mem{min}$ followed by phases of smaller entrainment
rates. The irregular periodicity of the entrainment fluctuation is of
the order of $100\mem{min}$. When subtracting the first $\approx 150$
to $200\mem{min}$ which represent the initial transient and the
settling into a steady-state convective flow, we are only left with
$\approx 300\mem{min}$ that can be used for the entrainment analysis,
since the highest-resolution run ($1536^3$) has been computed only to
$498\mem{min}$. Therefore, the time averaging of the entrainment rate
can be performed only over a small multiple of the typical period over
which the entrainment rate fluctuates, which introduces a
statistical error in determining the entrainment rates for each grid
size. One of the simulations, the $1024^3$ run, has a particularly
strong entrainment gust from $t=410\mathrm{min}$ to
$t=460\mathrm{min}$ preceded and followed by a period of lower than
average entrainment. Such individual events, and the way in which they
are included or excluded in the integration will inevitably introduce
some error. 
\begin{figure}[tb]            
   \includegraphics[width=0.5\textwidth]{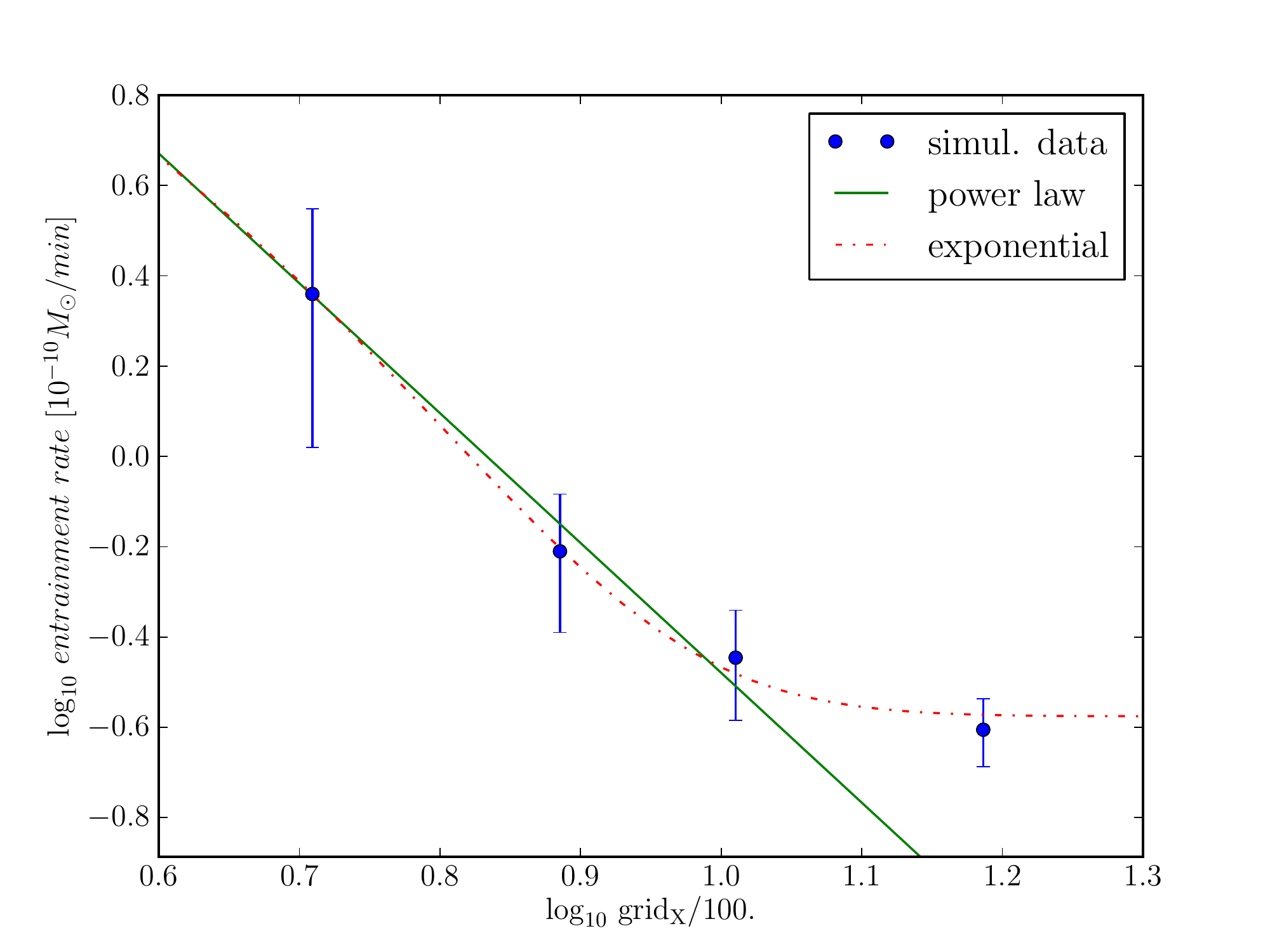}
   \caption{Logarithm of mean entrainment rates with error bars representing $99\%$
     confidence intervals (see text) as a function of grid size. The
     mean entrainment rates have been fitted with a power law and an
     exponential according to \glt{eq:exp}.}
   \label{fig:entrain_fit}
\end{figure}

This error is estimated by performing 12 fits of the entrainment rates
with different assumptions on the selection of the range of cycles
over which a linear least-square fit of the entrained mass as a
function of time is performed to determine the entrainment rate, the
choice of the weight of the zero-point of the entrainment evolution as
well as the upper boundary of the integration (see discussion at the
end of this section). Together these 12 cases represent the range of
reasonable choices that could be made when integrating the entrained
mass, and they represent the sample of measurements on which the
measurement error is based. The resulting entrainment rates are
provided in Table \ref{tab:entrain_rate} and plotted in
\abb{fig:entrain_fit}.
\begin{deluxetable}{llll}
  \tablecaption{\label{tab:entrain_rate} 
    Entrainment rates ($10^{-10}\mem{\msun/min}$) for each
    run with different grid resolution, and the asymptotic
    entrainment rate under the assumption of a fit according to
    \glt{eq:exp}. $c_\mem{entr}$ is the asymptotic limit variable
    given in \glt{eq:exp}. }
\tablehead{ \colhead{Grid size/case} & \colhead{Mean}& \colhead{Std.
    dev.}   & \colhead{99\% CI}    }
\startdata
 $ 512^3$          &$2.290$&$ 1.668$&$\pm 1.2420$ \\
 $ 768^3$          &$0.616$&$ 0.281$&$\pm 0.2089$ \\  
$  1024^3$        &$0.358$&$ 0.132$&$\pm 0.0979$ \\
$  1536^3$        &$0.248$&$ 0.057$&$\pm 0.0424$ \\
$  c_\mem{entr} $&$0.263$&$ 0.089$&$\pm 0.0659$ \\
\enddata
\end{deluxetable}

We adopt two simple expressions that represent the
non-convergence and the convergence case. If the entrainment rate does
not converge for the range of grids adopted here it may follow a
power-law $f_\mem{pow}(x)=ax^b $ as a function of the number of grid
points in one direction $x=grid_X$.  However, if the entrainment does
converge, the entrainment rate would instead show signs of approaching an
asymptotic limit. In this case, the dependence of the entrainment rate
on the grid may be represented by an
exponential plus a constant of the form
\begin{equation}
f_\mem{exp}(x)=a e^{b x} + c_\mem{entr}  \label{eq:exp} \punkt
\end{equation}
In that case the constant $c_\mem{entr}$ would represent an estimate
of the true entrainment rate based on the assumed functional form of
\glt{eq:exp}. \abb{fig:entrain_fit} shows both alternatives fitted to
the entrainment rate data provided in \tab{tab:entrain_rate}. The data
is better represented by the exponential function, and the asymptotic
limit approximated by $c_\mem{entr}$ agrees within the adopted
uncertainty with the entrainment rate of the $ 1536^3$ simulation, and
represents the converged entrainment rate (for our assumption on
the integration boundaries).

We repeated the same procedure just with the three lower resolution
runs, and find in that case $c_\mem{entr} = 0.311$. This is within the
$99\%$ CI of the asymptotic entrainment. But the fits do not
convincingly show convergence by a clear preference for the
exponential fit. However, if we use the knowledge that the simulations
do converge we
could obtain, for a similar case, a very good result with $\approx 1/5^\mathrm{th}$ of the
computational cost, which is the fraction of the cost of the three
lower resolution runs combined compared to the cost of all runs.

\paragraph{Limitations of the convergence analysis}
The entrained mass shown in
\abb{fig:entrained_mass} is determined by integrating up to
$29200\mathrm{km}$. This is $550\mathrm{km}$ below the formal
convection boundary defined as where in the initial setup the entropy
gradient becomes positive. This choice is necessary to exclude the actual
entrainment interface from the integration. This interface is not
resolved sufficiently in the lower-resolution runs. The horizontal
velocity profile, for example, depends on the resolution
(\abb{fig:profiles_velocity_resolution}). At the same time the region
nearest to the formal convection zone has the highest abundance of the
'H+He' material from the stable layer, and would therefore dominate
the integration of the entrained material. Including the boundary
layers would therefore rather measure the amount of material present
in the $550\mathrm{km}$ or so within the range of the
boundary. We cannot show convergence in the way described above for
the material in the $550\mathrm{km}$-layer inside the convection
boundary.  However, that material will not react with \czw\ and release
energy. We are therefore not interested whether or not the amount of
material in the boundary layer is converged. Instead,
we only ask if the material that actually enters the region below this
boundary layer, i.e.\ below $29200\mathrm{km}$ can be quantitatively
simulated. That, indeed, can be accomplished with simulations at the
resolutions used here.

\section{Discussion and conclusions}
\label{sec:concl}
The main goals of this paper are to investigate the properties of
He-shell flash convection in $4\pi$ geometry and determine the
entrainment properties of H-rich material at the top of the convection
zone. This entrainment process is the basis of the H-ingestion flash
in low-Z AGB stars and in post-AGB stars, and the results of this
investigation are the foundation of our future work on this topic
\citep[see][for preliminary results]{Herwig:2013vf}. The top
convection boundary in this He-shell flash convection situation is
much stiffer compared to the O-shell burning convection or the core
convection investigated by \citet{Meakin:2007dj}, as visual comparison
of their Fig.\,3 \& 10 and our \abb{fig:FV_ring} immediately
reveals. It is therefore very consistent that their entrainment rates
for O shell burning ($\natlog{1.1}{-4}\msun\mathrm{/s}$) and core
convection ($\natlog{2.72}{-7}\msun\mathrm{/s}$, with a driving
luminosity $10\times$ natural) are much larger compared to our
converged entrainment rate of $\natlog{4.38}{-13}\msun\mathrm{/s}$.
There exists a significant range of entrainment rates, even between
convection boundaries of different deep stellar interior convection
regimes. He-shell flash convection boundaries with H-ingestion are
much stiffer compared to core convection and in turn O shell
burning. Large entrainment rates have been reported as well by
\citet{Mocak:2010kj} based on a single 3-D simulation.

In order to demonstrate that the fidelity of our approach matches the
challenge of the problem, we describe key elements of our method that
we will apply in future work. The key capability that we demand from a
simulation is to reproduce the total amount and rate of entrainment of
material from above the convection zone into the unstable layer. It
will be this entrainment amount that, through the nuclear reactions
between protons and \czw\, generates energy at a rate that is large
enough to alter the global flow properties. The accretion of unburned
material through the upper convection boundary of the He-shell flash
convection zone will therefore determine all subsequent H-ingestion
flash simulation steps. Our simulations show that, despite the
enormously small entrainment rate, the H-rich material that enters the
He-shell flash convection zone is advected into the deeper layer in
distinct and coherent clouds. The visual inspection of the fractional
volume images demonstrate that these clouds reach the burning layer
where protons will react with \czw. This is an important result, as
without our simulations one may have plausibly assumed that strong
horizontal turbulence would shred any downward directed H-rich
advection. In such a case, a spherically symmetric simulation approach
of the H-ingestion flash may have seemed justifiable. Our simulations,
however, show significant inhomogeneities in the depth of the
convection zone where the $\czw + \p$ reaction would take place. Our
results therefore provide evidence that global $4\pi$ simulations are
indeed needed to simulate the H-ingestion flash accurately.

How does the entrainment rate we obtain compare to the H-ingestion
flash constraints we already have established for Sakurai's object?
\citet{herwig:10a} demonstrated that the anomalous observed abundance
distribution could be reproduced in the 1-dimensional multi-zone
nucleosynthesis models with a static mixing-length based convective
diffusion parameter if an entrainment rate of
$\natlog{5.3}{-10}\msun/\mathrm{s}$ is assumed. The hydrodynamic
entrainment rate that we have determined is significantly smaller than
the nucleosynthesis-constrained entrainment rate. We may therefore
conclude that the hydrodynamic feedback from the nuclear energy release
due to H ingestion will lead to enhanced entrainment rates.
\citet{herwig:10a} did also find that the neutron flux must be
terminated at some point, and speculated that this would be due to the
hydrodynamic feedback mechanism induced by the nuclear energy release
from the ingested protons. 

The ingestion rate of protons that is
necessary to cause significant flow variations to the convective
velocity field established by the underlying He-burning luminosity can
be estimated by noticing that for this to happen the luminosity from
$\czw + \p$ reactions must equal or exceed the He-burning luminosity
$L_\mathrm{He} = \natlog{4.2}{7}\lsun$. This is the case if
\begin{equation}
\dot{M}_\mathrm{entr} \apgeq \frac{L_\mathrm{He} m_\mathrm{p}}{Q} = \natlog{4.9}{-11}\msun\mathrm{/s} 
\end{equation}
where $m_\mathrm{p}$ is the mass of the proton and $Q$ is the energy
released in each $\czw(\p,\gamma)\ndr$ reaction. Our entrainment rate
is significantly lower than this limit, and we therefore do not expect
an immediate and dramatic effect of the burning of the ingested
protons. In fact, one may wonder how any relevant hydrodynamic
feedback can arise with such a low entrainment rate. As we will
demonstrate in detail in a forthcoming publication H-ingestion
entrainment simulations for Sakurai's object with energy feedback from
H ingestion taken into account show a long quiescent phase of
entrainment and nuclear burning initially without noticeable effect on
the hydrodynamic flow pattern, followed eventually by violent and
non-spherical eruptions that significantly elevate the entrainment
rate. The details of these events will be reported elsewhere.

A key property of the convective boundary layer is the entropy profile
immediately above the Schwarzschild boundary. In \abb{fig:A-top-boundary}
we observe a self-similar
evolving entropy profile. Thus, the numerical method is able after
a short while to establish an entropy profile that does not appreciably
change shape as it translates in response to the entrainment.

We give evidence that the entrainment rate is a quantitative
simulation property that shows convergence. However, some other
quantities cannot be considered to be converged. It seems that the
spherically averaged radial velocity components are rather similar for
all resolutions, while the horizontal velocity components still show
some dependence on resolution. Depending on how the horizontal
component is averaged, we roughly estimate that horizontal velocity
components may differ by $30\%$ between our runs with lowest vs.\ the
highest resolution. Although it would be desirable to get rid of these
differences as well, one should keep in mind that the modeling
uncertainties introduced by the traditional 1-D spherically symmetric
simulation approach based on the MLT and some convective boundary
mixing algorithm have been shown to be much larger and even
qualitative in some instances, as demonstrated via a
nucleosynthesis-based validation analysis by \citet{herwig:10a}.

The fact that there are aspects of the simulations that are
not fully converged is interesting but not important, as long as we
can show that convergence is or can be obtained for those quantities
that do matter. In our case this is the entrainment rate. Based on
this result, we may now move ahead and include the nuclear burn of H
with \czw\ and the associated energy release.

\acknowledgments PRW acknowledges support from contracts from the Los
Alamos (LANL) and Sandia National Laboratories and NSF PRAC grant
OCI-0832618.  The development of the PPB scheme for multifluid
fractional volume advection was supported by PRW's LANL contract
and DoE grant DEFG02-03ER25569.
FH acknowledges NSERC Discovery Grant funding.  The
hydrodynamics simulations were performed on the LCSE workstation
cluster, supported by NSF CRI grant CNS-0708822, on NSF's Kraken
supercomputer at the National Institute for Computational Sciences
(NICS), on the Itasca machine at the Minnesota Supercomputing
Institute (MSI), and the Canadian WestGrid Lattice and Orcinus
high-performance computers.

\appendix \section{The PPB Multifluid Advection Scheme}
\label{sec:app_method}

\subsection{PPM Gas Dynamics}   
Our simulation code for
the hydrogen ingestion flash is based on the Piecewise-Parabolic Method
\citep[PPM][]{woodward:81,woodward:84,colella:84,woodward:86,woodward:06b}.  
We use a version of this numerical scheme that has been
modified in a few minor, yet substantive, ways over the years.  It is
described in complete detail in \citet{woodward:06b}, and the effective
numerical viscosity of PPM has been quantitatively characterized by
\citet{porter:94}.  The most relevant modification of
this scheme from the present perspective is that in the interpolation
process that determines the parabolae we use to describe the subcell
structure, monotonicity constraints are applied only if the local
behavior of the function to be interpolated is judged not to be
sufficiently smooth.  A measure of function smoothness is constructed
that has a 5-cell stencil in the direction of the present 1-D pass.  In
our low-Mach-number flows, we do not expect to find shocks, so that we
expect the flow to be smooth under most conditions.  In this case, we do
not expect the elaborate interpolation of the values of the function at
each of the two interfaces of the cell with its nearest neighbors in the
direction of the 1-D pass to be modified in any way by monotonicity
constraints. These cell interface values are interpolated by evaluating
at these points the unique cubic polynomials that assume the prescribed
cell averages in the 2 grid cells to the left and the 2 to the right of
the interface in question.  These interface values are therefore one
full formal order more accurate than the parabola used to describe the
distribution of the interpolated variable inside the grid cell.  That
parabola, when not modified by monotonicity constraints, passes through
the two cell interface values and has the prescribed cell-averaged
value.  To find the amount of this variable that is advected across an
interface of the grid cell, we use this parabola in the upstream cell. 
For low Mach number flows, only a thin sliver of the cell is advected
across the interface.  Therefore the average value of the interpolated
variable within this sliver is nearly equal to the value of the variable
at the interface.  Because this value is one order more accurate than
the parabola as a whole, the PPM scheme in low Mach number flow regimes
has the very important property that it becomes more accurate as the
Mach number, and hence the Courant number, is decreased.  Even though
more time steps are then required to arrive at a given time level, the
higher formal accuracy completely offsets the loss in accuracy that we
would have expected at low Courant numbers.  This behavior is counter to
the behavior of most numerical schemes.  It is a feature of PPM that is
a result of its history.  PPM was developed from an earlier, more
accurate numerical scheme in which the cell interface values were stored
and updated as primary data for the method.  Because this data structure
was incompatible with large production codes at the Lawrence Livermore
National Lab in the late 1970s, the PPM scheme was designed as a
replacement.  It generated the previously independent cell interface
data by interpolation from the cell averages.  To retain as much of the
earlier method's accuracy as possible, this interpolation of cell
interface values was made as accurate as seemed practical or necessary
at the time.

In our stellar hydrodynamic flows, we are concerned that structure in
the flow with a characteristic scale size of 20 to 40 grid cells can be
advected in our circulating convection zone for, say, 100 scale lengths
with minimal amplitude damping and acceptable phase error.  We can
assess this capability using 1-D advection of a sine wave of 40 cells. 
Using a Courant number of 0.03125, advecting this sine wave for 100
wavelengths with our PPM scheme, we find that the wave amplitude is
damped by a factor of 0.9985, while the final phase is off by 0.078 cell
widths, which is just 1 part in 50,000.  A more stringent test is to
advect a Gaussian pulse with a full width at half maximum of 10 grid
cells a distance of 100 times this width, or 1000 cell widths, with a
Courant number of 0.03125.  Doing this results in a pulse with height
damped by a factor of 0.9383 and with a phase error of just 0.723 grid
cell widths, which is just 1 part in 1384.  The behavior of the scheme
with a Courant number of 1/3 has slightly greater amplitude error but
less phase error.  Running the same sine wave and Gaussian advection
tests at this Courant number, we find amplitude damping by factors of
0.9923 and 0.9320 with phase errors of 0.02 and 0.234 grid cell widths. 
It should therefore be clear that with PPM advection we suffer very
little accuracy loss in going from the typical range of Courant numbers
encountered in explicit gas dynamics simulations to those we encounter
in our present simulations of the hydrogen ingestion flash.

\subsubsection{PPB Multifluid Volume Fraction Advection}
Our multifluid PPM code uses the much more accurate
Piecewise-Parabolic Boltzmann (PPB, described below) scheme to advect
the critically important cell volume fraction occupied by the
hydrogen-rich fluid, which is located initially just above the top
boundary of the helium shell flash convection zone.  For comparison
with PPM, PPB running the same advection experiments just mentioned at
a Courant number of 0.03125 for the same sine wave and Gaussian
produces damping by factors of 0.999937 and 0.99961 with phase errors
of 0.938 and 0.02 cell widths.  The PPB phase error may seem large
until one realizes that, as discussed in
\citet{vanLeer:77,woodward:86}, PPB incurs one-time errors related to
going over to the piecewise-parabolic representation, after which
there is extremely little error accumulation.  Consequently, advecting
the same sine wave at the same small Courant number for an additional
100 wavelengths causes the phase error to increase from 0.938 cell
widths after 100 wavelengths to only 1.066 cell widths after a second
100 wavelengths.  As a general rule of thumb, PPB advection is about
as accurate as PPM advection if PPM is given 3 times more grid cells
(and 3 times more time steps) to work with.  This is because, in 1-D,
PPB has 3 times as much independent information on any given grid.
PPB advection has fifth-order formal accuracy
\citep{vanLeer:77,woodward:86}, while PPM advection is only
third-order accurate.  Nevertheless, our rule of thumb holds for
practical problems, since we are then essentially never in the range
of grid resolutions where strict formal convergence rates hold.  For
the advection of sharp jumps, PPB cannot yield a thinner numerical
representation of the jump than PPM, with its contact discontinuity
detection and steepening, because PPM's representation is the thinnest
possible when parabolae are used for the subgrid structures.
Nevertheless, PPB advection is still superior even in this case,
because PPB generates far fewer numerical glitches in rare
pathological circumstances.  Once again, this result is due to PPB
having much more independent data to work with.

PPB achieves its very high accuracy by conserving to machine accuracy
the 10 lower-order moments -- $\left\langle f\right\rangle $,
$\left\langle f\tilde{x}\right\rangle $, $\left\langle
  f\tilde{y}\right\rangle $, $\left\langle f\tilde{z}\right\rangle $,
$\left\langle f\tilde{x}\tilde{x}\right\rangle $, $\left\langle
  f\tilde{x}\tilde{y}\right\rangle $, $\left\langle
  f\tilde{x}\tilde{z}\right\rangle $, $\left\langle
  f\tilde{y}\tilde{y}\right\rangle $, $\left\langle
  f\tilde{y}\tilde{z}\right\rangle $, $\left\langle
  f\tilde{z}\tilde{z}\right\rangle $ -- of the fractional volume, $f$,
of a tracked fluid within a local volume element.  We define these
moments with respect to a set of Cartesian coordinates, $\tilde{x}$,
$\tilde{y}$, and $\tilde{z}$, centered on the grid cell of interest,
aligned with our grid directions, and for which the width of the cell
in each dimension is unity.  Using these cell-centered coordinates, we
define the moments as:
\[\left\langle f{\tilde{x}}^l{\tilde{y}}^m{\tilde{z}}^n\right\rangle =\int^{{1}/{2}}_{-{1}/{2}}{f\ {\tilde{x}}^l\ {\tilde{y}}^m\ {\tilde{z}}^n\ d\tilde{x}}\ d\tilde{y}\ d\tilde{z}\] 

Our use of the cell-centered
and normalized coordinates restricts us to the use of cubical grid
cells.  However, given the proliferation of adaptive mesh refinement
(AMR) techniques, we view this restriction as entirely acceptable for
all practical calculations.  For our work with stars, cubical grid
cells provide the least distortion of the underlying flow geometry
that is possible with a Cartesian mesh.  The centered and normalized
coordinates give us a tremendous advantage; they allow us to perform
our calculations at double speed in 32-bit precision.  On the very
fine grids permitted by the tremendous power of today's computing
systems \citep[we have run problems related to inertial confinement fusion
on over one trillion grid cells][]{woodward:12a}, it can easily consume 4 digits
of precision just for a grid cell to know where it is located in a
global sense.  Injecting this global location through moments of
\textit{x}, \textit{y}, and \textit{z} in place of $\tilde{x}$,
$\tilde{y}$, and $\tilde{z}$ is simply wasteful.  It forces the
machine to do twice as much work for no good reason, since the global
location of a grid cell is irrelevant in specifying the distribution
of a variable within it.  One might ask if the fifth-order formal
accuracy of the scheme, to whatever extent it really matters in a
practical computation, is preserved in 32-bit precision. Our
experience indicates that it is, with the single exception of the IBM
Cell processor, which did not perform rounding in its 32-bit
arithmetic. 

Our PPM scheme is directionally split.  We update the flow for each
time step in three 1-D passes, using a symmetrized sequence
\textit{xyzzyx} in each time step pair.  We therefore require a
directionally split version of the PPB advection scheme.  This
delivers an immense simplification.  The PPB scheme is built upon van
Leer's Scheme VI \citep{vanLeer:77}, which is a 1-D scheme with no
monotonicity or other constraints.  In the early 1980s, this scheme
was made into a 2-D, directionally split, and constrained scheme,
described in \citet{woodward:86}.  The version of PPB that we use in
our work is simplified from this early work, with a minimal set of
moments and a simplified method of updating them, described below.  It
is also enhanced by a more elaborate and useful method of constraining
each 1-D pass of the algorithm.  This scheme has only been described
in detail in internal reports \citep{woodward:05} and in broad terms
\citep{woodward:10b,woodward:08a,woodward:12a} to this date.  A code
module containing this PPB scheme for 3-D computation was delivered to
the Los Alamos XRAGE code in 2004, and this module, combined with our
version of PPM, was included in the official 2005 release of the Los
Alamos XRAGE code.  Here we give a complete, but brief, description of
the PPB advection scheme, adapted for use in the advection of a fluid
concentration, so that \textit{f} is forced to remain within the range
of values from 0 to 1.  A more voluminous description can be found in
the LCSE internal report \citep{woodward:05}.  The antecedents of this
PPB scheme go back to the 1970s and 1980s, and Woodward has taught it
in his course on numerical methods at the University of Minnesota in
the late 1980s and 1990s.  A parallel development of numerical
schemes, beginning around the same time as the work of van Leer but in
the finite element community, has been given the name Discontinuous
Galerkin \citep[cf.\ recent books, ][]{cockburn:00,hesthaven:08}.
These schemes are similar, but they involve more computational labor
as a result of their use of Runge Kutta techniques.  They also apply
the methodology over the full set of hydrodynamical equations,
although there is little evidence of which we are aware that this
yields any significant benefit for flows in which shocks and contact
discontinuities are involved, so that the very high formal order of
these techniques might be justified.  Here we apply a simpler form
stemming from van Leer's original work \citep{vanLeer:77}, and we cut
to the bone the computational labor involved.  We also apply an
elaborate set of constraints that address the special problem of
multifluid flow.  Despite the complexity of these constraints, our
formulation in 1-D passes allows us to apply these constraints only to
the 3 moments $\left\langle f\right\rangle $, $\left\langle
  fx\right\rangle $, and $\left\langle fxx\right\rangle $, where
\textit{x} is the direction of our 1-D pass.  This delivers an
enormous simplification of the scheme, to the extent that it now
involves only about 3 times the computational labor of the much less
accurate PPM advection scheme.

We can use the 10 moments of \textit{f} to construct a quadratic
polynomial describing the subgrid behavior of \textit{f} within a grid
cell:
\[f\left(\tilde{x},\tilde{y},\tilde{z}\right)=f_{000}+f_{100}\tilde{x}+f_{010}\tilde{y}+f_{001}\tilde{z}+f_{200}{\tilde{x}}^2+f_{110}\tilde{x}\tilde{y}+f_{101}\tilde{x}\tilde{z}+f_{020}{\tilde{y}}^2+f_{011}\tilde{y}\tilde{z}+f_{002}{\tilde{z}}^2\]
where:
\begin{alignat*}{3}
f_{000}&=4.75\ \left\langle f\right\rangle -15\left(\left\langle
    f\tilde{x}\tilde{x}\right\rangle +\left\langle
    f\tilde{y}\tilde{y}\right\rangle +\left\langle
    f\tilde{z}\tilde{z}\right\rangle \right) \\
f_{100}&=12\ \left\langle f\tilde{x}\right\rangle  
&f_{010}&=12\ \left\langle f\tilde{y}\right\rangle   
&f_{001}&=12\ \left\langle f\tilde{z}\right\rangle \\
f_{110}&=144\ \left\langle f\tilde{x}\tilde{y}\right\rangle
&f_{101}&=144\ \left\langle f\tilde{x}\tilde{z}\right\rangle
&f_{011}&=144\ \left\langle f\tilde{y}\tilde{z}\right\rangle \\
f_{200}&=15\ \left(12\ \left\langle f\tilde{x}\tilde{x}\right\rangle
  -\left\langle f\right\rangle \right)  
& \quad f_{020}&=15\ \left(12\ \left\langle f\tilde{y}\tilde{y}\right\rangle
  -\left\langle f\right\rangle \right)  
& \quad f_{002}&=15\ \left(12\ \left\langle f\tilde{z}\tilde{z}\right\rangle
  -\left\langle f\right\rangle \right)
\end{alignat*} 
From this single polynomial, we can construct the following separate
functions of  $\tilde{x}$:
\begin{alignat*}{3}
 f_x\left(\tilde{x}\right)&=f_{000}+f_{100}\tilde{x}+f_{200}{\tilde{x}}^2  
& \quad f_y\left(\tilde{x}\right)&=f_{010}\tilde{y}+f_{110}\tilde{x}\tilde{y}  
&\quad f_z\left(\tilde{x}\right)&=f_{001}\tilde{z}+f_{101}\tilde{x}\tilde{z}\\
 f_{yz}\left(\tilde{x}\right)&=f_{011}\tilde{y}\tilde{z}
& f_{yy}\left(\tilde{x}\right)&=f_{020}{\tilde{y}}^2
&f_{zz}\left(\tilde{x}\right)&=f_{002}{\tilde{z}}^2
\end{alignat*} 
In a 1-D pass, we perform separate advection operations on the above
functions of $\tilde{x}$.  Only the first of these 6 functions,
$f_x\left(\tilde{x}\right)$, requires the application of any
constraints.  The other 5 functions can be advected without concern
for constraints, because the results will be constrained in later 1-D
passes in the $\tilde{y}$ and $\tilde{z}$ directions.  We will discuss
the constraints last, and first give the formulae for updating the 10
moments of \textit{f} in the cell during an \textit{x}-pass.  We begin
with the most difficult formulae, which result from the advection of
$f_x\left(\tilde{x}\right)$.  Once we see how to advect
$f_x\left(\tilde{x}\right)$, all the remaining functions will be
trivial by comparison, although a slight complication results from the
dependence of $f_{020}$ and $f_{002}$ upon $\left\langle
  f\right\rangle $.  A further complication arises from changes in
volume of the grid cells during the time step.  Conceptually, we draw
each of our 6 functions of $\tilde{x}$ in the grid cells according to
the values of the coefficients in their polynomial representations
given above.  Then we demand that the value of each function at a
particular point must remain unchanged as that point moves with its
time-averaged velocity in \textit{x} over the time step. This
construction is illustrated for $f_x\left(\tilde{x}\right)$ in the
left panel of 
\abb{fig:PPB}.  Finally, we integrate the resulting function
segments over the new grid cell volume to obtain the new
\textit{x}-moments for each of our 6 functions, which are then the new
10 low-order moments of \textit{f} in the new grid cell.
\begin{figure}[tb]
  \includegraphics[width=0.45\textwidth]{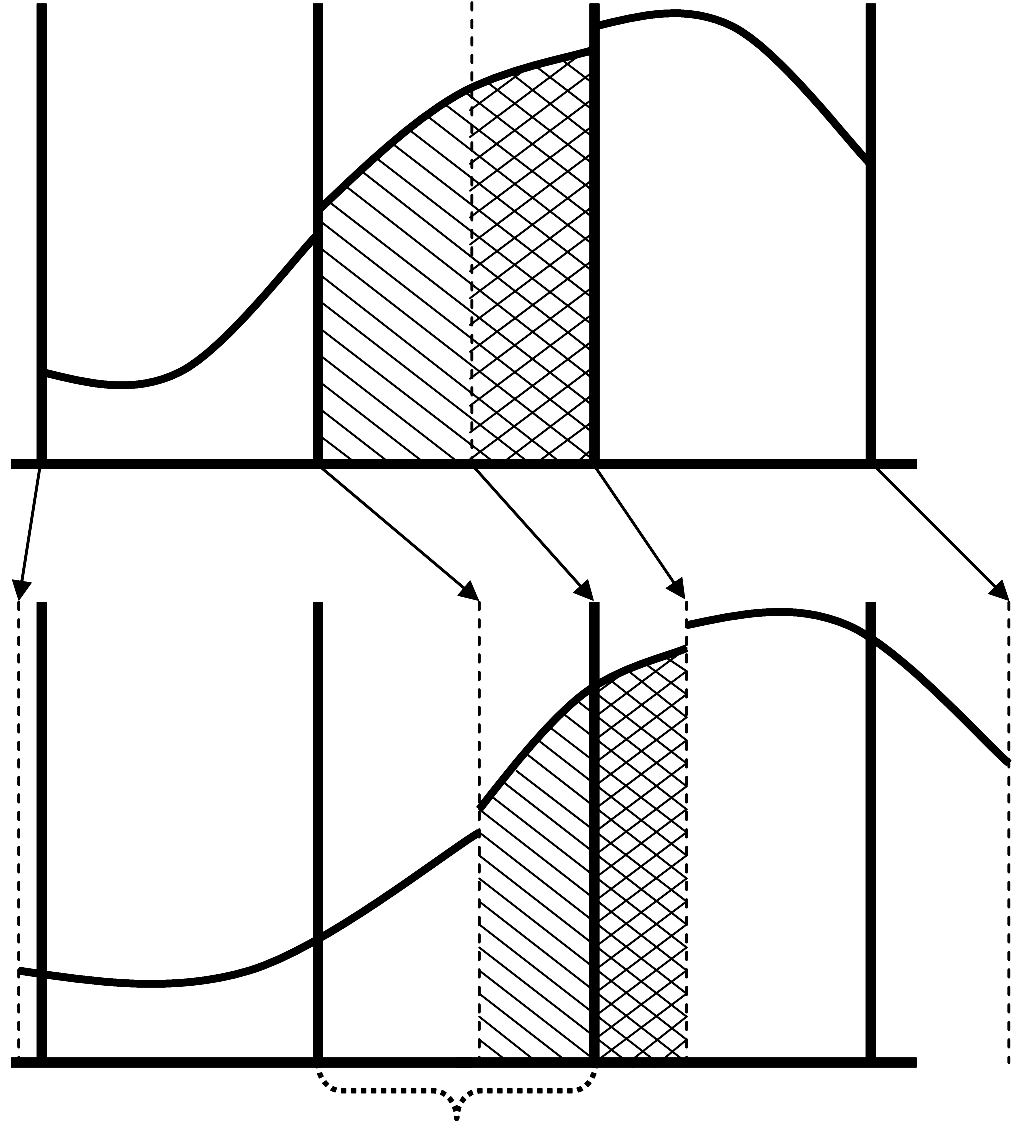}
  \includegraphics[width=0.55\textwidth]{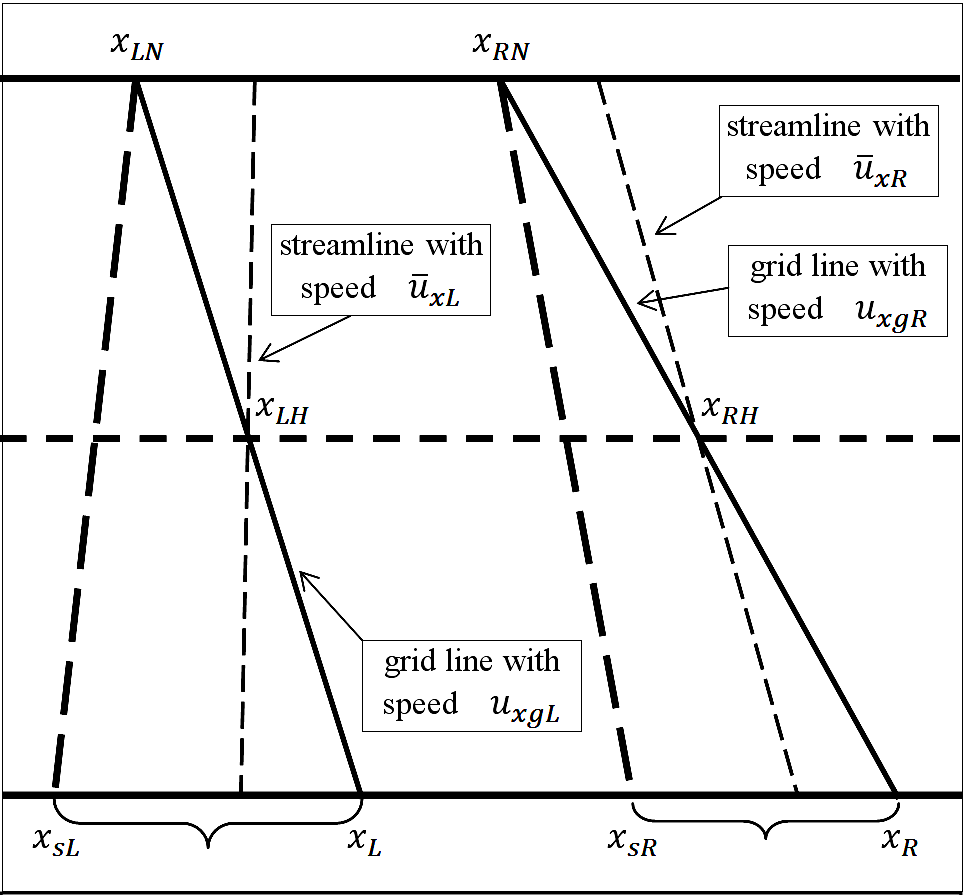}
  \caption{\emph{Left panel:} The interpolation parabolae,
     $f_x\left(\tilde{x}\right)$, for 3 grid cells are shown at the
     beginning of the x-pass in the upper part of the figure. The
     motion of the cell interfaces is indicated, and the new stretched
     or squashed parabolae are shown in the lower part of the
     figure. The two portions of the central cell that become parts of
     the new central cell and of its neighbor on the right are
     indicated by the diagonal and cross-hatched shading patterns in
     both parts of the figure. To obtain the new interpolation
     parabola for the central cell, we must evaluate the moment
     integrals over the cell domain in x, which is indicated by the
     bracket at the bottom of the figure. \emph{Right panel:}
     Space-time diagram showing time levels, moving grid trajectories
     and fluid streamlines, with coordinate and speed values indicated
     that are used in the evaluation of moment integrals over the
     advected portions of grid cells, which are marked with brackets.}
   \label{fig:PPB}
\end{figure}

Many applications require a moving (radially collapsing or
expanding) Eulerian grid, such as the rapid expansion of the outer
convection zone in a H-ingestion flash during the first thermal pulse
of a low-Z AGB star. Although we have not used this feature in the
entrainment simulations presented here, it is an integral part of our
PPB implementation and we prefer to describe here the general
formalism that allows for a co-moving grid, and that will be used in
future stellar convection simulations. The diagram in the right
panel of \abb{fig:PPB} is intended to clarify the coordinates that we
use.  Each grid cell interface has a constant velocity proportional to
its distance from the origin.  This is indicated in the diagram.  Our
PPM gas dynamics scheme computes the time-averaged velocity,
${\overline{u}}_{xL}$, of the fluid at each moving grid cell interface
\textit{L}.  This is indicated in the diagram by the slopes of the
fluid streamlines crossing the cell interfaces at the time level
half-way through the time step (subscript ``H'').  We denote the new
time level, at the end of the time step, by the subscript ``N.''

We perform these moment integrals with respect to the new grid cell's
own cell-centered and scaled coordinates.  To do this, we relate the
new cell's coordinates to those in the old cell.
\[
{\tilde{x}}_N=\left(x_N-x_{MN}\right)\ /\ 
\Delta x_N\ =\ \left[\left(x_M-x_{MN}\right)+\tilde{x}\ \Delta
  x+\left({\overline{u}}_{xM}+{\tilde{x}}_H{\rm \ }
\Delta {\overline{u}}_x\right)\Delta t\right]\ /\ \Delta x_N
\] 
Here we have used the subscript ``M'' to denote the middle of the
cell, so that
${\overline{u}}_{xM}=\left({\overline{u}}_{xL}+{\overline{u}}_{xR}\right)/2$
and $u_{xgM}=\left(u_{xgL}+u_{xgR}\right)/2$.  Also we denote the
difference in the time-averaged fluid velocity across the cell by
$\Delta
{\overline{u}}_x={\overline{u}}_{xR}-{\overline{u}}_{xL}$${}_{.}$ The
difference across the cell for the grid velocity is $\Delta
u_{xg}=u_{xgR}-u_{xgL}$.  We approximate each streamline as having a
constant velocity equal to its time average, interpolated linearly in
space at the half-time level.  Therefore,
${\tilde{x}}_H=\left(\tilde{x}+{\tilde{x}}_N\right)/2$.  After some
algebra, we write:
\begin{equation} \label{eqn:tildeXn}
\tilde{x}_N=f_{\rm exp} \tilde{x}+{\sigma }_M
\end{equation}
where                          
\[f_{\rm exp} =\left[1+0.5\ \Delta {\overline{u}}_x\ \Delta t/\Delta x\right]\ /\ \left[1+\left(\Delta u_{xg}-\Delta {\overline{u}}_x/2\right)\left(\Delta t/\Delta x\right)\right]\]
and               
\[{\sigma }_M=\left({\overline{u}}_{xM}-u_{xgM}\right)\left(\Delta
  t/\Delta x\right)\ /\ \left[1+\left(\Delta u_{xg}-\Delta
    {\overline{u}}_x/2\right)\left(\Delta t/\Delta x\right)\right]
\punkt \] 

In the spirit of conservation laws, we generate fluxes at the
interfaces of the grid cells.  Extra terms arise, because each cell
has its own set of coordinates.  We assume a uniform grid of cells
with widths $\Delta x$. Thus for streamlines such as those shown in
the right panel of \abb{fig:PPB} that cross grid cell interfaces, we
must either add or subtract 1 from the expression for ${\tilde{x}}_N$
\glp{eqn:tildeXn}.  As stated earlier, this extra complexity is
necessary to permit us to use 32-bit arithmetic in the computation.
We now focus our attention on the cell interface \textit{L}, which is
the left-hand interface of our cell of interest.  The velocity
${\overline{u}}_x$ at the cell interface that is \textit{upstream}
from interface \textit{L} is ${\overline{u}}_{xLup}$, and the sign of
the velocity ${\overline{u}}_{xL}$ is $s_L$.  Thus, $s_L=1$ if
${\overline{u}}_{xL}\ge 0$, and $s_L=-1$ otherwise.  In the cases
$s_L=\pm 1$, the streamline that reaches the cell interface at the end
of the time step begins at ${\tilde{x}}_{sLup}=\left(s_L/2\ -{\sigma
  }_{MLup}\right)\ /\ f_{{\rm exp}Lup}$.  Here we have used
the subscript \textit{Lup} to denote more clearly that
${\tilde{x}}_{sLup}$ is evaluated using the scaled coordinate at the
beginning of the time step in the grid cell upstream from interface
\textit{L}.  To denote the cell downstream from interface \textit{L}
we will use the subscript \textit{Ldn}.  We run subscripts together in
a Germanic style, because they translate naturally into suffixes for
variable names in a code.  We will need to evaluate moment increments
that result from advection of the segments indicated in Figure 2 into
adjacent cells.  We denote these moment increments by $d{{\mathcal
    M}}_{kL}$.  These will be integrals over the indicated segments of
the upstream cells at the beginning of the time step, but using weight
factors corresponding to the new grid cell's scaled coordinates at the
new time.  Thus:
\[d{{\mathcal M}}_{kLdn}=s_L\ \int^{s_L/2\ }_{{\tilde{x}}_{sLup}}{{d\tilde{x}}_{Lup}}\left({d\tilde{x}}_{NLdn}{/d\tilde{x}}_{Lup}\right)\ f_{Lup}\left({\tilde{x}}_{Lup}\right){\ \tilde{x}}^k_{NLdn}\] 
We now recognize that
$\left({d\tilde{x}}_{NLdn}{/d\tilde{x}}_{Lup}\right)$  is just  
$f_{{\rm exp}Lup}$  and that 
\[{\tilde{x}}_{NLdn}={\tilde{x}}_{NLup}-s_L=f_{{\rm exp}Lup}{\rm \ }{\tilde{x}}_{Lup}+{\sigma }_{MLup}-s_L\] 
We define the Courant number for the cell interface \textit{L} as ${\sigma }_L=1/2-{s_L\tilde{x}}_{sLup}$.  We should be careful to advect this same fraction of the upstream cell across the interface when, in the PPM algorithm, computing the fluxes of conserved quantities.  We now write for the moment contributions:
\[d{{\mathcal M}}_{kLdn}=s_L\ f_{{\rm exp}Lup} \int^{s_L/2\ }_{{\tilde{x}}_{sLup}}{{d\tilde{x}}_{Lup}}\ f_{Lup}\left({\tilde{x}}_{Lup}\right){\ \ \left(f_{{\rm exp}Lup}{\rm \ }{\tilde{x}}_{Lup}+{\sigma }_{MLup}-s_L\right)}^k\] 
It is easy to express these integrals in terms of simpler ones that are more readily evaluated:
\[d{\mathfrak{M}}_{kLup}=s_L\ \int^{s_L/2\ }_{{\tilde{x}}_{sLup}}{{d\tilde{x}}_{Lup}}\ f_{Lup}\left({\tilde{x}}_{Lup}\right)\ \ {\tilde{x}}^k_{Lup}\ =\ \ {\sigma }_L\ \left(s^k_Lf_{0Lup}D_{Lk}+s^{k+1}_Lf_{1Lup}D_{Lk+1}+s^{k+2}_Lf_{2Lup}D_{Lk+2}\right)\] 
The  $f_{0Lup}$, $f_{1Lup}$, and $f_{2Lup}$ are the coefficients of the parabola representing  \textit{f}  as a function of  ${\tilde{x}}_{Lup}$,  and the constants  $D_{Lk}$  can be evaluated recursively via   $D_{L0}=1$   and  
\[D_{Lk}=\ \left[2^{-k}+{\ s}_L\ k\ {\tilde{x}}_{sLup}{\ D}_{Lk-1}\right]\ /\ \left[k+1\right]\] 
We may now express the desired moment integrals as follows:
\begin{alignat*}{1}
               d{{\mathcal M}}_{0Ldn} &=f_{{\rm exp}Lup} d{\mathfrak{M}}_{0Lup}
\komma \qquad   d{{\mathcal M}}_{1Ldn} =f^2_{{\rm exp}Lup} d{\mathfrak{M}}_{1Lup}
            +\left({\sigma }_{MLup}-{\ s}_L\right)\ d{{\mathcal M}}_{0Ldn} \\ 
                d{{\mathcal M}}_{2Ldn}&=f^3_{{\rm exp}Lup} d{\mathfrak{M}}_{2Lup}
            +\left({\sigma }_{MLup}-{\ s}_L\right) \left( f^2_{{\rm exp}Lup}
                d{\mathfrak{M}}_{1Lup}\ +\ d{{\mathcal M}}_{0Ldn} \right)
 \end{alignat*} 
In these formulae we take special care to expose a highly efficient
strategy for evaluating these quantities in a computer
code. 

We are evaluating contributions to conservation laws, even though this
is somewhat obscured by our use of different coordinates in different
cells and at different times.  Nevertheless, it is easiest to evaluate
the moment integrals corresponding to the portions of the fluid that
remain in their original cells by instead evaluating total moment
integrals over these cells and then subtracting off the advected
portions.  We denote these total moment integrals by  
$d{{\mathcal M}}_{kT}$:
\[d{{\mathcal M}}_{kT}=f_{{\rm exp}}{\rm \ }\int^{1/2\ }_{-1/2}{d\tilde{x}}\ f\left(\tilde{x}\right){\ \ \left(f_{{\rm exp}}{\rm \ }\tilde{x}+{\sigma }_M\right)}^k\] 
These are quite easily evaluated.  Expressing the results recursively, we find:
\begin{alignat*}{1} 
d{{\mathcal M}}_{0T}&=f_{{\rm exp}}\left\langle f\right\rangle
\komma \qquad          
d{{\mathcal M}}_{1T}=f^2_{{\rm exp}}\ \left\langle f\tilde{x}\right\rangle +\ {\sigma }_M\ d{{\mathcal M}}_{0T}\\
d{{\mathcal M}}_{2T}&=f^3_{{\rm exp}}\ \left\langle f{\tilde{x}}^2\right\rangle +\ {\sigma }_M\ \left(f^2_{{\rm exp}}\
  \left\langle f\tilde{x}\right\rangle \ +\ d{{\mathcal
      M}}_{1T}\right)
\end{alignat*} 
To obtain the contributions to the new moments in this original cell from the fluid that remains there, we need to subtract from the above total moment integrals the portions corresponding to the fluid that is advected into neighboring cells.  These integrals are very closely related to those, $d{{\mathcal M}}_{kLdn}$,  evaluated above.  They only lack the term $-s_L$ in the bracket in the integrand that is raised to the \textit{k}${}^{th}$ power.  Thus:
\[d{{\mathcal M}}_{kLup}=s_L\ f_{{\rm exp}Lup}\ \int^{s_L/2\ }_{{\tilde{x}}_{sLup}}{{d\tilde{x}}_{Lup}}\ f_{Lup}\left({\tilde{x}}_{Lup}\right){\ \ \left(f_{{\rm exp}Lup}{\rm \ }{\tilde{x}}_{Lup}+{\sigma }_{MLup}\right)}^k\] 
Because we have done most of the work in evaluating these slightly different integrals already, we find:
\begin{alignat*}{1}  
d{{\mathcal M}}_{0Lup}&=d{{\mathcal M}}_{0Ldn} \komma \qquad          
d{{\mathcal M}}_{1Lup}=d{{\mathcal M}}_{1Ldn}+\ s_L\ d{{\mathcal M}}_{0Lup} \\
d{{\mathcal M}}_{2Lup}&=d{{\mathcal M}}_{2Ldn}+\ s_L\ d{{\mathcal
    M}}_{1Ldn}+\ s_L\ d{{\mathcal M}}_{1Lup}
\end{alignat*} 
There are now 4 cases to consider in evaluating the moments in the cells at the new time level:
\[ \left\langle f{\tilde{x}}^k\right\rangle _N = 
    \begin{dcases*}
             d{{\mathcal M}}_{kT}-\ d{{\mathcal
                 M}}_{kLup}+d{{\mathcal M}}_{kRdn}
             & if $s_L=-1$ and $s_{R\ }=-1$ \\
             d{{\mathcal M}}_{kT}-\ d{{\mathcal
                 M}}_{kLup}-d{{\mathcal M}}_{kRup}
             & if $s_L=-1$ and $s_{R\ }=+1$ \\
             d{{\mathcal M}}_{kT}+\ d{{\mathcal M}}_{kLdn}+d{{\mathcal
                 M}}_{kRdn}
             & if $s_L=+1$ and $s_{R\ }=-1$ \\
             d{{\mathcal M}}_{kT}+\ d{{\mathcal M}}_{kLdn}-d{{\mathcal M}}_{kRup}
             & if $s_L=+1$ and $s_{R\ }=+1$ 
    \end{dcases*}
\]
All these computations are easily implemented in 100\% vectorized
loops making use of vector logic (the \textit{cvmgm} or conditional
move hardware instructions).

We have now described the most difficult of the 6 advection
calculations.  It is easy to see how the moments of the functions
$f_y\left(\tilde{x}\right)$ and $f_y\left(\tilde{x}\right)$, defined
earlier, are computed.  We use the same method that was just described
for the function $f_x\left(\tilde{x}\right)$, but we simply drop the
highest-order terms.  Updating the moments of the functions
$f_{yz}\left(\tilde{x}\right)$, $f_{yy}\left(\tilde{x}\right)$, and
$f_{zz}\left(\tilde{x}\right)$ is simpler still, but we must take care
to handle the dependence of the last 2 of these functions on
$\left\langle f\right\rangle $.

We now discuss the very important topic of how we constrain the
initial functions that we advect as described above.  Because we will
perform 1-D passes in all 3 grid dimensions, there is no need to
constrain any of the functions in the \textit{x}-pass except for
$f_x\left(\tilde{x}\right)$.  Since van Leer introduced the 1-D
unconstrained version of this advection scheme in the mid 1970s, we
have learned a great deal about constraining interpolation parabolae.
The constraints that were introduced for the PPB scheme
\citep{woodward:86} have been augmented and improved over many years
of using PPB advection. The modifications have sought to eliminate the
generation of tiny bits of a fluid that can otherwise precede or
follow a moving region of a given fluid. The modifications also seek
to keep a sharp multifluid interface sharp when it moves perpendicular
to itself while letting the interface spread when that is
appropriate. Behavior of the numerical representation of the interface
between two fluids when that interface is physically unstable is
particularly important. This is a delicate business, and that is why
the algorithm described below is complex.

Constraints that we apply take two general forms.  The first, and the
simplest, is blending with $f_x\left(\tilde{x}\right)$ a fraction of
the constant function $\left\langle  f \right\rangle $, so that
an extremum of the composite function that was negative or exceeded
unity before this blending operation just attains the value 0 or 1
after blending.  We never blend in the constant function $\left\langle
 f\right\rangle $ for any other reason or in any other
circumstances.  This is very important.  The PPB advection scheme is
so accurate that any degradation arising from tampering with the
functions it advects can be devastating, removing essentially all
benefit of its elaborate advection computation.  Therefore we only
blend in constant functions in order to remove unphysical implied
values inside a grid cell.  Because we use PPB to advect multifluid
concentrations, values outside the interval from 0 to 1 must be
eliminated.  However, blending in portions of the constant function is
not the only way, and is often not the best way, to enforce the
specialness of the values 0 and 1.  Nevertheless, if $\left\langle
 f\right\rangle $ is negative or exceeds unity, we must set
the function $f_x\left(\tilde{x}\right)$ to a constant at either the
value 0 or 1 as appropriate.  This is the first constraint that we
apply, and we can see no alternative to it.

Second, we reset edge values that lie outside the allowed range from 0
to 1 (0.000002 to 0.999998 when using 32-bit arithmetic), and we mark
cells where we have done this.  We also reset the edge values to 0 or
1 if in the cell right across this cell interface we have pure fluid,
with $\left\langle f\right\rangle $ equal to 0 or 1, and we mark these
cells along with the previous set.  We need to keep track of these
marked cells, because in these cells and these cells only we will
constrain the implied parabolae in the cells so that they do not
assume minimum or maximum values within the cell.  A difficulty in
treating cells adjacent to pure fluid regions arises from the scheme's
tendency to insert in such cells interpolation parabolae that have
extrema in those cells.  Even if we flatten these parabolae to the
point that these extrema are constrained to lie in the allowed range
of values, we will have directly adjacent to the pure fluid region a
tiny pocket of ever so slightly mixed fluid.  When this tiny pocket
translates, the PPB scheme can begin to treat it with an interpolation
parabola having a permitted extremum within the cell.  In such an
event, the little pocket of mixed fluid will be preserved, which is an
undesired behavior.  We therefore wish to detect such cells and force
the scheme to use monotone interpolation parabolae in them for which,
if possible, the edge value adjacent to the pure fluid region is unity
or zero.  Our detection must be very carefully done, or else we will
mistakenly find such cells all over the grid, apply monotone
interpolations in them, and destroy the resolving power of the PPB
scheme.  Ideally, we would like the scheme to treat physical diffusion
of edges properly, while still keeping non-diffusing edges sharp.
This implies that we would like the scheme to sense the difference
between physical and numerical diffusion, allowing the former and
disallowing the latter.  This is a tall order.  We will assume that
edges, as opposed to truly smooth transitions, are sharper than the
Gaussian curve in their approach to either special value 0 or 1.  This
is a property that we can successfully detect.  We test this by
demanding that our constraints do not introduce sharp jumps
artificially into a Gaussian pulse of height unity and with full width
at half maximum that is 25\% of a periodic domain resolved by between
9 and 27 grid cells when this Gaussian is advected for 100 transits of
this periodic domain at a Courant number of 0.15 or more.  This is not
an empty requirement.  The PPM advection scheme, with its contact
discontinuity and steepening feature enabled, will turn such a
Gaussian pulse into a square wave under these circumstances.  Higher
grid resolution than this (6 cells across the pulse at half maximum)
is needed to distinguish the Gaussian from a square wave using PPM.
Disabling the discontinuity detection and steepening in PPM leaves the
Gaussian a Gaussian, but diminishes its amplitude substantially at low
grid resolutions.  PPB advection, with the constraints laid out below,
easily distinguishes Gaussians from square waves, and vice versa, and
performs the appropriate interpolations for each.  PPB needs only as
few as 6 cells across a square wave pulse to preserve its full
amplitude while advecting it 400 times its width.  Square waves with
widths of only 2 or 3 cells are turned into Gaussian-type pulses by
the PPB scheme.  This seems a more benign behavior than performing the
reverse metamorphosis, so long as the pulse amplitude is adequately
maintained.  For PPB this is the case, due to its subcell resolution
provided through its moment data.  In marginal cases, where grid
resolution is insufficient, we are forced to make a choice between
square and Gaussian-type pulse shapes.  Our constraints will choose
the square pulse shape in cases that cannot be clearly decided from
the data provided, but the subcell information PPB maintains allows us
to apply such arbitrary decisions exclusively on very highly
under-resolved signals.  Also, we arbitrarily favor sharp interpolated
features only when we have transitions to one of the special function
values of 0 or 1.  The design of our constraints below therefore means
that an interface that is originally sharp will remain so unless
sufficiently strong physical diffusion is applied in a single time
step.  We assume that this is a desired behavior.  Originally smooth
multifluid interface transitions will become sharp only if they are
stretched in the dimensions parallel to the transition surface, so
that Liouville's theorem implies that the transition region must
become thinner.  This, we believe, is also a highly desirable
behavior.  We note that our experience with many multifluid
applications of the PPB advection scheme indicates that very few
numerical glitches arise from the scheme's switching between
constrained and unconstrained behaviors \citep[cf.][]{woodward:12a,woodward:10a}, since this occurs only when the
multifluid interfaces involved have numerical representations of about
3 cells in thickness or less.

We will perform our testing and resetting of the left and right
interface values, $f_L$ and $f_R$, in a given cell according to the
following logical tree.  Note that because this resetting of both
values causes a monotone interpolation parabola to result, we can
avoid unnecessary later work in these cells by flagging them as
already handled.  Using subscripts \textit{ZL} and \textit{ZR} to
denote the zone (or grid cell) to the left or right of the cell of
interest, we execute the following logical tree:
\begin{alignat*}{4}  
\mathrm{if}\ \ {\left\langle f\right\rangle }_{ZL} &<0.001 \quad \mathrm{and} &\quad {\left\langle f\right\rangle }_{ZR}&>5\left\langle  f\right\rangle &\qquad \mathrm{then} \quad  f_L&=0 \quad \mathrm{and} &\quad f_R&=3\left\langle f\right\rangle \\
\mathrm{if}\ \ {\left\langle f\right\rangle }_{ZR} &<0.001 \quad\mathrm{and} &\quad {\left\langle f\right\rangle }_{ZL}&>5\left\langle  f\right\rangle   &\qquad \mathrm{then} \quad  f_R&=0 \quad \mathrm{and} &\quad f_L&=3\left\langle f\right\rangle \\
\mathrm{if}\ \ {\left\langle f\right\rangle }_{ZL} &>0.999 \quad\mathrm{and} &\quad {\left\langle f\right\rangle }_{ZR}&<5\left\langle  f\right\rangle -4 &\qquad \mathrm{then} \quad  f_L&=1 \quad \mathrm{and} &\quad f_R&=3\left\langle f\right\rangle -2\\
\mathrm{if}\ \ {\left\langle f\right\rangle }_{ZR} &>0.999 \quad\mathrm{and} &\quad {\left\langle f\right\rangle }_{ZL}&<5\left\langle  f\right\rangle -4 &\qquad \mathrm{then} \quad  f_R&=1 \quad \mathrm{and} &\quad f_L&=3\left\langle f\right\rangle -2\\
\end{alignat*}

We now locate cells containing extrema.  At this point in our sequence
of operations, we are working with the 3 quantities $f_L$, $f_R$, and
$\left\langle f\right\rangle $, which together are sufficient to
define our interpolation parabola.  We now compute the coefficients,
$f_0$, $f_1$ and $f_2$, of the interpolation parabola's terms in
${\tilde{x}}^0$, ${\tilde{x}}^1$, and ${\tilde{x}}^2$:
\[f_1=f_R-f_L \komma \qquad f_2=3\left(f_L+f_R-2\left\langle
    f\right\rangle \right)\quad \mathrm{and}\quad f_0=\left\langle
  f\right\rangle -\ f_2/12 \punkt
\] 
It is easily shown that we must
have a minimum inside the cell if $f_2>\left|f_1\right|$, and that we
must have a maximum inside the cell if $-f_2>\left|f_1\right|$ .  Such
an extremum must occur at ${\tilde{x}}_{ext}=-f_1/\left(2f_2\right)$,
where the parabola assumes the value 
\[f_{ext}=f_0\
-f^2_1/\left(4f_2\right)=\left\langle f\right\rangle -\
\left(f_2/12\right)\left[1+3{\left(f_1/f_2\right)}^2\right]
\punkt
\]
The reduction factor that we apply to $f_1$ and $f_2$ in these cells in
order to bring $f_{ext}$ to the desired, allowed value
${\left\langle f\right\rangle }_{OK}$ is then 
\[f_{reduce}\ =\ 12\
\left(\left\langle f\right\rangle -{\left\langle f\right\rangle
  }_{OK}\right)\ /\ \left(f_2\
  \left[1+3{\left(f_1/f_2\right)}^2\right]\right) \punkt
\] 
We can make this
process of testing computationally efficient, avoiding unnecessary
divisions and evaluations by first computing only $f_1$ and $f_2$ from
our values of $f_L$ and $f_R$, and testing and resetting according to
the following:
\[\mathrm{if} \quad \left|f_2\right|>\left|f_1\right| \quad \mathrm{and}
\quad f^2_2+3f^2_1>12\ f_2\ \left(\left\langle f\right\rangle
  -{\left\langle f\right\rangle }_{OK}\right) \qquad \mathrm{then} 
\qquad f_{reduce} = \frac{12\ f_2\ \left(\left\langle f\right\rangle
    -{\left\langle f\right\rangle
    }_{OK}\right)}{\left(f^2_2+3f^2_1\right)}
\] 
The reduction factor $f_{reduce}$ is applied to $f_1$ and $f_2$ only in these cells.  We can see that we do not actually need to evaluate the coefficient $f_0$.  We note that in the set of cells that we have already marked for the application later of constraints forcing the parabolae within these cells to be monotone, we must take care to assure that $f_{reduce}$ is set to unity.  For these cells, we will keep $f$ inside the range from 0 to 1 in a different fashion.

The idea here is that we reduce the magnitudes of  $f_1$ and $f_2$ together, flattening the interpolation parabola, in cases where our grid cell is likely to be located in between regions where $f$ is either  0  or  1.  Then we are likely to have in our cell a segment of a thin, unresolved strip in which the distribution  $f$ is either  1  or  0  (respectively).  It is entirely appropriate to describe this unresolved strip segment using a parabola that has an extremum inside our grid cell.  We must however be careful not to let the extreme value poke outside of our allowable range.  If, however, we have reset either edge value to one of our limiting allowable values, that is, to either  0  or  1,  then we expect that our cell is located next to a region in which the distribution $f$ is either  0  or  1  (respectively).  In this case, it would be inappropriate to simply flatten our interpolation parabola, which would cause the value at the cell edge to move away from the limiting value shared by the region adjacent to it.  Instead, we would like to keep the edge value at this limit, if that is possible.

In order to prevent the generation of inappropriate subcell extrema in cells located next to regions of roughly constant  $f$ where this constant value is neither  0  nor  1,  we will additionally demand that  $f_{reduce}$  be set to unity unless the cell average,  $\left\langle f\right\rangle $,  is an extremum relative to those on the left and right,  ${\left\langle f\right\rangle }_{ZL}$  and  ${\left\langle f\right\rangle }_{ZR}$,  or the inferred extremum within the grid cell was within the allowable range.   Here again the subscripts \textit{ZL} and \textit{ZR} signify ``zone on the left'' and ``zone on the right.''  Thus we demand that:
\[
\mathrm{if} \quad 
\left(\left\langle f\right\rangle -{\left\langle f\right\rangle
  }_{ZL}\right)\left({\left\langle f\right\rangle }_{ZR}-\left\langle
    f\right\rangle \right)\le 0 \quad  \mathrm{and} \quad
f_2>\left|f_1\right| \qquad \mathrm{then}\qquad f_{reduce}=1
\] 
We add cells reset in this way and where $f_{reduce}$ had previously been set to a value less than unity to the cells we have been marking as candidates for the application of PPM-style constraints that force the parabolae within these cells to be monotone.  We may now safely apply the factor $f_{reduce}$ to the present values of $f_1$ and $f_2$.  We will not need to evaluate $f_0$.

Finally, we apply the constraint that the parabolae in cells that we
have been marking must be monotone inside those cells.  Since our
constraint procedures are so complicated, it is worthwhile at this
point to remember which cells are now marked.  First, all cells for
which an implied edge value, either   $f_L$ or $f_R$, was outside the
allowable range and which, after these edge values were reset, had
extrema inside the cell have been marked.  Second, all cells with
internal extrema outside the allowable range and for which the cell
averages are not corresponding extrema relative to the cell averages
in neighboring cells have been marked.  Thus in all marked cells,  the
parabola describing the subcell structure is not monotone, but we will
demand that it be monotone.  In the marked cells, we therefore apply:
\begin{alignat*}{5}  
\mathrm{if} \enskip  f_L&=0 \enskip &\mathrm{or} \quad f_R&=1 \quad &\mathrm{and} \quad -f_2&>\left|f_1\right| \qquad &\mathrm{then} \qquad &f_2&=\ -f_1&=3\ \left({\left\langle f\right\rangle -f}_R\right)\\ 
\mathrm{if} \enskip  f_L&=1 \enskip &\mathrm{or} \quad f_R&=0 \quad &\mathrm{and} \quad f_2&>\left|f_1\right| \qquad &\mathrm{then} \qquad &f_2&=\ -f_1&=3\ \left({\left\langle f\right\rangle -f}_R\right)\\ 
\mathrm{if} \enskip  f_L&=1 \enskip &\mathrm{or} \quad f_R&=0 \quad &\mathrm{and} \quad -f_2&>\left|f_1\right| \qquad &\mathrm{then} \qquad &f_2&=\ f_1&=3\ \left({\left\langle f\right\rangle -f}_L\right)\\ 
\mathrm{if} \enskip  f_L&=0 \enskip &\mathrm{or} \quad f_R&=1 \quad
&\mathrm{and} \quad f_2&>\left|f_1\right| \qquad &\mathrm{then} \qquad
&f_2&=\ f_1&=3\ \left({\left\langle f\right\rangle -f}_L\right)
\end{alignat*}
At the end of this resetting procedure, we recompute the moments of the distribution via:
\[\left\langle f\tilde{x}\right\rangle =f_1/12\quad \mathrm{and}\quad \left\langle f\tilde{x}\tilde{x}\right\rangle \ =\ \left(\left\langle f\right\rangle +\ f_2/15\right)\ /\ 12\] 
With this final operation, the application of our constraints is complete.

One may compare the above PPB advection scheme with multiple other
variants of the scheme as well as with variations on PPM advection in
both 1-D and 2-D by experimenting with the MS Windows applications
provided on the LCSE Web site at 
\url{http://www.lcse.umn.edu/two-stream-test},
\url{http://www.lcse.umn.edu/Gas1D}, and
\url{http://www.lcse.umn.edu/WindTunnel}.  These Web URLs contain
example applications, executable on any Windows PC, that were
generated in support of courses taught in the 1990s.  The first solves
a 1-D gravitational two-stream instability problem in a 2-D phase
space.  Multiple variations on PPB, several more accurate than the
version we have described above and that we have used for the last
several years, are enabled in this Windows program.  The behavior of
PPB can be compared with that of PPM, in multiple flavors, as well as
a high-order Runge-Kutta advection scheme used in meteorological
codes.  The Gas1D and WindTunnel URLs host Windows applications that
combine PPB multifluid fractional volume advection with PPM gas
dynamics in much the way that they are combined in our present codes.
The WindTunnel code performs a multifluid variation on the classic 2-D
wind tunnel test problem \citet{woodward:84}.  We do not guarantee that
these downloadable applications will work on any PC, although they
might, nor that we will maintain them forever, nor that we will answer
inquiries concerning them.  We also assert that one downloads these
applications at his or her own risk, and they are provided ``as is.''
Nevertheless, they do now work for us, and they are both instructive
and fun to experiment with.  Examples of the use of our PPB scheme for
large-scale multifluid gas dynamics simulations with the PPM code can
be found in
\citep{woodward:12a,woodward:10a,woodward:08a,woodward:08b,herwig:10a}.

\subsubsection{Combining PPB advection with PPM gas
dynamics}   

Once we have computed the time-and-space averaged
velocities at the grid cell interfaces using our version of PPM, we
perform the PPB advection computation to obtain the new values of the 10
moments of $f$  without reference to the values of any other
variables.  This computation gives us advected volumes of the two
fluids, but not advected masses.  We obtain from PPB volumes of the two
gases within the cells \textit{at the beginning of the time step} that
cross the cell interfaces into neighboring cells.  To convert these
advected volumes into advected masses, needed for strict mass
conservation, we must introduce interpolations of the individual fluid
densities as functions of cell volume coordinates at the beginning of
the time step.  We find these interpolation parabolae using our standard
PPM procedure.

Here we invoke an important assumption.  We assume strict pressure and
temperature equilibrium inside each cell.  This implies that the two
fluids must have at each point in the cell a ratio of their densities
that is given by that of their mean molecular weights -- which we assume
is constant over the entire duration of the problem.  In our hydrogen
ingestion flash simulations \citep{herwig:10a}, we take this ratio to be $2.26$.  The
pressure and temperature equilibrium assumption permits us to derive the
individual densities of the two fluids given only the averaged density
of the mixture plus the mixing fraction,  $f$. This is a huge
simplification, because we need not store densities and internal
energies for each fluid.  When we interpolate a parabola to represent
the variation of the density of one of the fluids across a cell, this
implies such a parabola for the other fluid.  Together with our moments
for the distribution of the mixing fraction,  $f$,  we can derive
the implied distribution of density for the mixture.  Because careful
interpolation is very expensive, this represents a great saving in
computational labor.  There is still another advantage.  Our equilibrium
assumption means that even in a cell containing no gas of one type, a
reasonable average density for that gas is implied.  Consequently, we
have no need to interpolate gas density across a discontinuity at a
multifluid boundary:  the density of each gas is well-behaved across
such a boundary.  It is instead the mixing fraction,  \textit{f},  that
jumps suddenly across this boundary.  However, we have for  \textit{f} 
our very much more accurate PPB description, with its 10 moments in each
cell. This allows, as a practical matter, a smooth description of 
\textit{f}  across a multifluid boundary that is only about 2 grid cells
thick.  This representation is prevented from becoming too sharp, which
would introduce numerical oscillations or glitches, because it is forced
to consist of a parabola extending all the way across each grid cell. 
Our interpolated distribution of  \textit{f}  is therefore very sharp --
only a couple of grid cells in thickness -- and at the same time very
smooth, because it is defined by parabolae in these cells that are
determined by subcell information that is operationally equivalent, as a
rule of thumb, to a two- or three-fold grid refinement for PPM
\citep[cf.][]{woodward:05}
for just this single, all-important variable.

In slow-flow, Rayleigh-Taylor instability problems
\citep[cf.][]{woodward:10a,woodward:12a}, we find that the elaborate approach of
the PPB scheme is sufficient to essentially eliminate the appearance
of certain bad behaviors familiar to us from the PPM advection scheme
when applied to multifluid problems \citep[cf.\ for
example][]{bassett:95b,bassett:95a}.  PPB advection, with its formal
fifth-order accuracy, is capable of moving multifluid interfaces with
very detailed structure great distances through the mesh with no
noticeable diffusion.  This behavior is possible, because PPB
consistently treats the internal structure of the multifluid interface
transition from 0 to 1 in the mixing fraction, \textit{f}, as a smooth
transition.  Unlike PPM advection, it does not switch between
fundamentally different interpolation strategies dynamically.
Switching strategies in this way can cause PPM to introduce small
glitches, which can later become amplified by a physical instability
to form large glitches.  We have been using multifluid PPM+PPB for
Rayleigh-Taylor problems in the weakly compressible regime since 2004,
and find it very much superior to PPM alone for these problems, as the
results reported in \citet{woodward:10a,woodward:12a} attest.  \citet{almgren:10}
reported similar experience with high-order advection and
Rayleigh-Taylor problems using their CASTRO code.  This
experience, we find, does not carry over to the much more violent
inertial confinement fusion problems considered in \citet{woodward:12a}.  This can be
seen in 2-D for both our code and for CASTRO in code comparison work
\citep{joggerst:12}.  
We find that the interaction of strong shocks, as
they are handled in PPM, with our very carefully treated multifluid
interfaces produce familiar sorts of glitches, whose causes were
explained decades ago \citep{woodward:84}.  These issues are addressed and largely
resolved by \citet{woodward:12a}, which also describes our multifluid PPM code's highly
scalable parallel implementation in some detail.


\end{document}